\documentclass{BaylorPhD}

\usepackage{amssymb, amsmath, amsfonts, graphicx}

\usepackage{epstopdf}
\usepackage{latexsym}
\usepackage{multirow}
\usepackage[center]{subfigure}
\theoremstyle{plain}
\newtheorem{theorem}{Theorem}

\newtheorem{proposition}[theorem]{Proposition}

\theoremstyle{definition}

\numberwithin{equation}{chapter}
\numberwithin{theorem}{chapter}
\numberwithin{definition}{chapter}
\numberwithin{lemma}{chapter}
\numberwithin{corollary}{chapter}
\numberwithin{prop}{chapter}
\numberwithin{remark}{chapter}
\numberwithin{example}{chapter}
\numberwithin{table}{chapter}

\DeclareMathOperator{\myRe}{Re}

\renewcommand{\Re}{\myRe}

\def\Rem{{\textup{\Re}_\mu}}
\def\Imm{{\textup{Im}_\mu}}
\def\Rem*{{\textup{\Re}_{\mu_*}}}
\def\Imm*{{\textup{Im}_{\mu_*}}}

\title{Aspects of Black Holes in Gravitational Theories with Broken Lorentz and Diffeomorphism Symmetries}
\author{Satheeshkumar V H}
\degrees{B.Sc., M.Sc., M.Phil.} 

\mentor{Anzhong Wang, Ph.D.} 
\reader{Gerald B. Cleaver, Ph.D.} 
\readerThree{Jay R. Dittmann, Ph.D.}
\readerFour{Truell W. Hyde, Ph.D.}
\readerFive{Klaus Kirsten, Ph.D.}

\confDate{May 2015} 

\makeCopyrightPage

\graduateDean{J. Larry Lyon, Ph.D.}
\deptChair{Gregory A. Benesh, Ph.D.}

\abstract{
Since Stephen Hawking discovered that black holes emit thermal radiation, black holes have become the theoretical laboratories for testing our ideas on quantum gravity. This dissertation is devoted to the study of singularities, the formation of black holes by gravitational collapse and the global structure of spacetime. All our investigations are in the context of a recently proposed approach to quantum gravity, which breaks Lorentz and diffeomorphism symmetries at very high energies. 
}

\acknowledgements{
First of all I would like to thank Dr. Anzhong Wang for his mentorship and advice on various facets of research. I am grateful to him for listening to my crazy ideas and answering many of my often embarrassingly simple questions on basic concepts. He gave me enough freedom to explore the subject, which made this pursuit a lot more enjoyable. One thing that has amazed me repeatedly is his magical ability to reduce my ``philosophical'' questions to  mathematical problems. Besides his creativity and productivity, one day I wish to emulate his extraordinary demeanor. 

Dr. Gerald B. Cleaver has been a constant source of encouragement throughout. I took the highest number of courses at Baylor with him. I owe him most of my knowledge of Quantum Field Theory and String Theory. 

I learned a great deal of physics, mathematics and computational techniques from my collaborators Jared Greenwald, Jonatan Lenells and Jian-Xin Lu.

It was my distinct pleasure to collaborate with my friend and former classmate Douglas G. Moore. We spent a lot of time together studying physics, formulated our own problems and solved and published them too. Besides we shared many good laughs and enjoyed thoroughly whatever we did. It was a wonderful learning experience and I count it as one of the highlights of my graduate student life. 

Without the time and patience of Dr. Jay Dittmann, this dissertation would be riddled with spelling and grammatical errors. I sincerely thank him for that. Any surviving typographical mistakes are entirely due to me.

Thanks to my dissertation committee members Drs. Gerald~B.~Cleaver, Jay R. Dittmann, Truell~W.~Hyde and  Klaus~Kirsten for their valuable time.

I am indebted to all my teachers at Baylor, Drs G. B. Cleaver, K. Hatakeyama, J. S. Olafsen, A. Wang,  B. F. L. Ward, Y. Wu and Z. Zhang. 

Thanks to Qiang (Bob) Wu, Yongqing (Steve) Huang, Xinwen Wang, Bao-Fei Li, Tao Zhu, Raziyeh Yousefi, Ahmad Borzou, Bahram Shakerin, Kai Lin, Miao Tian and Ot\'{a}vio Goldoni. I have learned one or the other thing from these people in our group meetings. I also thank Michael Devin for many stimulating discussions.

The HEP Seminar Series has given me a chance to share and learn many good ideas. I thank Drs. Jay Dittmann, Kenichi Hatakeyama and B. F. L. Ward for that.

In the initial stages of my grad school experience, Dr. Walter Wilcox, who was the Graduate Program Director at that time, put up with my visa delays and made sure I was comfortable when I first arrived in the US. He also has given consolation when I needed the most and inspired me to give my best. 

Teaching undergraduate Astronomy and Physics labs has been an invigorating experience which I cherish forever. It would not have been so pleasant without the generous guidance from Mr. Randy Hall, Dr. Linda Kinslow, Dr. Ray Nazzario and Dr. Dwight Russell. I especially thank Dr. Russell for his encouragement in all my endeavors. I spent a good chunk of my graduate life in the Astronomy Lab and I am going to miss it the most when I leave Baylor.

I am grateful for the help I have received from Baylor University Libraries. They obtained every single article I asked for, often papers more than a century old, published in obscure foreign journals. 

My special thanks are due to Mrs. Chava Baker and Mrs. Marian Nunn-Graves for helping me with several administrative issues.

It was gratifying to interact with various researchers in my field at conferences, meetings, workshops and summer schools I attended at the Aspen Center for Physics, Caltech, Institute for Advanced Study at Princeton, Princeton University, Texas A~\&~M University at College Station, University of Mississippi at Oxford, University of Texas at Austin and University of Texas at Dallas. I thank all those organizers for giving me the opportunity.

When I first came to Baylor, it was Martin Frank who helped me to settle down and introduced me to a bunch of people. My social life in the first two years invariably involved BJ Enzweiler and Tim Renner, and was enriched by Cindy Calvert, James Creel, Jonathan Perry, Ashley Palmer, Cassidy Cooper, Chance Cooper, Liz Patterson Tews, Victor Land, Anne Land-Zandstra and John Mosher. 

For the last few years, my social circle has been centered around Liz and Travis Tews. I am going to miss them very dearly. I am also thankful to  Mr. and Mrs. Patterson for all their love and affection. 

Thanks to Yanbin Deng and Xinwen Wang for being there whenever I needed help. They have been great friends and fed me lots of delicious Chinese food.

I am grateful to my mentor from my high school years, Mr.~B.~K.~Muniswamy. He is like family to me and he has taken upon himself a lot of burden on my behalf so that I can concentrate on my research without distraction.  Along with my middle school teacher Mrs.~N.~C.~Yashoda, he shaped me in those formative years. I trace back my interest in Einstein's relativity theories to the eighth grade when he gave me a Kannada translation of the book {\it What Is Relativity?} by L. D. Landau and G. B. Rumer. I also thank Dr. P. Vivekananda Rao, Dr. P. K. Suresh and Dr. R. N. Iyengar for their untiring support.

Although neither my family nor my very close friends understand a single thing I do, they have never failed to let me know how proud they are on the other side of the world. My late mother would have been really happy to see me doing what I always wanted to do. I am thankful for my father V.~H.~Honnappa, my brother V.~H.~Sureshkumar, my in-laws Mrs.~Yashodamma and Mr.~Hanumantharayappa and Shivu. Thanks to my longtime friends David, Manju, Sand, Shanti and Yogesh for always being with me in spirit. Finally, no words can express my happiness ever since Pavi came into my life. She makes everything fun with her presence and gets me through the toughest times with her sense of humor.
}

\dedication{
\begin{center}
to\\the memory of my mother\\
\end{center}

\begin{center}
to\\Pavi -- the poetry of my life\\
\end{center}
}

\begin{document}

\newcommand{\nequation}{\setcounter{equation}{0}}
\renewcommand{\theequation}{\mbox{\arabic{section}.\arabic{equation}}}
\newcommand{\proofbegin}{\noindent{\it Proof.\,\,}}
\newcommand{\proofend}{\hfill$\Box$\bigskip}

\newcommand{\sech}{\text{\upshape \,sech}}
 \newcommand{\bq}{\begin{equation}}
 \newcommand{\eq}{\end{equation}}
 \newcommand{\bqn}{\begin{eqnarray}}
 \newcommand{\eqn}{\end{eqnarray}}
 \newcommand{\nb}{\nonumber}
 \newcommand{\lb}{\label}
\newcommand{\PRL}{Phys. Rev. Lett.}
\newcommand{\PL}{Phys. Lett.}
\newcommand{\PR}{Phys. Rev.}
\newcommand{\CQG}{Class. Quantum Grav.}

\pagenumbering{arabic}

\chapter{Introduction}
\renewcommand{\theequation}{1.\arabic{equation}} \setcounter{equation}{0}

Nature at the most fundamental level is described by four types of interactions, namely the gravitational, electromagnetic, weak and strong interactions. Gravity is best described by the theory of General Relativity, which was proposed in its final form a century ago on November 25, 1915 by Albert Einstein~\cite{Einstein:1915ca}. General Relativity is a classical field theory. The remaining three interactions are all described by quantum field theories. The theory of Quantum Electrodynamics was developed in the late 1940s and early 1950s, largely by Sin-itiro Tomonaga, Julian Schwinger and Richard Feynman~\cite{Schwinger:1958}.   The theory of electro-weak interactions was put together in the 1960s by Sheldon Glashow~\cite{Glashow:1961tr}, Steven Weinberg~\cite{Weinberg:1967tq} and Abdus Salam~\cite{Salam:1968rm}. The theory of strong interactions known as Quantum Chromodynamics was discovered in 1973 by David Gross and Frank Wilczek~\cite{Gross:1973id}, and independently by David Politzer~\cite{Politzer:1973fx}. All these theories are experimentally well tested. But the unsatisfactory fact is that gravity is described by a classical theory while the rest of the fundamental interactions are described by quantum theories, hence the fundamental laws of nature do not constitute one logically consistent system. Despite the fact that this disparity does not affect physics as long as one stays safely below the Planck scale, one is forced into such a high energy regime when studying the very early universe or the final stages of star collapse. This requires a quantum field theory of gravity and perhaps a unification of all the fundamental interactions. The latter goal is pursued by String Theory, while there are quite a few different approaches to quantize gravity.  

\section{Problems of Quantizing Gravity}

Soon after Quantum Field Theory was invented and applied to the electromagnetic field by Werner Heisenberg and Wolfgang Pauli~\cite{HeisenbergPauli}, it was realized that perturbative computation of the electromagnetic self-energy for an electron led to infinities. Provoked by Pauli, who thought  the infinities could be eliminated if gravitational effects were included, L\'{e}on Rosenfeld applied Quantum Field Theory to study the gravitational field in 1930~\cite{Rosenfeld}. He showed that the gravitational self-energy of a photon is quadratically divergent in the lowest order of perturbation theory. Later in 1950, using a manifestly gauge-invariant method due to Julian Schwinger, Bryce DeWitt~\cite{DeWittThesis} showed that this result implied merely a renormalization of electric charge rather than a non-zero mass of photon and it also implied that the one-loop contribution is zero.

\subsection{Unitarity}

In 1962, while studying the one-loop behavior of General Relativity quantized in a covariant gauge, Richard Feynman noticed that the diagrams are unitary only if fictitious particles are added besides the graviton ~\cite{Feynman:1963ax}. They are fictitious because they were anticommuting bosonic particles circulating in the loop. Such nonphysical fields that are introduced in a manifestly Lorentz covariant quantization procedure to maintain unitarity at one-loop order of perturbation theory are usually referred to as ``ghost'' fields.  DeWitt extended this to the two-loop case in 1964~\cite{DeWitt:1964yg} and to all orders in 1966~\cite{DeWitt:1967yk, DeWitt:1967ub, DeWitt:1967uc}. Ghosts were independently introduced  by Ludwig D. Faddeev and Victor N. Popov in 1967~\cite{Faddeev:1967fc}, who gave a prescription for quantizing Yang-Mills gauge theories in general gauges.

\subsection{Renormalizability}

Even early on, Heisenberg noticed that the gravitational coupling constant $G$ has the dimension of a negative power of mass and consequently the divergences in the perturbation expansion of General Relativity will be different from renormalizable quantum field theories.

In 1974, 't Hooft and Veltman  investigated all the one-loop divergences in General Relativity. They discovered no physically relevant divergences in the one-loop \mbox{S-matrix}. However, the finiteness of pure gravity  was destroyed by coupling a single scalar to it~\cite{'tHooft:1974bx}. Soon after that, essentially all of the known matter couplings were investigated and it was found that none of the matter fields would share the one-loop finiteness of pure gravity. In 1986, Marc Goroff and Augusto Sagnotti made it clear that our conventional quantization techniques do not work for gravity~\cite{Goroff:1985th}.

Non-renormalizability means simply that the theory is effective and its predictions can be trusted only at sufficiently low energies. For example, the Schr\"{o}dinger equation is non-renormalizable, yet its predictions are experimentally found to be correct. The same is true for the Four-Fermi Theory of weak interactions. Also, theories that are perturbatively non-renormalizable can be non-perturbatively renormalizable. If there is any hope at all for the renormalizability of General Relativity, that should come from a non-perturbative approach.

\section{Semi-classical Approach}

At the one-loop level, the contribution of the gravitational loop is of the same order as the contributions of matter fields. At usual energies less than the $10^{16}\,$GeV scale, the contributions of additional gravitational loops are highly suppressed. Therefore, a semi-classical concept applies when the quantum matter fields together with the linearized perturbations of the gravitational field interact with the background gravitational field~\cite{Kibble:1980ia, Duff:1980ix, Moore:2013sra}.

Not withstanding the difficulties of quantizing gravity, Jacob Bekenstein  and Stephen Hawking took up semi-classical studies of black holes in the early 1970's. In such an approach, the gravitational field is treated as a classical background on which quantum fields are studied.  It was shown that black holes behave like black bodies and emit thermal radiation~\cite{Hawking:1974rv, Hawking:1974sw}. Hence one can assign temperature and entropy to a black hole. In fact, they showed entropy is proportional to the area of the event horizon~\cite{Bekenstein:1973ur, Bekenstein:1974ax}. This started the modern era of quantum gravity.

The semi-classical theory of the early universe called Inflationary theory~\cite{Guth:1980zm} has been very successful too. While black holes are the theoretical laboratories for quantum gravity, Inflation is the experimental laboratory that might give us the first glimpses of quantum gravity~\cite{Krauss:2013pha, Zhu:2014wda}. The recent observational constraints~\cite{Ade:2015tva} on the inflationary gravitational waves imprinted in the cosmic microwave background are strengthening our hopes that experimental quantum gravity is in sight.






\section{Common Ground}

In the absence of experimental data, it is pragmatic to see what common features are shared by different approaches to solve the problems in quantum gravity. Today, reproducing the black hole entropy formula has become a standard test for any theory of quantum gravity. String Theory has successfully reproduced these results, although only in certain extremal black hole cases~\cite{Strominger:1996sh}. In Loop Quantum Gravity, the black hole entropy has been computed for the Schwarzschild black hole, except for a factor of one fourth in the formula, which can be obtained with an appropriate choice of a free parameter of the theory~\cite{Ashtekar:1997yu}.  It is often argued that another common feature shared by theories of quantum gravity is the similar fractal behavior~\cite{Carlip:2012md}. It is shown that many, if not all, approaches to quantum gravity predict a spectral dimension of 2 in ultraviolet regime. However, it has recently been shown that Bosonic String Theory  behaves differently~\cite{Moore:2014kua}. One has to investigate such a spectral behavior of Superstring Theory before drawing any final conclusions.







\section{Current Status of Quantum Gravity}

To the best of our knowledge today, one cannot have a quantum field theoretic description of General Relativity without an idea from outside the conventional framework. One such idea that has been pursued for many decades is supersymmetry. In 1983, it was established that ${\cal N} = 4$ super-Yang-Mills theory in four dimensions is ultraviolet finite to all-loop orders~\cite{Mandelstam}. The supersymmetric version of General Relativity called supergravity showed improved ultraviolet behavior in the early days although it was thought to diverge in the third loop. Now, contrary to all the power-counting arguments, it is shown that ${\cal N} = 8$ supergravity in four dimensions is finite to seventh-loop order in perturbation theory~\cite{Bern:2014sna}. That is a remarkable achievement.

Another area of active research is String Theory whose most fundamental result to date is that under certain caveats, gravity is dual to certain conformal field theories~\cite{Maldacena:1997re}. This result might contain some deeper clues to quantum gravity. 

The latest progress comes from a new way of describing gravity within the framework of quantum field theory by breaking Lorentz and diffeomorphism symmetries. This was proposed by Petr Ho\v{r}ava in 2009~\cite{Horava:2009uw}. In the relatively short span of its existence, it has been shown that this theory is a part of String Theory and at the same time makes contact with other seemingly different approaches to quantum gravity. 

Despite making enormous progress in our understanding of the high energy behavior of gravity, our quest for quantum gravity is far from over.


\chapter{Gravitational Theories with Anisotropic Scaling}
\renewcommand{\theequation}{2.\arabic{equation}} \setcounter{equation}{0}

\section{Introduction}

In the previous chapter, we listed the core problems in quantizing gravity and argued that in order to achieve this feat, we need ideas from outside the conventional framework. The investigations in this dissertation are concerned with the latest of such ideas, inspired by anisotropic scaling in the theories of condensed matter physics. Before diving into the details of gravitational theories with anisotropic scaling, we discuss some important ingredients of the theory and why they are essential.

\subsection{Higher-order Terms}

The non-renormalizability of General Relativity means that it is an effective theory and the Einstein-Hilbert action contains only the terms relevant at low energies. One is naturally tempted to add higher-order curvature terms to the action, thereby making the theory applicable at high energies. This possibility was first explored in 1962 by Ryoyu Utiyama and Bryce S. DeWitt~\cite{Utiyama:1962sn}. They noticed that the action of quantum gravity should contain functionals of higher derivatives of the metric tensor besides the Einstein-Hilbert action. But is such a theory renormalizable? This question was answered affirmatively in 1977 by Kellogg Stelle~\cite{Stelle:1976gc}. He showed that the theory is renormalizable with quadratic curvature invariants. However due to the presence of higher time derivatives, such a theory has the negative norm state called ``ghosts'' which allow the probability to be negative and hence breaks  unitarity. In fact, as far back as in 1850, Mikhail Ostrogradsky showed that the presence of time derivatives higher than two generally leads to the problem of ghosts~\cite{Ostrogradsky}. Combining all these ideas, Ho\v{r}ava added only the terms containing higher spacial derivatives while keeping the time derivatives to the second order. Also, power-counting renormalizability restricts the number of spacial derivatives to at least six. This means that space and time are treated differently and hence the local Lorentz and diffeomorphism symmetries are broken.

\subsection{Lorentz Invariance}

Invariance under the Lorentz symmetry group is a cornerstone of modern physics, and is strongly supported by observations. In fact, all the experiments carried out so far are consistent with it ~\cite{Liberati13}, and no evidence shows that such a symmetry must be  broken at certain energy scales, although  it is arguable that such constraints in  the gravitational sector are much weaker than those in the matter sector   ~\cite{LZbreaking}. It should be emphasized  that the breaking of Lorentz invariance can have significant effects on low-energy physics  through the interactions between gravity and matter, no matter how high the scale of symmetry breaking is~\cite{Collin04, Pola}.  The question of how to prevent the propagation of the Lorentz violations into the Standard Model of particle physics remains challenging~\cite{PS}.  Nevertheless, it is known that demanding  renormalizability and unitarity of gravity often lead to the violation of Lorentz invariance~\cite{CarlipBook}. There are benefits of violating Lorentz invariance~\cite{Visser:2009fg}. Recently, a mechanism of supersymmetry breaking by coupling a Lorentz-invariant supersymmetric matter sector to non-supersymmetric gravitational interactions with Lifshitz scaling was shown to lead to a consistent HL gravity~\cite{PT14}.

\subsection{Diffeomorphism Invariance}

In many theories of quantum gravity and especially in String Theory, it is believed that the concept of space and time is an emergent notion that is not present in the fundamental theory~\cite{Seiberg:2006wf}. Another way of looking at it is that one does not have to respect the symmetries of spacetime that are thought to be fundamental at known energies, such as the local Lorentz and diffeomorphism symmetries, especially given the fact that our understanding of spacetime at the Planck scale is highly limited.

\subsection{Ultraviolet Fixed Point}

We know from quantum field theory that the  value of coupling strengths of fundamental interactions depend on the energy scale at which they are measured. This variation of coupling strengths with energy scale is called the running of the couplings. This has been experimentally seen for the Standard Model interactions. The essential tool for studying this is renormalization group flow, developed by Kenneth G. Wilson~\cite{Wilson:1973jj}, which is a sophisticated version of the general theory of thermodynamic phase transitions due to Lev Landau from 1937~\cite{Landau}. This is closely related to critical phenomena in statistical mechanics~\cite{Zinn-Justin}, originally developed by E. M. Lifshitz in 1941~\cite{Lifshitz}. A quantum field theory is said to have an ultraviolet fixed point if its renormalization group flow approaches a fixed point at very high energies. Theories with the trivial or Gaussian ultraviolet fixed point are said to be asymptotic free theories. They may be, but not necessarily are, strongly-coupled theories whose study require nonperturbative methods. The famous example is QCD. Ho\v{r}ava-Lifshitz Theory is also thought to be of this kind. Theories with a nontrival or non-Gaussian fixed point are said to be asymptotic safe theories. A notable example is asymptotic safety in quantum gravity proposed by Steven Weinberg~\cite{Weinberg:1980gg}.

\section{Ho\v{r}ava-Lifshitz Theory}

\subsection{Original Version}

Ho\v{r}ava achieved all the above mentioned objectives in 2009~\cite{Horava:2009uw} by borrowing from condensed matter physics the idea of anisotropic scalings~\cite{Lifshitz} of space and time, given by 
\bq
\lb{1.1}
{\bf x} \rightarrow \ell {\bf x}, \;\;\;  t \rightarrow \ell^{z} t,
\eq
where $z$ is called the dynamical  critical exponent.  This was first used to construct scalar field theory by E. M. Lifshitz, hence the theory is often referred to as the Ho\v{r}ava-Lifshitz (HL)  Theory. The anisotropic scaling  provides  a crucial mechanism: the gravitational action can be constructed in such a way that only  higher-dimensional spatial, but not time, derivative operators are included, so that the UV behavior of the theory  is dramatically improved. In $(3+1)$-dimensional spacetimes, HL theory is power-counting  renormalizable provided that  $z \ge 3$~\cite{Horava:2009uw, Visser:2009ys, Fujimori:2015wda}.   In this dissertation, we will assume that $z =3$. At low energies, the theory is expected to flow  to \mbox{$z=1$}. In this limit the Lorentz invariance is ``accidentally restored." The exclusion of high-dimensional time derivative operators,  on the other hand,   prevents the ghost instability, whereby the unitarity of the theory is assured. 

The anisotropy between time and space mentioned above is conveniently expressed in terms of the Arnowitt-Deser-Misner (ADM)~\cite{Arnowitt:1962hi} variables: the lapse function $N$, the shift vector $N^{i}$ and the three-dimensional metric defined on the leaves of constant time $ g_{ij}$  where $i, \; j = 1, 2, 3$.
In the metric form, it is given by
 \bqn
 \lb{1.2}
ds^{2} &=& - N^{2}c^{2}dt^{2} + g_{ij}\left(dx^{i} + N^{i}dt\right)
     \left(dx^{j} + N^{j}dt\right), ~~~~~~~~  (i, \; j = 1, 2, 3).
 \eqn
 Under the rescaling (\ref{1.1})   with   $z  = 3$,   $N, \; N^{i}$ and $g_{ij}$ scale  as 
 \bq
 \lb{1.3}
  N \rightarrow  N ,\;\;\;  N^{i}
\rightarrow {\ell}^{-2} N^{i},\;\;\; g_{ij} \rightarrow g_{ij}.
 \eq

The requirement that the foliation defined by these leaves be preserved by any gauge symmetry implies that the theory is covariant only under the action of the group Diff($M, \; {\cal{F}}$) of foliation-preserving diffeomorphisms,  
\bq
\lb{1.4res}
\delta{t} =  - f(t),\; \;\; \delta{x}^{i}  =   - \zeta^{i}(t, {\bf x}),
\eq
for which $N, \; N^{i}$ and $g_{ij}$ transform as
\bqn
\lb{1.5}
\delta{g}_{ij} &=& \nabla_{i}\zeta_{j} + \nabla_{j}\zeta_{i} + f\dot{g}_{ij},\nb\\
\delta{N}_{i} &=& N_{k}\nabla_{i}\zeta^{k} + \zeta^{k}\nabla_{k}N_{i}  + g_{ik}\dot{\zeta}^{k}
+ \dot{N}_{i}f + N_{i}\dot{f}, \nb\\
\delta{N} &=& \zeta^{k}\nabla_{k}N + \dot{N}f + N\dot{f},
\eqn
where $\dot{f} \equiv df/dt$,  $\nabla_{i}$ denotes the covariant  derivative with respect to the 3-metric $g_{ij}$, and  $N_{i} = g_{ik}N^{k}$, etc. From these expressions one can see that the lapse function $N$ and the shift vector $N_{i}$ play the role of gauge fields of the Diff($M, \; {\cal{F}}$) symmetry. Therefore, it is natural to assume that $N$ and $N_{i}$ inherit the same dependence on spacetime as the corresponding generators
\bq
\lb{1.6}
N = N(t), \;\;\; N_{i} = N_{i}(t, x),  
\eq
which is   clearly preserved by Diff($M, \; {\cal{F}}$) and usually referred to as the projectability condition.

The non-projectable version of HL gravity~\cite{Blas:2009qj} is  self-consistent and passes all the solar system, astrophysical  and cosmological  tests. In fact, in the infrared the theory can be identified with the hypersurface-orthogonal Einstein-aether theory~\cite{EA} in a particular gauge~\cite{Jacob, Wang:2012nv, LACW}, whereby the consistence of the theory with observations  can be deduced.

\subsubsection{Problems.}  Due to the restricted diffeomorphisms, one more degree of freedom appears in the gravitational sector, i.e.,  the spin-0 graviton. This is potentially dangerous, and needs to decouple  in the infrared in order to be consistent with observations~\cite{Horava:2011gd}. In particular,  the spin-0 mode is not stable in the Minkowski background  in the original version of the HL theory ~\cite{Horava:2009uw} and in the Sotiriou, Visser  and Weinfurtner (SVW) generalization~\cite{SVW,WM}. However, in the SVW setup, it is stable in the {de Sitter} background  ~\cite{HWW}. In addition, non-perturbative analysis showed that  it indeed decouples in the vacuum spherical static ~\cite{Mukreview} and  cosmological~\cite{WWb} spacetimes.  

Another potential complication of the HL theory is that the theory becomes strongly coupled when the energy is very low~\cite{Blas:2009yd}. However, as long as the theory is consistent with observations when the nonlinear effects are taken into account, this is not necessarily a problem, at least not classically. A careful analysis shows that the theory is consistent with observations in the vacuum spherically symmetry static  case~\cite{Mukreview} and in the cosmological setting~\cite{IM, Wang:2009rw, GMW}.

\subsection{Extended Symmetry}

One way to overcome the above problems is to introduce an extra local $U(1)$ symmetry, so that the total symmetry of the theory is extended to ~\cite{HMT}
\bq
\lb{symmetry}
 U(1) \ltimes {\mbox{Diff}}(M, \; {\cal{F}}).
\eq 
This is achieved by introducing a gauge field $A$ and a Newtonian prepotential $\varphi$.
One consequence of the $U(1)$ symmetry is that the spin-0 gravitons are eliminated 
\cite{HMT,WWb}.  
Effectively, the spatial diffeomorphism symmetries of General Relativity are kept intact, but  the
 time reparametrization symmetry   is contracted to a local gauge symmetry~\cite{TH}. The restoration of general covariance, characterized by  Eq.~(\ref{symmetry}),  nicely maintains the special status of time,  so that the anisotropic scaling (\ref{1.1}) with $z > 1$ can still be  realized. As a result, all problems related to them, such as  instability, strong coupling, and different propagation speeds in the gravitational sector, are resolved.

 Under the Diff($M, \; {\cal{F}}$), the fields $A$ and $\varphi$ transform as
\bqn
\lb{2.2}
\delta{A} &=& \zeta^{i}\partial_{i}A + \dot{f}A  + f\dot{A},\nb\\
\delta \varphi &=&  f \dot{\varphi} + \zeta^{i}\partial_{i}\varphi,
\eqn
while under the local $U(1)$,  the gauge fields together with $N$, $N_i$ and $g_{ij}$  transform as
\bqn
\lb{2.3}
\delta_{\alpha}A &=&\dot{\alpha} - N^{i}\nabla_{i}\alpha,\nb \\
\delta_{\alpha}\varphi &=& - \alpha,\nb\\ 
\delta_{\alpha}{N} &=& 0,\nb\\
\delta_{\alpha}N_{i} &=& N\nabla_{i}\alpha,\nb\\
\delta_{\alpha}g_{ij} &=& 0, 
\eqn
where $\alpha$ is   the generator  of the local $U(1)$ gauge symmetry.

A remarkable by-product of this ``non-relativistic general covariant" setup is that it forces the coupling constant $\lambda$, introduced originally  to characterize the deviation of the kinetic part of the action from General Relativity~\cite{Horava:2009uw},
to take exactly  its relativistic value $\lambda = 1$. A different view   can be found in~\cite{daSilva:2010bm}. The $U(1)$ symmetry was initially introduced in the case of $\lambda = 1$, but the formalism was soon extended to the case of any $\lambda$~\cite{daSilva:2010bm,HW,LWWZ}.
 In the presence of a $U(1)$ symmetry, the consistency of the HL theory with solar system tests and cosmology was systematically studied in~\cite{GSW,AP, cosmo}. In particular, it was shown in~\cite{LMW} that, in order for the theory to be consistent with solar system tests, the gauge field $A$ and the Newtonian prepotential $\varphi$ must be part of the metric in the infrared limit. This ensures that the line element $ds^2$ is a scalar not only under $ {\mbox{Diff}}(M, \; {\cal{F}})$ but also under the local $U(1)$ symmetry.  
In this model, the formation of black holes by gravitational collapse was first studied in~\cite{Greenwald:2013kja}.


Another possibility is to give up the projectability condition (\ref{1.6}). This allows for new operators to be included in the action, in particular, operators involving $a_i \equiv N_{,i}/N$. In this way, all the problems mentioned above can be avoided by properly choosing the coupling constants.  However, since this leads to a theory with more than 70 independent coupling constants~\cite{KP}, it makes the predictive power of the theory questionable, although only five coupling constants are relevant in the infrared. 

A non-trivial generalization of the enlarged symmetry (\ref{symmetry}) to the nonprojectable, $N = N(t, x)$, case was discovered by Zhu, Wu, Wang and Shu~\cite{ZWWS,ZSWW}. It was shown that, as in General Relativity, the only degree of freedom of the model in the gravitational sector is the spin-2 massless graviton. Moreover, thanks to the elimination of the spin-0 gravitons, the physically viable range for the coupling constants is considerably enlarged, in comparison  with the original non-projectable version, where the extra $U(1)$ symmetry is absent.    Furthermore, the number of independent coupling constants is dramatically reduced from more than 70 to 15.  The consistency of the model with cosmology was recently established in~\cite{ZSWW,Zhu:2012zk,WWZZ}. In the case with spherical symmetry, the model was shown to be consistent with solar system tests ~\cite{LW}. In contrast to the projectable case, the consistency can be achieved without taking the gauge field $A$ and  Newtonian prepotential $\varphi$ to be part of the metric. Finally, the duality between this version of the HL theory and a non-relativistic quantum field theory was analyzed in~\cite{Janiszewski:2012nb},  and its embedding  in String Theory was constructed in~\cite{Janiszewski:2012nf}. For other examples see~\cite{Jeong:2009aa}.   

It is remarkable to note that, despite of the stringent observational  constraints on the Lorentz violation~\cite{Liberati13}, the non-relativistic  general covariant  Ho\v{r}ava-Lifshitz gravity constructed in~\cite{ZWWS} is consistent with all the solar system tests~\cite{LMWZ, Will} and cosmology ~\cite{cosmo,ZHW}.  

\begin{table*}[htdp]
\vspace*{.4cm}
\begin{center}
\label{default0}
\caption{A summary of various versions of the Ho\v{r}ava-Lifshitz Theory}
\begin{tabular}{lccccc}
\hline
\hline
\multirow{ 2}{*}{Property of the theory} & \multicolumn{2}{ c }{${\mbox{Diff}}(M,{\cal{F}})$} & & \multicolumn{2}{ c }{$U(1)\ltimes{\mbox{Diff}}(M,{\cal{F}})$}\\
\cline{2-3}
\cline{5-6}
 & $N=N(t)$ & $N=N(t,x)$ & & $N=N(t)$ & $N=N(t,x)$\\
\hline
Independent Couplings & 11 & $>$75 & & 11 & 15\\
No Spin-0 Graviton & \text{\sffamily X} & \text{\sffamily X} & & $\checkmark$ & $\checkmark$ \\
No instability of $M_4$ & \text{\sffamily X} & $\checkmark$ &  & $\checkmark$ & $\checkmark$ \\
No Strong Coupling &  \text{\sffamily X} & $\checkmark$ & & $\checkmark$ & $\checkmark$\\
Solar System & $\checkmark$ & $\checkmark$ & & $\checkmark$ & $\checkmark$\\
Cosmology & $\checkmark$ & $\checkmark$ & & \text{\sffamily ?} & $\checkmark$\\
String Theory Embedding & \text{\sffamily ?} & \text{\sffamily ?} & & \text{\sffamily ?} & $\checkmark$\\
\hline \hline
\end{tabular}
\end{center}
\vspace*{-.5cm}
\end{table*}

\section{Non-relativistic General Covariant HL Theory}

In this section, we shall  give a very brief introduction to the non-relativistic general covariant theory of gravity with projectability for  an arbitrary coupling constant $\lambda$ and the enlarged symmetry (\ref{symmetry}). All our discussions in this dissertation are based on this version of the theory.
  
The total action is given by
 \bqn \lb{2.4}
S &=& \zeta^2\int dt\, d^{3}x\, N \sqrt{g} \Big({\cal{L}}_{K} -
{\cal{L}}_{{V}} +  {\cal{L}}_{{\varphi}} +  {\cal{L}}_{{A}} +  {\cal{L}}_{{\lambda}} 
\left. + {\zeta^{-2}} {\cal{L}}_{M} \right),
 \eqn
where $g={\rm det}\,g_{ij}$, and
 \bqn \lb{2.5}
{\cal{L}}_{K} &=& K_{ij}K^{ij} -   \lambda K^{2},\nb\\
{\cal{L}}_{\varphi} &=&\varphi {\cal{G}}^{ij} \Big(2K_{ij} + \nabla_{i}\nabla_{j}\varphi\Big),\nb\\
{\cal{L}}_{A} &=&\frac{A}{N}\Big(2\Lambda_{g} - R\Big),\nb\\
{\cal{L}}_{\lambda} &=& \big(1-\lambda\big)\Big[\big(\nabla^{2}\varphi\big)^{2} + 2 K \nabla^{2}\varphi\Big].
 \eqn
Here   the coupling constant $\Lambda_{g}$, which acts like a three-dimensional cosmological
constant, has the dimension of (length)$^{-2}$. The 
Ricci and Riemann terms all refer to the three-metric $g_{ij}$.
 $K_{ij}$ is the extrinsic curvature, and ${\cal{G}}_{ij}$ is the 3-dimensional ``generalized"
Einstein tensor defined  by
 \bqn \lb{2.6}
K_{ij} &=& \frac{1}{2N}\left(- \dot{g}_{ij} + \nabla_{i}N_{j} +
\nabla_{j}N_{i}\right),\nb\\
{\cal{G}}_{ij} &=& R_{ij} - \frac{1}{2}g_{ij}R + \Lambda_{g} g_{ij}.
 \eqn
${\cal{L}}_{M}$ is the
matter Lagrangian density and  
${\cal{L}}_{{V}}$ denotes the potential part of the action given by
 \bqn \lb{2.5a} 
{\cal{L}}_{{V}} &=& \zeta^{2}g_{0}  + g_{1} R + \frac{1}{\zeta^{2}}
\left(g_{2}R^{2} +  g_{3}  R_{ij}R^{ij}\right)\nb\\
& & + \frac{1}{\zeta^{4}} \left(g_{4}R^{3} +  g_{5}  R\;
R_{ij}R^{ij}
+   g_{6}  R^{i}_{j} R^{j}_{k} R^{k}_{i} \right)\nb\\
& & + \frac{1}{\zeta^{4}} \left[g_{7}R\nabla^{2}R +  g_{8}
\left(\nabla_{i}R_{jk}\right)
\left(\nabla^{i}R^{jk}\right)\right],  ~~~~
 \eqn 
which preserves the parity,  where the coupling  constants $ g_{s}\, (s=0, 1, 2,\dots, 8)$  are all dimensionless. The relativistic limit in the IR
 requires that
 \bq
 \lb{2.5b}
 g_{1} = -1,\;\;\; \zeta^2 = \frac{1}{16\pi G},
 \eq
where $G$ denotes the Newtonian constant.

Variation of the total action (\ref{2.4}) with respect to the lapse function $N(t)$  yields the
Hamiltonian constraint
 \bqn \lb{eq1}
 \int{ d^{3}x\sqrt{g}\left[{\cal{L}}_{K} + {\cal{L}}_{{V}} - \varphi {\cal{G}}^{ij}\nabla_{i}\nabla_{j}\varphi 
- \big(1-\lambda\big)\big(\nabla^{2}\varphi\big)^{2}\right]}
= 8\pi G \int d^{3}x {\sqrt{g}\, J^{t}},
 \eqn
where
 \bq \lb{eq1a}
J^{t} = 2 \frac{\delta\left(N{\cal{L}}_{M}\right)}{\delta N}.
 \eq
 
Variation of the action with respect to the shift $N^{i}$ yields the
super-momentum constraint
 \bqn \lb{eq2}
& & \nabla^{j}\Big[\pi_{ij} - \varphi  {\cal{G}}_{ij} - \big(1-\lambda\big)g_{ij}\nabla^{2}\varphi \Big] = 8\pi G J_{i},
 \eqn
where the super-momentum $\pi^{ij} $ and matter current $J^{i}$
are defined as
 \bqn
 \lb{eq2a}
\pi^{ij} & \equiv &  - K^{ij} + \lambda K g^{ij},\nb\\
J^{i} & \equiv & - N\frac{\delta{\cal{L}}_{M}}{\delta N_{i}}.
 \eqn
Similarly, variations of the action with respect to $\varphi$ and $A$ yield, respectively, 
\bqn
\lb{eq4a}
 {\cal{G}}^{ij} \Big(K_{ij} + \nabla_{i}\nabla_{j}\varphi\Big) + \big(1-\lambda\big)\nabla^{2}\Big(K + \nabla^{2}\varphi\Big)
 &=& 8\pi G J_{\varphi}, \\
\lb{eq4b}
 R - 2\Lambda_{g} &=&    8\pi G J_{A},
\eqn
where
\bq
\lb{eq5}
J_{\varphi} \equiv - \frac{\delta{\cal{L}}_{M}}{\delta\varphi},\;\;\;
J_{A} \equiv 2 \frac{\delta\left(N{\cal{L}}_{M}\right)}{\delta{A}}.
\eq
On the other hand, variation with respect to $g_{ij}$ leads to the
dynamical equations
 \bqn \lb{eq3}
&&
\frac{1}{N\sqrt{g}}\Bigg\{\sqrt{g}\Big[\pi^{ij} - \varphi {\cal{G}}^{ij} - \big(1-\lambda\big) g^{ij} \nabla^{2}\varphi\Big]\Bigg\}_{,t} 
 = -2\left(K^{2}\right)^{ij}+2\lambda K K^{ij} \nb\\
& &  ~~~~~ + \frac{1}{N}\nabla_{k}\left[N^k \pi^{ij}-2\pi^{k(i}N^{j)}\right] 
- 2\big(1-\lambda\big) \Big[\big(K + \nabla^{2}\varphi\big)\nabla^{i}\nabla^{j}\varphi + K^{ij}\nabla^{2}\varphi\Big]\nb\\
& & ~~~~~ + \big(1-\lambda\big) \Big[2\nabla^{(i}F^{j)}_{\varphi} - g^{ij}\nabla_{k}F^{k}_{\varphi}\Big] 
 +  \frac{1}{2} \Big({\cal{L}}_{K} + {\cal{L}}_{\varphi} + {\cal{L}}_{A} + {\cal{L}}_{\lambda}\Big) g^{ij} \nb\\
& &  ~~~~~    + F^{ij} + F_{\varphi}^{ij} +  F_{A}^{ij} + 8\pi G \tau^{ij},
 \eqn
where $\left(K^{2}\right)^{ij} \equiv K^{il}K_{l}^{j},\; f_{(ij)}
\equiv \left(f_{ij} + f_{ji}\right)/2$, and
 \bqn
\lb{eq3a} 
F_{A}^{ij} &=& \frac{1}{N}\left[AR^{ij} - \Big(\nabla^{i}\nabla^{j} - g^{ij}\nabla^{2}\Big)A\right],\nb\\ 
F_{\varphi}^{ij} &=&  \sum^{3}_{n=1}{F_{(\varphi, n)}^{ij}},\nb\\
F^{ij} &\equiv& \frac{1}{\sqrt{g}}\frac{\delta\left(-\sqrt{g} {\cal{L}}_{V}\right)}{\delta{g}_{ij}}
 = \sum^{8}_{s=0}{g_{s} \zeta^{n_{s}} \left(F_{s}\right)^{ij} }, \qquad
 \eqn
with 
$n_{s} =(2, 0, -2, -2, -4, -4, -4, -4,-4)$.  The  3-tensors $ \left(F_{s}\right)_{ij}$ and 
$F_{(\varphi, n)}^{ij}$ are given below.
  \bqn 
(F_{0})_{ij}&=& -\frac12g_{ij},\nb\\
(F_{1})_{ij}&=&-\frac12g_{ij}R+R_{ij},\nb\\
(F_{2})_{ij} &=&-\frac12g_{ij}R^2+2RR_{ij}-2\nabla_{(i}\nabla_{j)}R 
 +2g_{ij}\nabla^2R,\nb\\
(F_{3})_{ij}&=&-\frac12g_{ij}R_{mn}R^{mn}+2R_{ik}R^k_j-2\nabla^k\nabla_{(i}R_{j)k} 
 +\nabla^2R_{ij}+g_{ij}\nabla_m\nabla_nR^{mn},\nb\\
(F_{4})_{ij}&=&-\frac12g_{ij}R^3+3R^2R_{ij}-3\nabla_{(i}\nabla_{j)}R^2 
 +3g_{ij}\nabla^2R^2,\nb\\
(F_{5})_{ij}&=&-\frac12g_{ij}RR^{mn}R_{mn}+R_{ij}R^{mn}R_{mn} 
 + 2RR_{ki}R^k_j  -\nabla_{(i}\nabla_{j)}\left(R^{mn}R_{mn}\right)\nb\\
                    && - 2\nabla^n\nabla_{(i}RR_{j)n}  +g_{ij}\nabla^2\left(R^{mn}R_{mn}\right) 
  + \nabla^2\left(RR_{ij}\right)   +g_{ij}\nabla_m\nabla_n\left(RR^{mn}\right),\nb\\
(F_{6})_{ij}&=&-\frac12g_{ij}R^m_nR^n_pR^p_m+3R^{mn}R_{ni}R_{mj} 
 +\frac32\nabla^2\left(R_{in}R^n_j\right) + \frac32g_{ij}\nabla_k\nabla_l\left(R^k_nR^{ln}\right)\nb\\
                      & & -3\nabla_k\nabla_{(i}\left(R_{j)n}R^{nk}\right),\nb\\
(F_{7})_{ij}&=&\frac12g_{ij}(\nabla R)^2- \left(\nabla_iR\right)\left(\nabla_jR\right) + 2R_{ij}\nabla^2R 
 -2\nabla_{(i}\nabla_{j)}\nabla^2R+2g_{ij}\nabla^4R,\nb\\
(F_{8})_{ij}&=& -\frac12g_{ij}\left(\nabla_p R_{mn}\right)\left(\nabla^p R^{mn}\right) -\nabla^4R_{ij} 
 + \left(\nabla_i R_{mn}\right)\left(\nabla_j R^{mn}\right)\nb\\
&& +2\left(\nabla_p R_{in}\right)\left(\nabla^p R^n_j\right)  +2\nabla^n\nabla_{(i}\nabla^2R_{j)n}+2\nabla_n\left(R^n_m\nabla_{(i}R^m_{j)}\right) \nb\\
&&
 -2\nabla_n\left(R_{m(j}\nabla_{i)}R^{mn}\right)-2\nabla_n\left(R_{m(i}\nabla^nR^m_{j)}\right) 
 -g_{ij}\nabla^n\nabla^m\nabla^2R_{mn},
 \eqn

\bqn 
  \lb{A.2}
F_{(\varphi, 1)}^{ij} &=& \frac{1}{2}\varphi\left\{\Big(2K + \nabla^{2}\varphi\Big) R^{ij}  
- 2 \Big(2K^{j}_{k} + \nabla^{j} \nabla_{k}\varphi\Big) R^{ik} \right.
 - 2 \Big(2K^{i}_{k} + \nabla^{i} \nabla_{k}\varphi\Big) R^{jk}\nb\\
& & \left. - \Big(2\Lambda_{g} - R\Big) \Big(2K^{ij} + \nabla^{i} \nabla^{j}\varphi\Big)\right\},\nb\\
F_{(\varphi, 2)}^{ij} &=& \frac{1}{2}\nabla_{k}\left\{\varphi{\cal{G}}^{ik}  
\Big(\frac{2N^{j}}{N} + \nabla^{j}\varphi\Big) 
+ \varphi{\cal{G}}^{jk}  \Big(\frac{2N^{i}}{N} + \nabla^{i}\varphi\Big) 
-  \varphi{\cal{G}}^{ij}  \Big(\frac{2N^{k}}{N} + \nabla^{k}\varphi\Big)\right\}, \nb\\   
F_{(\varphi, 3)}^{ij} &=& \frac{1}{2}\left\{2\nabla_{k} \nabla^{(i}f^{j) k}_{\varphi}  
- \nabla^{2}f_{\varphi}^{ij}   - \left(\nabla_{k}\nabla_{l}f^{kl}_{\varphi}\right)g^{ij}\right\},
\eqn
where
\bqn
\lb{A.3}
f_{\varphi}^{ij} &=& \varphi\left\{\Big(2K^{ij} + \nabla^{i}\nabla^{j}\varphi\Big) 
- \frac{1}{2} \Big(2K + \nabla^{2}\varphi\Big)g^{ij}\right\}. 
\eqn

The stress 3-tensor $\tau^{ij}$ is defined as
 \bq \label{tau}
\tau^{ij} = {2\over \sqrt{g}}{\delta \left(\sqrt{g}
 {\cal{L}}_{M}\right)\over \delta{g}_{ij}}.
 \eq
 
The matter quantities $(J^{t}, \; J^{i},\; J_{\varphi},\; J_{A},\; \tau^{ij})$ satisfy the
conservation laws
 \bqn \lb{eq5a} & &
 \int d^{3}x \sqrt{g} { \left[ \dot{g}_{kl}\tau^{kl} -
 \frac{1}{\sqrt{g}}\left(\sqrt{g}J^{t}\right)_{, t}  
 +   \frac{2N_{k}}  {N\sqrt{g}}\left(\sqrt{g}J^{k}\right)_{,t}
  \right.  }  
\left.   - 2\dot{\varphi}J_{\varphi} -  \frac{A} {N\sqrt{g}}\left(\sqrt{g}J_{A}\right)_{,t}
 \right] = 0, \nb \\
\eqn
\bqn 
\lb{eq5b} & & \nabla^{k}\tau_{ik} -
\frac{1}{N\sqrt{g}}\left(\sqrt{g}J_{i}\right)_{,t}  - \frac{J^{k}}{N}\left(\nabla_{k}N_{i}
- \nabla_{i}N_{k}\right)   
- \frac{N_{i}}{N}\nabla_{k}J^{k} + J_{\varphi} \nabla_{i}\varphi - \frac{J_{A}}{2N} \nabla_{i}A = 0. \nb \\
\eqn

In General Relativity,  the four-dimensional energy-momentum tensor is defined as
\bq
\lb{EMT}
T^{\mu\nu} = \frac{1}{\sqrt{-g^{(4)}}} \frac{\delta\left(\sqrt{-g^{(4)}}{\cal{L}}_{M}\right)}{\delta g^{(4)}_{\mu\nu}},
\eq
where $\mu, \nu = 0, 1, 2, 3$, and 
\bq
\lb{4Dmetric}
g^{(4)}_{00} = -N^{2} + N^{i}N_{i}, \;\;\; g^{(4)}_{0i} = N_{i}, \;\;\; g^{(4)}_{ij} = g_{ij}. 
\eq
Introducing the normal vector $n_{\mu}$ to the hypersurface $ t = $ constant by
\bq
\lb{4Dnorm}
n_{\mu} = N\delta^{t}_{\mu}, \;\;\; n^{\mu} = \frac{1}{N} \left(- 1,   N^{i} \right),
\eq
one can decompose $T_{\mu\nu}$ as follows~\cite{Ann}:
\bqn
\lb{4DEM}
\rho_{H} &\equiv& T_{\mu\nu} n^{\mu} n^{\nu},\;\;\;
s_{i}  \equiv -  T_{\mu\nu} h^{(4)\mu}_{i} n^{\nu},\nb\\ 
s_{ij}  &\equiv&  T_{\mu\nu} h^{(4)\mu}_{i} h^{(4)\nu}_{j},
\eqn
where $h^{(4)}_{\mu\nu}$ is the projection operator defined by $h^{(4)}_{\mu\nu} \equiv g^{(4)}_{\mu\nu}
+ n_{\mu}n_{\nu}$. In the relativistic  limit, one may make the following  identification: 
\bq
 \left(J^{t},\; J_{i}, \; \tau_{ij}\right) = \left(- 2\rho_{H},\; - s_{i},\; s_{ij}\right).
 \eq

\section{Summary}

In this chapter, we discussed how Ho\v{r}ava-Lifshitz Theory is constructed and the reasons for including certain features into the theory. After outlining the aspects of various versions of the theory, we have given a brief introduction to the projectable version with extended symmetry for an arbitrary value of the coupling constant. The rest of the dissertation deals only with this version of the theory.

\chapter{Spherically Symmetric Spacetimes}
\renewcommand{\theequation}{3.\arabic{equation}} \setcounter{equation}{0}

\begin{center}
\begin{singlespace}
\noindent This chapter published as:  J.~Greenwald, V.~H.~Satheeshkumar and A.~Wang,
  ``Black holes, compact objects and solar system tests in non-relativistic general covariant theory of gravity,'' JCAP {1012}, 007 (2010).  
\end{singlespace}
\end{center}

\section{Introduction}

In this chapter,  we  investigate systematically the spherically symmetric spacetimes and develop the
general formulas of such spacetimes 
filled with an anisotropic fluid with heat flow. We study the vacuum solutions. Although the solutions are not unique, when we apply them to the solar system, we find that these tests 
 pick up the Schwarzschild solution generically. Later, we consider the junction conditions across the surface of a compact object with the minimal requirement that the matching is mathematically meaningful. In this chapter we work with $\lambda = 1$.

\section{Spherically Symmetric Static Spacetimes}

The metric for   spherically symmetric static spacetimes that obey anisotropic scaling and the projectability condition can be cast in the form  
\bq
\lb{Ch3-3.1b}
ds^{2} = - c^{2}dt^{2} + e^{2\nu} \left(dr + e^{\mu - \nu} dt\right)^{2}  + r^{2}d^2\Omega,  
\eq
in the spherical coordinates $x^{i} = (r, \theta, \phi)$, where  $d^2\Omega = d\theta^{2}  + \sin^{2}\theta d\phi^{2}$,
and 
\bq
\lb{Ch3-3.1}
 \mu = \mu(r),\;\;\; \nu = \nu(r),\;\;\; N^{i} = \left\{e^{\mu - \nu}, 0, 0\right\}.
 \eq
The corresponding timelike Killing vector is  $\xi = \partial_{t}$. For the  above metric,  one finds
\bqn
\lb{Ch3-3.3a}
K_{ij} &=& e^{\mu+\nu}\Big(\mu'\delta^{r}_{i}\delta^{r}_{j} + re^{-2\nu}\Omega_{ij}\Big),\nb\\
R_{ij} &=&  \frac{2\nu'}{r}\delta^{r}_{i}\delta^{r}_{j} + e^{-2\nu}\Big[r\nu' - \big(1-e^{2\nu}\big)\Big]\Omega_{ij},\nb\\
{\cal{L}}_{K} &=& - \frac{2}{r^{2}} e^{2(\mu-\nu)}\left(2r\mu' + 1\right),  \nb\\
{\cal{L}}_{\varphi} &=& \frac{\varphi e^{-4\nu}}{r^2} \Bigg\{\Big[e^{2\nu}\left(\Lambda_{g} r^2 - 1\right) + 1 \Big]  \Big(\varphi '' - \nu'\varphi' + 2e^{\mu + \nu}\mu'\Big) \nb\\
    & & ~~~~ ~~~~ ~~~ - 2\Big(\nu' -  \Lambda_{g} re^{2\nu}\Big)\Big(\varphi' + 2e^{\mu +\nu}\Big)\Bigg\},   \nb\\
{\cal{L}}_{A} &=&  \frac{2A}{r^2} \Big[e^{-2\nu}\left(1 - 2r \nu'\right) + \left(\Lambda_{g} r^2 - 1\right)\Big],\nb\\
{\cal{L}}_{V} &=& \sum_{s=0}^{3}{{\cal{L}}_{V}^{(s)}},
\eqn
where  a prime denotes the ordinary derivative with respect to its indicated argument,
  $\Omega_{ij} \equiv \delta^{\theta}_{i}\delta^{\theta}_{j}  + \sin^{2}\theta\delta^{\phi}_{i}\delta^{\phi}_{j}$,
  and ${\cal{L}}_{V}^{(s)}$'s are given by 
  
\bqn
\lb{LVs}
{\cal{L}}_{V}^{(0)} &=& 2\Lambda - \frac{2e^{-2\nu}}{r^{2}}\Big[2r\nu' -\big(1-e^{2\nu}\big)\Big],\nb\\
{\cal{L}}_{V}^{(1)} &=&  \frac{2e^{-4\nu}}{\zeta^{2}r^{4}}\Bigg\{2g_{2}\Big[2r\nu' -\big(1-e^{2\nu}\big)\Big]^{2} + g_{3}\Big[3r^{2}\nu'^{2} 
 - 2r\big(1-e^{2\nu}\big)\nu' +\big(1-e^{2\nu}\big)^{2}\Big]\Bigg\},\nb\\
{\cal{L}}_{V}^{(2)} &=&  \frac{2e^{-6\nu}}{\zeta^{4}r^{6}}\Bigg\{4g_{4}\Big[2r\nu' -\big(1-e^{2\nu}\big)\Big]^{3}  + 2g_{5}\Big[6r^{3}\nu'^{3} -7r^{2}\big(1-e^{2\nu}\big)\nu'^{2}  + 4r\big(1-e^{2\nu}\big)^{2}\nu' \nb\\
 & &
 -  \big(1-e^{2\nu}\big)^{3}\Big] 
 + g_{6}\Big[5r^{3}\nu'^{3}  -3r^{2}\big(1-e^{2\nu}\big)\nu'^{2}  + 3r\big(1-e^{2\nu}\big)^{2}\nu' 
 -  \big(1-e^{2\nu}\big)^{3}\Big]\Bigg\},\nb\\
{\cal{L}}_{V}^{(3)} &=&  \frac{2e^{-6\nu}}{\zeta^{4}r^{6}}\Bigg\{4g_{7}\Bigg[2r^{4}\nu'\big(\nu''' - 7\nu'\nu'' + 6 \nu'^{3}\big) 
 - r^{3}\Big[\big(1-e^{2\nu}\big)\nu''' - \big(9-7e^{2\nu}\big)\nu'\nu''\nb\\
 & & ~~~~~~~~~ ~~~  + 2\big(5-3e^{2\nu}\big)\nu'^{3}\Big] 
 - r^{2}\Big[\big(1-e^{2\nu}\big)\nu'' + 4\nu'^{2}\Big] 
 + r\big(1-e^{2\nu}\big)^{2}\nu' + \big(1-e^{2\nu}\big)^{2}\Bigg]\nb\\  
 & &  + g_{8}\Bigg[3r^{4}\Big[\big(\nu'' - 4\nu'^{2}\big)\nu'' + 4 \nu'^{4}\Big] 
 -2 r^{3}\big(\nu'' - 2\nu'^{2}\big)\nu' 
 + r^{2}\Big[4\big(1-e^{2\nu}\big)\nu'' - \big(3-8e^{2\nu}\big)\nu'^{2}\Big]\nb\\    
                             & & ~~~~ + 8r\big(1-e^{2\nu}\big)\nu' + 6\big(1-e^{2\nu}\big)^{2}\Bigg]\Bigg\}.  
\eqn

 The Hamiltonian constraint (\ref{eq1}) reads
 \bq 
 \lb{Ham-const}
\int{\left( {\cal{L}}_{K} + {\cal{L}}_{V} - {\cal{L}}_{\varphi}^{(1)} - 8 \pi G J^{t} \right) e^{\nu} r^{2} dr}
= 0,
 \eq 
 where 
\bqn
{\cal{L}}_{\varphi}^{(1)} &=& \frac{\varphi e^{-4\nu}}{r^2} \Bigg\{\Big[e^{2\nu}\left(\Lambda_{g}r^2 -1\right) + 1\Big] (\varphi'' - \nu' \varphi') 
  - 2\left(\nu' -    \Lambda_{g} re^{2\nu}\right) \varphi'\Bigg\},  
\eqn 
 while the momentum constraint (\ref{eq2}) yields
 \bqn
 \lb{mom-const}
  & & 2 r \nu' + e^{-(\mu + \nu)} \Big[e^{2\nu}\left(\Lambda_{g}r^2 - 1\right) + 1 \Big] \varphi' 
  =  - 8\pi G r^2 e^{2(\nu -\mu)}  v,
 \eqn
 where 
 $J^{i} = e^{-(\mu + \nu)}\big(v, 0, 0\big)$.
 It can be also shown that Eqs.~(\ref{eq4a}) and (\ref{eq4b}) now read
 \bqn
 \lb{phi-const}
\Big[e^{2\nu}\left(\Lambda_{g}r^2 -1\right) + 1\Big]\Big(\varphi'' - \nu' \varphi' + e^{\mu + \nu}\mu'\Big) 
 -2\Big(\nu' - \Lambda_{g}re^{2\nu}\Big)\Big(\varphi' + e^{\mu + \nu}\Big)\nb \\
  = 8\pi G r^{2} e^{4\nu} J_{\varphi},
\eqn
 \bqn
\lb{A-const}
 2 r \nu'  - \Big[e^{2\nu}\left(\Lambda_{g}r^2 - 1\right) +  1\Big] =   4\pi G r^2 e^{2\nu}  J_{A}.
\eqn
 The dynamical equations (\ref{eq3}), on the other hand, yield
 \bqn
 \lb{Ch3-3.3g}
 & & 2\big(\mu' + \nu'\big) + \frac{1}{r} + \frac{1}{2}re^{2(\nu-\mu)}\left({\cal{L}}_{\varphi} + {\cal{L}}_{A}\right) \nb\\
 & &  ~~~~ ~~~~
 = - re^{-2\mu}\Big(F_{rr} + F^{\varphi}_{rr} + F^{A}_{rr} + 8\pi G e^{2\nu}p_{r}\Big),
 \eqn
 \bqn
 \lb{Ch3-3.3h}
 & & \mu'' + \big(2\mu' - \nu'\big)\left(\mu' + \frac{1}{r}\right)  + \frac{1}{2}e^{2(\nu-\mu)}\left({\cal{L}}_{\varphi} + {\cal{L}}_{A}\right) \nb\\
 & &  ~~~~ ~~~~
 = -\frac{e^{ 2(\nu -\mu)}}{r^{2}}\Bigg(F_{\theta\theta} + F^{\varphi}_{\theta\theta} + F^{A}_{\theta\theta} + 8\pi G r^{2}p_{\theta}\Bigg),
 \eqn
 where
 \bqn
 \lb{Ch3-3.3i}
 \tau_{ij} = e^{2\nu}p_{r}\delta^{r}_{i}\delta^{r}_{j} + r^{2}p_{\theta}\Omega_{ij},~~~~~ ~~~~~ ~~~~~ ~~~~~\nb\\ 
F^{A}_{ij} = \frac{ 2}{r}\big(A' + A\nu'\big) \delta^{r}_{i}\delta^{r}_{j}  + \frac{\Omega_{ij}}{e^{2\nu}} \Big[r^{2}\big(A'' - \nu'A'\big) 
 + r\big(A' + A\nu'\big)  - A\Big(1 - e^{2\nu}\Big)\Big],
\eqn
and the $F_{ij}$'s 
and  $F_{(\varphi,s)}^{ij}$ are given below. 
\bqn
\lb{A.1}
\left(F_{0}\right)_{ij} &=& - \frac{1}{2} e^{2\nu}\delta^{r}_{i}\delta^{r}_{j} 
                                         - \frac{1}{2}  r ^{2} \Omega_{ij},\nb\\ 
%
\left(F_{1}\right)_{ij} &=&  \frac{1}{r^{2}} \big(1- e^{2\nu}\big)\delta^{r}_{i}\delta^{r}_{j} 
                                        -  r  e^{-2\nu} \nu' \Omega_{ij},\nb\\ 
\left(F_{2}\right)_{ij} &=& \frac{e^{-2\nu}}{r^{4}}\Big[4r^{2}\big(2\nu'' - 3\nu'^{2}\big) 
 + \big(1- e^{2\nu}\big)\big(7 + e^{2\nu}\big)\Big]\delta^{r}_{i}\delta^{r}_{j}\nb\\
& &  + \frac{2e^{-4\nu}}{r^{2}}\Bigg[4r^{3}\big(2\nu''' - 7\nu'\nu'' + 6\nu'^{3}\big) 
- 2r \nu'\big(7 - 3e^{2\nu}\big) \nb
-  \big(1 - e^{2\nu}\big)\big(7 + e^{2\nu}\big)\Bigg]\Omega_{ij},\nb\\
\left(F_{3}\right)_{ij} &=&  \frac{e^{-2\nu}}{r^{4}}\Big[3r^{2}\big(2\nu'' - 3\nu'^{2}\big) 
+ \big(1- e^{2\nu}\big)\big(5 + e^{2\nu}\big)\Big]\delta^{r}_{i}\delta^{r}_{j}\nb\\
& &  + \frac{e^{-4\nu}}{r^{2}}\Bigg[3r^{3}\big(\nu''' - 7\nu'\nu'' + 6\nu'^{3}\big) \nb
- 2r \nu'\big(5 - 2e^{2\nu}\big) \nb
-  \big(1 - e^{2\nu}\big)\big(5 + e^{2\nu}\big)\Bigg]\Omega_{ij},\nb\\
\left(F_{4}\right)_{ij} &=& \frac{4e^{-4\nu}}{r^{6}}\Big[16r^{3}\nu'\big(3\nu'' - 5\nu'^{2}\big) \nb
- 12r \big(1- e^{2\nu}\big)\big(2r\nu'' - 3r\nu'^{2} - 4\nu'\big) \nb\\
& & ~~~~~ ~~~ - \big(1- e^{2\nu}\big)\big(23 - 22 e^{2\nu}-  e^{4\nu}\big)\Big]\delta^{r}_{i}\delta^{r}_{j}\nb\\
& &  + \frac{4e^{-6\nu}}{r^{4}}\Bigg[24r^{4}\Big[\nu'\nu''' + \big(\nu''  - 11\nu'^{2}\big)\nu'' + 10\nu'^{4}\Big] \nb\\
& & ~~ - 4r^{3}\big(17 - 18e^{2\nu}\big)\nu'^{3} -  12r^{2}\big(15 - 11e^{2\nu}\big)\nu'^{2}\nb\\
& &  ~~ -   \big(1 - e^{2\nu}\big)\Big[12r^{3}\Big(\nu''' - 7\nu'\nu''\Big)\nb
- 48 r^{2}\nu'' + 3r\Big(1 + 7e^{2\nu}\Big)\nu' \nb\\
& & ~~~~~~~~~  
- 2\big(1- e^{2\nu}\big)\big(23 + e^{2\nu}\big)\Big]\Bigg]\Omega_{ij},\nb\\
\left(F_{5}\right)_{ij} &=&   \frac{2e^{-4\nu}}{r^{6}}\Bigg\{12r^{4}\Big[\nu'\big(\nu''' -11 \nu'\nu'' + 10 \nu'^{3}\big) + \nu''^{2}\Big] \nb\\
& & ~~~~~~~~~~~ - 4r^{3}\Big[\big(1 - e^{2\nu}\big)\big(\nu''' - 7\nu'\nu'' + 6\nu'^{3}\big)\nb
 - \nu'\big(3\nu'' -2  \nu'^{2}\big)\Big]\nb\\
& & ~~~~~~~~~~~ + r^{2}\big(1 - e^{2\nu}\big)\big(2\nu'' - 15\nu'^{2}\big)\nb
+ 4r\big(1 - e^{2\nu}\big)\big(13 - 2 e^{2\nu}\big)\nu'\nb\\
& &  ~~~~~~~~~~~ + \big(1 - e^{2\nu}\big)^{2}\big(23 -  e^{2\nu}\big)\Bigg\} \delta^{r}_{i}\delta^{r}_{j}\nb\\
& &  + \frac{2e^{-6\nu}}{r^{4}}\Bigg\{18r^{4}\Big[\nu'\nu''' + \nu''^{2}  - \nu'^{2}\big(11 \nu'' - 10\nu'^{2}\big)\Big] \nb\\
& &  ~~~~~~~~~~~~ - r^{3}\Big[7\big(1- e^{2\nu}\big)\nu'''   - \big(53 -49e^{2\nu}\big)\nu'\nu'' \nb
 + \big(45 - 42e^{2\nu}\big)\nu'^{3}\Big]\nb\\
& &  ~~~~~~~~~~~~ + r^{2}\Big[24\big(1- e^{2\nu}\big)\nu''  - \big(97 -69e^{2\nu}\big)\nu'^{2}\Big] \nb\\
& &  ~~~~~~~~~~~~ + r \big(1- e^{2\nu}\big)\big(13- 15e^{2\nu}\big)\nu' \nb
 + 2 \big(1- e^{2\nu}\big)^{2}\big(13 + e^{2\nu}\big)\Bigg\}\Omega_{ij}, \nb
\eqn

\bqn
\left(F_{6}\right)_{ij} &=&  \frac{e^{-4\nu}}{r^{6}}\Bigg\{10r^{3}\big(3\nu'' - 5\nu'^{2}\big)\nu'
     - 3r^{2}\Big[2\big(1-e^{2\nu}\big)\nu'' \nb\\
     & & ~~~~~~~~~ + 3 e^{2\nu}\nu'^{2}\Big]
     + 12r \big(1-e^{2\nu}\big)\nu' \nb
 -   \big(1-e^{2\nu}\big)^{2} \big(14 + e^{2\nu}\big)\Bigg\}\delta^{r}_{i}\delta^{r}_{j} \nb\\
& &  + \frac{e^{-6\nu}}{r^{4}}\Bigg\{15r^{4}\Big[\nu'\nu''' + \big(\nu''  - \nu'^{2}\big)\big(\nu'' - 10\nu'^{2}\big)\Big] \nb\\
& & ~~~~~~~~~~ - r^{3}\Big[3\big(1 - e^{2\nu}\big)\nu''' -  3\big(1 - 7e^{2\nu}\big)\nu'\nu''\nb
-   \big(25  + 18 e^{2\nu}\big) \nu'^{3}\Big]\nb\\
& &~~~~~~~~~~  + 3 r^{2}\Big[4\big(1 - e^{2\nu}\big)\nu'' -  \big(12 - 11e^{2\nu}\big)\nu'^{2}\Big]\nb\\
& &~~~~~~~~~~  + 12r\big(1 - e^{2\nu}\big)\big(2 - e^{2\nu}\big)\nu'\nb
 + 2\big(1 - e^{2\nu}\big)^{2}\big(14 + e^{2\nu}\big)\Bigg\}\Omega_{ij},\nb\\ 
\left(F_{7}\right)_{ij} &=&   \frac{8e^{-4\nu}}{r^{6}}\Bigg\{r^{4}\Big[-2 \nu^{(4)} + 20 \nu'\nu''' \nb
 + \big(15\nu'' - 82\nu'^{2}\big)\nu'' + 40\nu'^{4}\Big] \nb\\
& & ~~~~~~~~~ + 2r^{2}\Big[2\big(3 - e^{2\nu}\big)\nu'' - 3\big(5 - e^{2\nu}\big)\nu'^{2}\Big]\nb\\
& & ~~~~~~~~~ -8r \big(3 - e^{2\nu}\big)\nu' \nb
 - \big(1 - e^{2\nu}\big)\big(7 + e^{2\nu}\big)\Bigg\}\delta^{r}_{i}\delta^{r}_{j}\nb\\
& &  + \frac{8e^{-6\nu}}{r^{4}}\Bigg\{r^{5}\Big[- \nu^{(5)} + 16 \nu'\nu^{(4)} \nb
+  \big(25\nu''  - 101\nu'^{2}\big) \nu'''\nb\\
& & ~~~~~~~~~~~~~ -  \big(127\nu'' - 326 \nu'^{2}\big)\nu'\nu'' - 120\nu'^{5}\Big]\nb\\
& & ~~~~~ +  2r^{3}\Big[\big(3 - e^{2\nu}\big)\nu''' - \big(33 - 7e^{2\nu}\big)\nu'\nu''\nb
 + \big(45 - 6e^{2\nu}\big)\nu'^{3}\Big]\nb\\
& & ~~~~~ -  2r^{2}\Big[4\big(3 - e^{2\nu}\big)\nu'' - \big(51 - 11e^{2\nu}\big)\nu'^{2}\Big]\nb\\
& & ~~~~~ + r\big(57 - 24e^{2\nu}- e^{4\nu}\big)\nu'\nb
+ 2\big(1 - e^{2\nu}\big)\big(7+ e^{2\nu}\big)\Bigg\}\Omega_{ij},\nb
\eqn

\bqn
\left(F_{8}\right)_{ij} &=& \frac{e^{-4\nu}}{r^{6}}\Bigg\{r^{4}\Big[6 \nu^{(4)} - 68 \nu'\nu''' \nb
 - \big(59\nu'' - 358\nu'^{2}\big)\nu'' -224\nu'^{4}\Big] \nb\\
& & ~~~~~~~ + 2r^{3} \big(13\nu'' - 29\nu'^{2}\big)\nu'  \nb
- r^{2}\Big[8 \big(5 - 2 e^{2\nu}\big)\nu'' -  7 \big(13- 4 e^{2\nu}\big)\nu'^{2}\Big] \nb\\
& & ~~~~~~~ + 16r\big(4 - e^{2\nu}\big) \nu'  \nb
 + 6\big(1 - e^{2\nu}\big)\big(1+ 3 e^{2\nu}\big)\Bigg\}\delta^{r}_{i}\delta^{r}_{j} \nb\\
& &  + \frac{e^{-6\nu}}{r^{4}}\Bigg\{3r^{5}\Big[\nu^{(5)} - 16 \nu'\nu^{(4)}  \nb
 - \big(25\nu'' - 101\nu'^{2}\big)\nu''' \nb\\
& & ~~~~~~~~~~~~~  + \big(127\nu'' - 326\nu'^{2}\big)\nu'\nu'' + 120 \nu'^{5}\Big] \nb\\
& & ~~~~~~~ + r^{4}\Big[\nu'\nu''' - \big(5\nu'' - 13\nu'^{2}\big)\nu'' -  14\nu'^{4}\Big] \nb\\
& & ~~~~~~~ - r^{3}\Big[2\big(7 - e^{2\nu}\big)\nu''' - 2\big(78 - 7e^{2\nu}\big)\nu'\nu''\nb
+2\big(107 - 6e^{2\nu}\big)\nu'^{3}\Big]\nb\\
& & ~~~~~~~ + r^{2}\Big[8\big(7 - e^{2\nu}\big)\nu'' - \big(277 - 30e^{2\nu}\big)\nu'^{2}\Big]\nb\\
& & ~~~~~~~ - 16 r\big(13 - 7e^{2\nu}\big)\nu' 
- 6 \big(1 - e^{2\nu}\big)\big(11 - 3e^{2\nu}\big)\Bigg\}\Omega_{ij},
\eqn

\bqn
\lb{B.2}
\left[F_{(\varphi, 1)}\right]_{ij} &=&\frac{\varphi e^{-2\nu}}{r^{2}}\Bigg\{\Big[e^{2\nu}\Big(1-\Lambda_{g}r^{2}\Big) - \big(1 + r\nu'\big)\Big]\varphi''\nb\\
& & ~~ + \Big[e^{2\nu}\Big(\Lambda_{g}r^{2} -1\Big)  + \big(3 + r\nu'\big)\Big]\nu' \varphi'\nb\\
& &~~  - 2e^{\mu + \nu}\Big[e^{2\nu}\Big(\Lambda_{g}r^{2} -1\Big)+ 1\Big]\mu'\nb
+ 2e^{\mu+\nu}\left(2 - r\mu'\right) \nu' \Bigg\}\delta^{r}_{i}\delta^{r}_{j}\nb\\
& &  + \frac{1}{2}\varphi e^{-4\nu}\Bigg\{\Big[r\nu' - \Big(1- e^{2\nu}\Big)\Big]\varphi'' \nb
+ \Big(3 - r\nu' - e^{2\nu}\Big)\nu' \varphi' - 2 \Lambda_{g}r e^{2\nu} \varphi'\nb\\
& & ~~ - 2e^{\mu+\nu}\Big(2 - r\nu' - e^{2\nu}\Big)\mu' \nb
+ 4e^{\mu+\nu}\Big(\nu' - \Lambda_{g}r e^{2\nu}\Big)\Bigg\}\Omega_{ij},\nb
\eqn
\bqn
\left[F_{(\varphi, 2)}\right]_{ij} &=&\frac{ e^{-2\nu}}{2r^{2}}\Bigg\{\Big[e^{2\nu}\Big(\Lambda_{g}r^{2}-1\Big) + 1\Big]\Big[\varphi\varphi''
\nb
+ \Big(\varphi' + \varphi\nu'\Big)\varphi' + 2 e^{\mu+\nu} \Big(\varphi' + \varphi\mu'\Big)\Big] \nb\\
& &~~ + 4\varphi e^{\mu+\nu}\Big(\nu' - \Lambda_{g}r e^{2\nu}\Big)\nb
 - 2\Lambda_{g}r\varphi e^{2\nu}\varphi'\Bigg\} \delta^{r}_{i}\delta^{r}_{j}\nb\\
& &  + \frac{1}{2} e^{-4\nu}\Bigg\{r\varphi\Big(\nu' - \Lambda_{g}r e^{2\nu}\Big)\varphi'' \nb
 + r\varphi\Big(\varphi' + 2 e^{\mu +\nu}\Big)\nu''\nb
+   r\Big(\nu' - \Lambda_{g}r e^{2\nu}\Big)\varphi^{'2} \nb\\
& & ~~ - r\varphi\Big(3\varphi' + 4 e^{\mu +\nu}\Big)\nu^{'2}\nb
-  \Big(\varphi - 2r e^{\mu + \nu}  - \Lambda_{g} r^{2}\varphi e^{2\nu}\Big)\nu' \varphi'\nb\\
& &~~ - 2e^{\mu+\nu}\Big(\Lambda_{g}r^{2} e^{2\nu}\varphi' + \varphi\nu'\Big)\nb
+ 2r\varphi e^{\mu + \nu}\Big(\nu' - \Lambda_{g}r e^{2\nu}\Big)\mu'\Bigg\}\Omega_{ij},\nb
\eqn
\bqn
\left[F_{(\varphi, 3)}\right]_{ij} &=&\frac{ e^{-2\nu}}{r^{2}}\Bigg[r\varphi\nu'\varphi'' + \varphi^{'2}
- \varphi\big(r\nu' + 2\big)\nu'\varphi'\nb
 + 2e^{\mu +\nu}\Big(\varphi' + r\varphi\nu'\mu' -2\varphi\nu'\Big)\Bigg] \delta^{r}_{i}\delta^{r}_{j}\nb\\
& &  + \frac{1}{2} e^{-4\nu}\Bigg\{2r\Big(\varphi' - \varphi\nu' + e^{\mu + \nu}\Big)\varphi'' \nb
-  r\varphi\Big(\varphi' + 2 e^{\mu +\nu}\Big)\nu'' \nb
 + \Big[4r\big(\varphi\nu' -\varphi'\big) + \varphi\Big]\nu'\varphi'\nb\\
&& ~~ + e^{\mu +\nu}\Big[4r\varphi\big(\nu' - \mu'\big)\nu' + 2\varphi\nu' \nb
+ 2r\big(\mu' -3\nu'\big)\varphi'\Big]\Bigg\}\Omega_{ij}.
\eqn
Here we define a fluid with $p_{r} = p_{\theta}$ as a perfect fluid, which in general conducts heat flow along the radial direction~\cite{Santos85}. 
 
Since the spacetime is static, one can see that now the energy conservation law (\ref{eq5a}) is satisfied identically,
while the momentum conservation law (\ref{eq5b}) yields
\bq
\lb{Ch3-3.3j}
v\mu' - \big(v' - p_{r}'\big) - \frac{2}{r}\big(v - p_{r} + p_{\theta}\big) + J_{\varphi} \varphi' - \frac{1}{2}J_{A} A'   = 0.
\eq

To relate   the quantities $J^{t},\; J^{i}$ and $\tau_{ij}$   to the ones often used in General Relativity,  
 one can first introduce the unit normal vector $n_{\mu}$ to the hypersurfaces $t =$ constant, and then 
the spacelike unit vectors $\chi_{\mu}, \; \theta_{\mu}$ and
$\phi_{\mu}$, are defined as~\cite{Ann}
\bqn
\lb{emt1}
 n_{\mu} &=& \delta^{t}_{\mu}, \;\;\; n^{\mu} = - \delta_{t}^{\mu} + e^{\mu-\nu}\delta_{r}^{\mu},\nb\\
 \chi^{\mu} &=& e^{-\nu}\delta^{\mu}_{r} , \;\;\; \chi_{\mu} = e^{\mu}\delta^{t}_{\mu} + e^{\nu} \delta^{r}_{\mu},\nb\\
 \theta_{\mu} &=& r\delta^{\theta}_{\mu},\;\;\; \phi_{\mu} = r\sin\theta \delta^{\phi}_{\mu}.
\eqn
In terms of these four unit vectors,    
the energy-momentum tensor for an anisotropic fluid with heat flow can be written as
\bqn
T_{\mu\nu} &=& \rho_{H}n_{\mu}n_{\nu} + q \big(n_{\mu} \chi_{\nu} + n_{\nu} \chi_{\mu} \big) 
 + p_{r}\chi_{\mu} \chi_{\nu}  + p_{\theta}\big(\theta_{\mu}\theta_{\nu} + \phi_{\mu}\phi_{\nu}\big)
\eqn
where $\rho_{H}, \; q,\; p_{r}$ and $p_{\theta}$ denote, respectively, the energy density, the heat flow 
along the radial direction and the radial and tangential pressures, measured by the observer with the four-velocity
$n_{\mu}$. 
Then,   one can see that such a decomposition is consistent with 
 the quantities $J^{t}$ and $J^{i}$, defined by
 \bq
 \rho_{H} = -2 J^{t},\;\;\; v= e^{\mu} q. 
\eq
  It should be noted that the definitions of the energy density $\rho_H$, 
 the radial pressure $p_r$ and the heat flow $q$ are different from the ones defined in a comoving frame 
 in General Relativity. 

 Finally, we note that in writing the above equations, we leave the choice of the $U(1)$ gauge open. From
 Eq.~(\ref{2.3}) one can see that it can be used to set  one (and only one) of the three functions $A,\; \varphi$ 
 and $N_{r}$  to zero.  To compare our results  with the one obtained in~\cite{HMT}, 
  without loss of the generality, we shall choose the gauge
 \bq
 \lb{phi-gauge}
 \varphi =0.
 \eq
Then, we find that
\bq
\lb{Lphi}
{\cal{L}}_{\varphi} = 0, \;\;\; F^{ij}_{(\varphi, n)} = 0, \; (n = 1, 2, 3).
\eq

\section{Vacuum Solutions}

 In the vacuum case, we have $J^t = v = p_{r} = p_{\theta} = J_{A} = J_{\varphi} = 0$.
 With the gauge (\ref{phi-gauge}), from the momentum constraint (\ref{mom-const})   we immediately obtain
 $\nu = $ constant, 
 while Eq.~(\ref{A-const}) further requires
 \bq
 \lb{Ch3-20}
 \nu = 0,\;\;\; \Lambda_{g} = 0.
 \eq
This is different from  the solutions presented  in~\cite{HMT}, where $\nu \not= 0,\; \mu = -\infty$.
Inserting the above into Eq.~(\ref{phi-const}), it can be shown that it is satisfied identically. Since $\nu = 0$, from the expressions of $\left(F_{s}\right)_{ij}$, we find that  $\left(F_{s}\right)_{ij} = 0$ for $s \not=0$,  and
$\left(F_{0}\right)_{ij} = -g_{ij}/2$, so that 
\bq
\lb{Ch3-21}
F_{ij} = - \Lambda g_{ij},
\eq
where $\Lambda \equiv \zeta^2 g_0/2$.
Substituting Eqs.~(\ref{Ch3-3.3i}) and (\ref{Lphi})--(\ref{Ch3-21}) into Eqs.~(\ref{phi-const}) and (\ref{A-const}), we find that
\bqn
\lb{Ch3-22}
& & \Big(2r\mu' + 1\Big) e^{2\mu} = \Lambda r^{2} - 2r A',\\
\lb{Ch3-23}
& & \mu'' + 2\mu'\Big(\mu' + \frac{1}{r}\Big) =  \frac{e^{-2\mu}}{r}\Big[\Lambda r  - \big(rA'\big)'\Big].
\eqn
It can be shown that Eq.~(\ref{Ch3-23}) is not independent, and can be obtained from Eq.~(\ref{Ch3-22}). Therefore,
the solutions are not uniquely determined, since now we have only one equation, (\ref{Ch3-22}), for  two unknowns
$\mu$ and $A$. In particular, for any given $A$, from Eq.~(\ref{Ch3-22}) we find that
\bq
\lb{Ch3-24}
\mu = \frac{1}{2}\ln\Bigg[\frac{2m}{r} + \frac{1}{3}\Lambda r^2 - 2A(r) + \frac{2}{r}\int^{r}{A(r') dr'}\Bigg].
\eq
On the other hand, we also have 
\bq
\lb{Ch3-25}
{\cal{L}}_{K} = 
 \frac{4A'}{r} - 2\Lambda,\;\;\;
{\cal{L}}_{V} = 2 \Lambda.
\eq
Inserting it into the Hamiltonian constraint (\ref{Ham-const}), we find that
\bq
\lb{Ch3-26}
\int_{0}^{\infty}{r A'(r) dr} = 0.
\eq
Therefore, for any given function $A$, subject to the above constraint, the solutions given by Eqs.~(\ref{Ch3-20}) 
and (\ref{Ch3-24}) represent the vacuum solutions of the HL theory.  Thus, in contrast to General Relativity, the
vacuum solutions in the HMT setup are not unique.

When $A$ is a constant (without loss of generality, we can set $A = 0$), from the above we find that
\bq
\lb{4.4b}
 \mu = \frac{1}{2}\ln\left(\frac{2m}{r} + \frac{1}{3}\Lambda r^2 \right),\; (A = 0),
\eq
which is exactly the Schwarzschild (anti-) de Sitter solution, written in the Gullstrand-Painleve coordinates~\cite{GP}.
It is interesting to note that when $m = 0$  we must assume that $\Lambda > 0$, in order to have $\mu$ real. 
That is, the anti-de 
Sitter solution cannot be written in the static ADM form (\ref{3.1b}).

\section{Solar System Tests}

 The solar system tests are usually written in terms of the Eddington parameters, by following the so-called
 ``parametrized post-Newtonian" (PPN) approach, introduced initially by Eddington~\cite{Edd}. The gravitational
 field, produced by a  pointlike and motionless particle with mass $M$,  is often described by the  
 form of metric~\cite{Hartle},
 \bq
 \lb{Ch3-6.1}
 ds^{2} = - e^{2\Psi}c^{2}dt^2 + e^{2\Phi}dr^2 + r^{2}d^2\Omega,
 \eq
 where $\Psi$ and $\Phi$ are functions of the dimensionless quantity $\chi \equiv GM/(rc^2)$ only. For the solar
 system, we have $GM_{\bigodot}/c^2 \simeq 1.5$~km, so that in most cases we have $\chi \ll 1$. Expanding
 $\Psi$ and $\Phi$ in terms of  $\chi$, we have~\cite{Hartle}
 \bqn
 \lb{Ch3-6.2}
 e^{2\Psi} &=& 1 - 2\left(\frac{GM}{c^2 r}\right) + 2\big(\beta - \gamma\big) \left(\frac{GM}{c^2 r}\right)^2 + ...,\nb\\
 e^{2\Phi} &=& 1 + 2\gamma \left(\frac{GM}{c^2 r}\right) +  ..., 
 \eqn
where $\beta$ and $\gamma$ are the Eddington parameters. General Relativity predicts $\beta = 1 = \gamma$
strictly, while the current radar ranging of the Cassini probe~\cite{BT}, and the procession of lunar laser ranging data
\cite{WTB} yield, respectively, the bounds~\cite{RJ}
\bqn
\lb{6.3}
\gamma - 1&=& (2.1 \pm 2.3)\times 10^{-5},\nb\\
\beta - 1&=& (1.2 \pm 1.1)\times 10^{-4},
\eqn
which are consistent with the predictions of General Relativity. 

To apply the solar system tests to the HL theory, we need first to transfer the above bounds to the metric coefficients $\mu$
and $\nu$. The relations between $(\Phi,\; \Psi)$ and $(\mu, \; \nu)$ have been worked out explicitly,
and are given by
\bq
\lb{Ch3-6.4}
\mu = \frac{1}{2}\ln\Bigg[c^{2}\Big(1 - e^{2\Psi}\Big)\Bigg],\;\;\; \nu = \Phi + \Psi,
\eq
or inversely,
\bqn
\lb{Ch3-6.5}
\Phi &=& \nu - \frac{1}{2}\ln\Big( 1 - \frac{1}{c^{2}}e^{2\mu}\Big),\nb\\
\Psi &=&   \frac{1}{2}\ln\Big( 1 - \frac{1}{c^{2}}e^{2\mu}\Big).
\eqn 
Inserting Eq.~(\ref{Ch3-6.2}) into Eq.~(\ref{Ch3-6.4}), we find that
\bqn
\lb{Ch3-6.6}
\mu &=& \frac{1}{2}\ln\left\{2c^{2}\Bigg[\left(\frac{GM}{c^{2}r}\right) - \big(\beta - \gamma\big)\left(\frac{GM}{c^{2}r}\right)^{2} + ...\Bigg]\right\},\nb\\
\nu &=& \big(\gamma - 1\big)  \left(\frac{GM}{c^{2}r}\right)  + ... .
\eqn
Comparing Eq.~(\ref{Ch3-6.6}) with Eq.~(\ref{Ch3-24}) for $\Lambda = 0$, we find that in order to be consistent with solar system tests, we must assume that
\bq
\lb{Ch3-6.7}
A(r) = {\cal{O}}\left[\left(\frac{GM}{c^{2}r}\right)^2\right].
\eq
Together with the Hamiltonian constraint (\ref{Ch3-26}), we find that this is impossible unless 
$A = 0$.
 Therefore, although the vacuum solution in the HMT setup is not unique, the solar system tests seemingly require that it must be the Schwarzschild
vacuum solution. 

It should be noted that by choosing $A(r)$ in very  particular forms, the  condition $A = 0$ could be relaxed~\cite{AP}. But, such chosen $A$ is not analytic (in terms
of the dimensionless quantity $\chi$), and it is not clear how to  expand it in the form of (\ref{Ch3-6.6}). Thus,   in this chapter we simply discard those possibilities.

\section{Perfect Fluid Solutions}

In this section, let us consider a perfect fluid without heat flow, that is,
\bq
\lb{Ch3-5.1}
p_{r} = p_{\theta} = p,\;\;\; v = 0.
\eq
Then, together with the gauge choice (\ref{phi-gauge}), from Eq.~(\ref{mom-const}) we find that $\nu =$ constant. 
 However,
to be matched with the vacuum solutions outside of the fluid, we must set this constant to  zero,
\bq
\lb{Ch3-5.2}
\nu = 0,
\eq
from which we immediately find that $R_{ij} = 0$,  $F_{ij}$ is still given by Eq.~(\ref{A.1}), and
\bqn
\lb{Ch3-5.3}
{\cal{L}}_{A} &=& 2\Lambda_{g}A,\nb\\
F_{ij}^{A} &=& \frac{ 2A'}{r} \delta^{r}_{i}\delta^{r}_{j}  + r\big(rA'\big)'  \Omega_{ij}.
\eqn
Inserting the above into Eqs.~(3.8)--(3.12), we find that
\bqn
\lb{Ch3-5.3aa}
 J_{\varphi} &=& \frac{\Lambda_{g}}{8\pi G r^{2}}\left(r^{2}e^{\mu}\right)_{,r},\\
\lb{Ch3-5.3ab}
 J_{A} &=& -\frac{\Lambda_{g}}{4\pi G},\\
\lb{Ch3-5.3a}
 \big(rf\big)' + 2rA' + \Lambda_{g}r^2 A - \Lambda r^2 &=& -8\pi G r^2 p, ~~\\
\lb{Ch3-5.3b}
 \frac{1}{2}r f'' + f'  + \big(rA'\big)' + \Lambda_{g}r A - \Lambda r &=& -8\pi G r p,   ~~ 
\eqn
where $f \equiv e^{2\mu}$. From the last two equations, we find that
\bq
\lb{Ch3-5.3c}
r^2 f'' -2 f  = -2r^3 \left(\frac{A'}{r}\right)'.
\eq
 On the other hand, the conservation law of momentum (\ref{Ch3-3.3j}) now reduces to
 \bq
 \lb{Ch3-5.4}
 p' + \frac{\Lambda_{g}}{8\pi G} A' = 0,
 \eq
 which has the solution,
 \bq
 \lb{Ch3-5.5}
 p =   p_{0} - \frac{\Lambda_{g}}{8\pi G} A,
 \eq
where $p_{0}$ is an integration constant. 

Substituting it into Eq.~(\ref{Ch3-5.3a}), and then taking a 
derivative of it, we find that the resulting equation is exactly given by Eq.~(\ref{Ch3-5.3c}). 
Thus, both Eqs.~(\ref{Ch3-5.3c}) and (\ref{Ch3-5.3b}) are not independent, and can all be derived from Eqs.~(\ref{Ch3-5.3a})
and (\ref{Ch3-5.5}). Then, in the present case there are   five independent equations, the Hamiltonian 
constraint (\ref{3.3b}), and Eqs.~(\ref{Ch3-5.3aa}), (\ref{Ch3-5.3ab}), (\ref{Ch3-5.3a}) and (\ref{Ch3-5.5}). 
However,  we have six unknowns, $A,\; \mu,\; p,\; J^{t},\; J_{\varphi},\; 
J_{A}$. Therefore, the problem now is not uniquely determined. As in the vacuum case, we
can express all these quantities in terms of the gauge field $A$. In particular, substituting Eq.~(\ref{Ch3-5.5})
into Eq.~(\ref{Ch3-5.3a}) and then integrating it, we obtain 
 \bq
 \lb{Ch3-5.7a}
\mu = \frac{1}{2}\ln\Bigg[\frac{2m}{r} + \frac{1}{3}\Big(\Lambda - {8\pi G p_{0}}\Big)r^2 - 2A 
+ \frac{2}{r}\int^{r}{A(r') dr'}\Bigg].
 \eq
Then, we find that
\bq
\lb{Ch3-5.6a}
{\cal{L}}_{\varphi} = \frac{4A'}{r} - 2\left(\Lambda - 8\pi G p_{0}\right),\;\;\;
{\cal{L}}_{V} = 2\Lambda.
\eq
Inserting the above into Eq.~(\ref{Ham-const}), we find that it can be cast in the form,
\bq
\lb{Ch3-5.7c}
\int^{\infty}_{0}{\tilde{\rho}(r) dr} = 0,
\eq
where
\bq
\lb{Ch3-5.7d}
J^{t} = \frac{1}{2\pi G}\Bigg(4\pi G p_{0} + \frac{A'(r)}{r} - \frac{\tilde{\rho}(r)}{r^2}\Bigg).
\eq

From the above one can see that once $A$ is given, one can immediately obtain all the rest. By properly
choosing it (and $\tilde{\rho}(r)$), it is not difficult to see that one can construct non-singular
solutions representing stars made of a perfect fluid without heat flow. To see this explicitly, 
let us consider the following two particular cases. 
 
\subsection{$\Lambda_{g} = 0$}
 
 When $\Lambda_{g} = 0$, we have
\bq
\lb{Ch3-5.7g}
J_{\varphi} = J_{A} = 0,\;\;\; p = p_{0},
\eq
while $\mu$ and $J^{t}$ are still given by Eqs.~(\ref{Ch3-5.7a}) and (\ref{Ch3-5.7d}), respectively.
To have a physically acceptable model, we require that the fluid be non-singular in the center. Since $R_{ij} = 0$, one can see that any
quantity built from the Riemann and Ricci tensor vanishes in the present case. Then, possible singularities can only come from the kinetic part,
  $K_{ij}$, where the very first quantity   is 
\bqn
K &=& g^{ij}K_{ij} = \frac{e^{\mu}}{r}\left(r\mu' +2\right)\nb\\
&=& \frac{e^{-\mu}}{r}\left(\frac{3m}{r} + \left(\Lambda - 8\pi Gp_{0}\right)r^{2} - rA' - 3A 
 +\frac{3}{r}\int{A(r')dr'}\right).
 \eqn
Assuming that near the center $A$ is dominated by the term $r^{\alpha}$, we find that $K$ is non-singular only when
\bq
\lb{Ch3-5.7f}
m = 0,\;\;\; \alpha \ge 2.
\eq
For such a function $A$, Eq.~(\ref{Ch3-5.7d}) shows that $J^{t}$ is non-singular, as long as $\tilde{\rho}(r) \simeq {\cal{O}}(r^{2})$.

 \subsection{$A = A_{0}$}
 
 When $A$ is a constant,    from Eq.~(\ref{Ch3-5.5}) we can see that the pressure $p$ is also a constant. Then, the integration of Eq.~(\ref{Ch3-5.3a}) yields,
  \bq
 \lb{Ch3-5.7b}
\mu = \frac{1}{2}\ln\Bigg\{\frac{2m}{r} + \frac{1}{3}\Big(\Lambda   - {8\pi G p_{0}}\Big)r^2\Bigg\}.
 \eq
Inserting it into Eq.~(\ref{Ch3-5.3c}) we find that it is satisfied identically, while the Hamiltonian constraint (\ref{Ham-const}) can also be cast in the form
of Eq.~(\ref{Ch3-5.7c}), but now with  
\bq
\lb{Ch3-5.7e}
J^{t} = \frac{1}{8\pi G}\Bigg(16\pi G p_{0}    - \frac{\tilde{\rho}(r)}{r^2}\Bigg).
\eq
Thus, the solutions of Eqs.~(\ref{Ch3-5.2}) and (\ref{Ch3-5.7b}) represent 
a perfect fluid with a constant pressure, $p = p(A_{0})$, given by Eq.~(\ref{Ch3-5.5}).  
In this case, it can be shown that $K$ is free of any spacetime singularity at the center only when $m = 0$. 

It should be noted that in~\cite{IM} it was shown that
non-singular static solutions of a perfect fluid without heat flow do not exist. Since their conclusions
only come from the conservation law of momentum, one might expect that this is also true in the 
current setup. However,  from Eq.~(\ref{Ch3-3.3j}) we can see that in the present case the conservation law
contains  two extra terms, $J_{\varphi}$ and $J_{A}$.  Only when both of them vanish can one obtain the above
conclusions. Since in general one can only choose one of them to be zero by using the  gauge freedom,
it is expected that non-singular static stars can be constructed by properly choosing the gauge field $A$. 

It should be also noted that the arguments presented in~\cite{IM} do not apply to the  case where the pressure 
is a constant. Therefore, when $p' = 0$ non-singular stars without heat flow
can be also constructed in other versions of the HL theory, although  when $p' \not=0$ this is possible
only in the HMT setup.  In addition, the definitions of the quantities $\rho,\; p$ and $v \; (\equiv q e^{\mu})$ used
in this chapter are different from the ones usually used in General Relativity. 

\section{Junction Conditions}

To consider the junction conditions across the hypersurface of a compact object, let us first divide the whole spacetime into three
regions, $V^{\pm}$ and $\Sigma$, where $V^{-} \; (V^{+})$ denotes the internal (external) region of the star, and $\Sigma$ is 
the surface of the star.  Once  the metric is cast in the form (\ref{Ch3-3.1b}), the 
coordinates $t$ and $r$ are all uniquely defined, so that  the coordinates defined in $V^{+}$ and $V^{-}$ must
 be the same,  
$ \big\{x^{+\mu}\big\} = \big\{x^{-\mu}\big\} = (t, r, \theta,\phi)$. 
Since the quadratic terms of the highest derivatives of the metric coefficients $\mu$ and $\nu$ are only terms of the forms,
$\nu''^{2},\; \nu''\nu'''$ and $\mu'^{2}$,  the minimal requirements for these two functions are that   
$\nu(r)$ and $\mu(r)$ are respectively at least $C^{1}$  and $C^{0}$
across the surface $\Sigma$,  
and that  
they are at least $C^{4}$ and $C^{1}$ elsewhere. For detail, we refer readers to~\cite{GPW}. Similarly, 
the quadratic terms of the Newtonian pre-potential are only involved with   the forms,
 $\varphi\varphi'',\; \varphi^{2}_{,r}$, and $\varphi\varphi'$. Therefore, the minimal requirement for $\varphi$ is
to be at least $C^{0}$ across the surface $\Sigma$. On the other hand,  the gauge field $A$ 
 and its derivatives all appear linearly. Thus, mathematically it can even be non-continuous across  $\Sigma$. However, in this chapter
 we shall require that $A$ be  at least $C^{0}$ too across $\Sigma$. Elsewhere,  $A$ and $\varphi$ are at least $C^{1}$. 
Then,  we can write $A$ and $\varphi$ in the form,
\bq
\lb{7.1}
E(r) = E^{+}(r) H\left(r-r_0\right) + E^{-}(r)\left[1 - H\left(r-r_0\right)\right], 
\eq
where $E = (A, \varphi)$, $r_{0}$ is the radius of the star, and $H(x)$ denotes the Heaviside function, defined as
\bqn
\lb{7.2}
H(x) = 
\begin{cases}
1,\,\,\,\,\,  x > 0,\\
0,\,\,\,\,\,  x < 0.
\end{cases}
\eqn
Since $A$ and $\varphi$ are continuous ($C^{0}$) across $r = r_0$, we must have  
\bq
\lb{7.3}
{\mbox{limit}}_{r \rightarrow r_{0}^{+}}{E^{+}(r) } = {\mbox{limit}}_{r \rightarrow r_{0}^{-}}{E^{-}(r)}.
\eq
Then, we find that
\bqn
\lb{7.4}
E'(r)&=& E^{D}_{,r}(r), \nb\\
  E''(r) &=& E^{D}_{,rr}(r) + \left[E'\right]^{-}\delta\left(r -r_0\right), 
\eqn
where  
\bqn
\lb{7.5}
 \left[E'\right]^{-} &\equiv &
 {\mbox{limit}}_{r \rightarrow r_{0}^{+}}{E^{+}_{,r}(r) } -  {\mbox{limit}}_{r \rightarrow r_{0}^{-}}{E^{-}_{,r}(r)},\nb\\
 E^{D}(r)  &\equiv& E^{+}H(r-r_0) + E^{-}\Big[1- H(r-r_0)\Big].
 \eqn
Combining the above with Eq.~(6.6), we find that  
\bqn
\lb{7.6a}
{\cal{L}}_{K} &=& {\cal{L}}_{K}^{D},\;\;\; {\cal{L}}_{A} = {\cal{L}}_{A}^{D},\nb\\
{\cal{L}}_{V} &=&  {\cal{L}}_{V}^{ D}  +  {\cal{L}}_{V}^{ Im}\delta(r - r_0), \nb\\
{\cal{L}}_{\varphi} &=&  {\cal{L}}_{\varphi}^{ D}   +  {\cal{L}}_{\varphi}^{Im} \delta(r - r_0),
\eqn
where  
\bqn
\lb{7.6b}
{\cal{L}}^{Im}_{V} &\equiv&   \frac{8g_{7}e^{-6\nu}}{\zeta^{4}r^{3}}   
\Big[2r\nu' - \big(1 - e^{2\nu}\big)\Big]   \left[\nu''\right]^{-},\nb\\
{\cal{L}}_{\varphi}^{Im} &\equiv&  \frac{\varphi e^{-4\nu}}{ r^{2}}   
\Big[\Lambda_{g}r^2 e^{2\nu}   + \big(1 - e^{2\nu}\big)\Big]  \left[\varphi'\right].~~~
\eqn
Setting
\bq
\lb{7.7}
J = J^{D} + J^{Im}\delta(r -r_0),
\eq
where $
J \equiv \{J^{t},\; v, \; J_{\varphi},\; J_{A}\}$,  and
$J^{Im}$ has support only on $\Sigma$,
we find that the Hamiltonian constraint
(\ref{Ham-const}) can be written as
 \bqn
 \lb{7.8}
& & \int^{D}{\left( {\cal{L}}_{K} + {\cal{L}}_{V} - {\cal{L}}_{\varphi}^{(1)} - 8 \pi G J^{t} \right) e^{\nu} r^{2} dr}\nb\\
& & ~~~~ =  \frac{1}{4\pi} \Big(8\pi G J^{t, Im} + {\cal{L}}^{Im}_{\varphi} -  {\cal{L}}^{Im}_{V}\Big),
 \eqn
where 
\bq
\lb{7.9}
 \int^{D}{I(r) dr} = {\mbox{limit}}_{\epsilon \rightarrow 0}\left(\int_{0}^{r_0 - \epsilon}{I(r) dr}
 + \int_{r_0 + \epsilon}^{\infty}{I(r) dr}\right).
 \eq
It should be noted that in writing Eq.~(\ref{7.8}), we had used the conversion
\bq
\lb{7.10}
\int{\sqrt{g}d^{3}x  f(r) \delta(r - r_0)} = f(r_0).
\eq

The momentum constraint (\ref{mom-const}) will take the same form in Regions $V^{\pm}$, while on the surface
$\Sigma$ it yields
\bq
\lb{7.11}
v^{Im} = 0.
\eq
That is, the surface does not support impulsive heat flow  in the radial direction. Similarly,  
Eqs.~(\ref{phi-const}) and (\ref{A-const}) take the same forms in Regions $V^{\pm}$. While on $\Sigma$ they reduce,
respectively, to
\bqn
\lb{7.12a}
& & \Big[\Lambda_{g} r^2 e^{2\nu} + \big(1 - e^{2\nu}\big)\Big]\left[\varphi'\right]^{-} = 8\pi G r^2 e^{4\nu} J_{\varphi}^{Im}, ~~~~~ \\
\lb{7.12b}
& & J_{A}^{Im} = 0,\; (r = r_0).
\eqn

 On the other hand, the dynamical equations (\ref{Ch3-3.3g}) and (\ref{Ch3-3.3h}) take the same forms in Regions $V^{\pm}$, 
 and  on the surface $\Sigma$
 they yield, 
 \bqn
 \lb{7.13a}
& &\Bigg\{\varphi e^{-2\nu} \Big[\Lambda_{g} r^2 e^{2\nu} + \big(1 - e^{2\nu}\big)\Big]\left[\varphi'\right]^{-}\nb
 + 2 r^{2}F^{\varphi, Im}_{rr}  \Bigg\}\delta(r -r_0) \nb\\
& & ~~ =   -2 r^{2}\Big(F^{Im}_{rr}  + 8\pi G e^{2\nu} p^{Im}_{r}\Big),\\
 \lb{7.13b}
& & \Bigg\{\left[\mu'\right]^{-} + \frac{1}{2} e^{2(\nu-\mu)} {\cal{L}}_{\varphi}^{Im} \nb
 + \frac{e^{2(\nu - \mu)}}{r^{2}}\Big(F^{\varphi, Im}_{\theta\theta} + F^{A, Im}_{\theta\theta}\Big)\Bigg\}\delta(r - r_0)\nb\\
& & ~~ = - \frac{e^{2(\nu - \mu)}}{r^{2}}\Big(F^{Im}_{\theta\theta} + 8\pi Gr^2 p^{Im}_{\theta}\Big),
\eqn
where $F^{Im}_{ij}$ are given below.

\bqn
 F^{Im}_{rr} &=& \frac{2e^{-4\nu}}{\zeta^{4}r^{2}}\Bigg\{\Bigg[4g_{5}\bigg(3\nu' - \frac{1}{r}\big(1 - e^{2\nu}\big)\bigg)[\nu'']^{-}\nb
 - 8g_{7} \bigg([\nu^{(3)}]^{-}  - 10 \nu' [\nu'']^{-}\bigg)\nb\\
 & & ~~~~~~~~~~~~~
 + g_{8} \bigg(3[\nu^{(3)}]^{-}  - 34 \nu' [\nu'']^{-}\bigg)\Bigg]\delta(x)\nb
 - \big(8g_{7} - 3g_{8}\big)  [\nu'']^{-}\delta'(x)\Bigg\},\nb\\
  F^{Im}_{\theta\theta} &=& \frac{(16g_{2} + 3g_{3}) re^{-4\nu}}{\zeta^{2}}[\nu'']^{-}\delta(x)\nb
+ \frac{e^{-6\nu}}{r\zeta^{4}}\Bigg\{48g_{4} \Big[2r\nu' - \big(1 -e^{2\nu}\big)\Big]\left[\nu''\right]^{-}\nb\\
 & &  + 2g_{5} \Big[18r\nu' - 7 \big(1 - e^{2\nu}\big)\Big] \left[\nu''\right]^{-}\nb
 + 3g_{6} \Big[5r\nu' - \big(1 - e^{2\nu}\big)\Big]\left[\nu''\right]^{-}\nb\\
 & &   - 8g_{7}\Big[r^{2}\big[\nu^{(4)}\big]^{-} - 16r^{2}\nu' \big[\nu^{(3)}\big]^{-}\nb
 - r^{2}\big(25 \left\{\nu''\right\}^{+} - 101\nu'^{2}\big)\big[\nu''\big]^{-}\nb
 - 2\big(3 - e^{2\nu}\big)\big[\nu''\big]^{-}\Big] \nb\\
  & &  + g_{8}\Big[3r^{2}\big[\nu^{(4)}\big]^{-} - 48r^{2}\nu' \big[\nu^{(3)}\big]^{-}\nb
 - 3r^{2}\big(25 \left\{\nu''\right\}^{+} - 101\nu'^{2}\big)\big[\nu''\big]^{-}\nb\\
 & & ~~~~~~~~~~~~~~~ 
 + \big(r\nu' -14 + 2e^{2\nu}\big)\big[\nu''\big]^{-}\Big]\Bigg\}\delta(x)\nb\\
 & & ~ -  \frac{(8g_{7} - 3g_{8})r}{\zeta^{4}e^{6\nu}}\Bigg(\big[\nu^{(3)}\big]^{-} - 16\nu' \big[\nu''\big]^{-}\Bigg)\delta'(x)\nb
 -  \frac{(8g_{7} - 3g_{8})r}{\zeta^{4}e^{6\nu}}  \big[\nu''\big]^{-} \delta''(x).
  \eqn

From Eqs.~(\ref{7.1}) and (\ref{7.4}) we find that $\left[F_{(\varphi, n)}\right]_{ij}$ given by
Eq.~(\ref{B.2}) takes the form
\bq
\lb{C.0}
\left[F_{(\varphi, n)}\right]_{ij} = \left[F_{(\varphi, n)}\right]_{ij}^{D} + \left[F^{Im}_{(\varphi, n)}\right]_{ij}\delta(r-r_0),
\eq
where
\bqn
\lb{C.1}
\left[F^{Im}_{(\varphi, 1)}\right]_{ij} &=&\frac{\varphi e^{-2\nu}}{r^{2}}\Bigg[e^{2\nu}\Big(1-\Lambda_{g}r^{2}\Big) - \big(1 + r\nu'\big)\Bigg] \nb
\left[\varphi'\right]^{-}\delta^{r}_{i}\delta^{r}_{j}\nb\\
& &  + \frac{1}{2}\varphi e^{-4\nu}\Bigg[r\nu' - \Big(1- e^{2\nu}\Big)\Bigg]\left[\varphi'\right]^{-}\Omega_{ij},\nb\\
%
%
\left[F^{Im}_{(\varphi, 2)}\right]_{ij} &=&\frac{\varphi e^{-2\nu}}{2r^{2}}\Bigg[e^{2\nu}\Big(\Lambda_{g}r^{2}-1\Big) + 1\Bigg]
\left[\varphi'\right]^{-} \delta^{r}_{i}\delta^{r}_{j}\nb\\
& &  + \frac{1}{2}r\varphi  e^{-4\nu}\Bigg[\Big(\nu' - \Lambda_{g}r e^{2\nu}\Big)\left[\varphi'\right]^{-} 
  + \Big(\varphi' + 2 e^{\mu +\nu}\Big)\left[\nu'\right]^{-}\Bigg]\Omega_{ij},\nb
\eqn
\bqn
\left[F^{Im}_{(\varphi, 3)}\right]_{ij} &=&\frac{\varphi\nu'  e^{-2\nu}}{r} \left[\varphi'\right]^{-} \delta^{r}_{i}\delta^{r}_{j}\nb
%
  + \frac{1}{2} r e^{-4\nu}\Bigg[2\Big(\varphi' - \varphi\nu' + e^{\mu + \nu}\Big)\left[\varphi'\right]^{-} \nb\\
& &~~ -  \varphi\Big(\varphi' + 2 e^{\mu +\nu}\Big)\left[\nu'\right]^{-} \Bigg]\Omega_{ij}.
\eqn

\bqn
\lb{7.14}
F^{A,Im}_{ij} &=& r^{2}e^{-2\nu} \left[A'\right]^{-}\Omega_{ij},\nb\\
L &=& L^{D} + L^{Im},
\eqn
with $L \equiv (p_{r},\; p_{\theta})$. We can see that $L^{Im}$ in general takes the form,
\bq
\lb{7.15}
L^{Im} = L^{(0) Im} \delta(r - r_0) +  L^{(1) Im} \delta'(r - r_0) + L^{(2) Im} \delta''(r - r_0).
\eq
The above represents the general junction conditions of a spherical compact object made of a fluid with heat flow, in which a thin matter  shell 
appears on $\Sigma$. 

In the following we shall consider the matching of the perfect fluid solutions.  Since 
\bq
\lb{7.16}
\nu^{\pm} = 0,\;\;\; \varphi^{\pm} = 0,
\eq
we immediately obtain
\bqn
\lb{7.17}
R^{\pm}_{ij} &=& 0,\;\;\;   {\cal{L}}_{V}^{\pm} = 2\Lambda_{\pm},\;\;\;  {\cal{L}}_{V}^{Im} =0,\nb\\
F^{\pm}_{ij} &=& - \Lambda_{\pm}g^{\pm}_{ij}, \;\;\; F^{Im}_{ij} = 0, \nb\\
{\cal{L}}_{\varphi}^{\pm} &=& {\cal{L}}_{\varphi}^{Im} =0, \;\;\;
\left(F_{\varphi}^{\pm}\right)_{ij}   = \left(F_{\varphi}^{Im}\right)_{ij} = 0.  
\eqn
Then, from Eqs.~(\ref{7.11}), (\ref{7.12a}), (\ref{7.12b}) and (\ref{7.13a}) we find
\bq
\lb{7.17a}
v^{Im} = J^{Im}_{\varphi} = J^{Im}_{A} = p^{Im}_{r} = 0.
\eq
That is, the radial pressure of the thin shell must vanish. This is similar to what happened in the relativistic case~\cite{Santos85}.

In the external region, $V^{+}$, the spacetime is vacuum, 
and the general solutions are given by Eq.~(\ref{4.4})  
\bq
\lb{7.18}
\mu^{+} = \frac{1}{2}\ln\Bigg[\frac{2m}{r} + \frac{1}{3}\Lambda_{+} r^2 - 2A^{+}(r) + \frac{2}{r}\int^{r}{A^{+}(r') dr'}\Bigg],
\eq
for which we have
\bq
\lb{7.19}
{\cal{L}}_{K}^{+}  = \frac{4 A^{+}_{,r}}{r} - 2\Lambda_{+},\;\;\; {\cal{L}}_{A}^{+} = 0.
\eq

In the internal region, $V^{-}$, two classes of solutions of perfect fluid without heat flow are found and given, respectively, by
Eqs.~(\ref{Ch3-5.7a}) and (\ref{Ch3-5.7b}), which can be written as 
 \bqn
 \lb{7.20a}
 \mu^{-} &=& \frac{1}{2}\ln\Bigg\{ \frac{1}{3}\Big(\Lambda_{-} - {8\pi G p_{0}}\Big)r^2 - 2A^{-}(r)\nb
 + \frac{2}{r}\int^{r}{A^{-}(r') dr'}\Bigg\}
 \eqn
 for $\Lambda^{-}_{g} = 0$, and
 \bq
 \lb{7.20b} 
 \mu^{-} = \frac{1}{2}\ln\Bigg\{\frac{1}{3}\Big[\Lambda_{-} - \Lambda^{-}_{g}A_{0} - {8\pi G p}\Big]r^2\Bigg\}
 \eq
 for  $A^{-} = A_{0}$, where $p \equiv \Lambda^{-}_{g}A_{0}^{2}/2 + p_{0}$. Then, we find that
 \bqn
 \lb{7.21}
 {\cal{L}}_{K}^{-} &=& 
\begin{cases}
 2\big(8\pi G p_{0} - \Lambda_{-}\big) + \frac{4A^{-}_{,r}}{r}, &  \Lambda_{g}^{-} = 0,\\
  2\big(8\pi G p - \Lambda_{-} + \Lambda_{g}^{-}A_{0}\big), & A^{-} = A_{0},
\end{cases}  
  \nb\\
  {\cal{L}}_{A}^{-} &=& 2\Lambda_{g}^{-} A^{-}(r).
 \eqn
 To further study the junction conditions, let us consider the two cases $\Lambda_{g}^{-} = 0$ and
 $A^{-} = A_{0}$ separately.
 
 \subsection{$\Lambda_{g}^{-} = 0$}
 
 In this case, the continuity  conditions of $\mu$ and $A$ across $\Sigma$ read,
 \bqn
 \lb{7.22}
  {m} + \frac{1}{6}\Delta\Lambda r^{3}_{0} + \int^{r_{0}}{\Delta{A}(r) dr} &=& - \frac{4\pi G }{3} p_{0} r^{3}_{0},\nb\\
  A^{+}(r_{0}) &=& A^{-}(r_{0}),
 \eqn
where $ \Delta\Lambda \equiv \Lambda_{+} - \Lambda_{-}$ and $\Delta{A} = A^{+} - A^{-}$.
Then, the Hamiltonian constraint (\ref{7.8}) becomes
\bq
\lb{7.23}
\int^{r_{0}}_{0}{\tilde{\rho}(r) dr} + \int^{\infty}_{r_{0}}{r A^{+}_{,r}(r) dr} = \frac{1}{2} G J^{t, Im},
\eq
while the dynamical equation (\ref{7.13b}) reduces to
\bq
\lb{7.24}
\Delta\Lambda = - 8 \pi G \Big(p_{0} + 2 p^{(0) Im}_{\theta}\Big),
\eq
where $p^{Im}_{\theta} \equiv p^{(0) Im}_{\theta}\delta(r-r_{0})$ [cf. Eq.~(\ref{7.15})]. 

When the matter thin shell
does not exist, we must set $J^{t, Im} = p^{(0) Im}_{\theta} = 0$, and Eqs.~(\ref{7.22})--(\ref{7.24}) become the matching conditions 
for the constants $\Lambda_{\pm},\; m,\; p_{0}$ and the functions $A^{\pm}(r)$ and $\tilde{\rho}(r)$.

 \subsection{$A = A_{0}$}
 
 In this case, it can be shown that the continuity  conditions for $\mu$ and $A$  become
 \bqn
 \lb{7.22a}
  {m} + \frac{1}{6}\Delta\Lambda r^{3}_{0} + \int^{r_{0}}{{A}^{+}(r) dr} &=& r_{0}A_{0} \nb
 - \frac{1}{6}\big(\Lambda^{-}_{g}A_{0} + 8\pi G  p\big) r^{3}_{0},\nb\\
  A^{+}(r_{0}) &=& A_{0},
 \eqn
while the Hamiltonian constraint (\ref{7.8}) reduces to
\bq
\lb{7.23a}
\int^{r_{0}}_{0}{\tilde{\rho}(r) dr} + 4 \int^{\infty}_{r_{0}}{r A^{+}_{,r}(r) dr} = {2} G J^{t, Im}.
\eq
The dynamical equation (\ref{7.13b}), on the other hand,  yields
\bq
\lb{7.24a}
\Delta\Lambda =  - \Lambda^{-}_{g}A_{0} - 8 \pi G \Big(p + 2 p^{(0) Im}_{\theta}\Big). 
\eq
 
In all the above cases, one can see that the matching is possible even without a thin matter
shell on the surface of the star, $J^{t, Im} = 0 = p^{(0) Im}_{\theta}$, by properly choosing the
free parameters.

 \section{Conclusions}

 In this chapter, we have systematically studied spherically symmetric static spacetimes generally
filled with an anisotropic 
fluid with heat flow along the radial direction. When the spacetimes are vacuum, we have found  solutions,
given explicitly by Eqs.~(\ref{Ch3-20}) and (\ref{Ch3-23}), from which one can see that the solution is not unique because the gauge 
field $A$ is  undetermined. When $A = 0$, the solutions reduce to the Schwarzschild (anti-) de Sitter solution.
We have also studied the solar system tests and found the constraint on the choice of  $A$.
 
 It should be noted that we have adopted a different point of view  of the gauge 
 field $A$ in the IR limit than that adopted in~\cite{HMT}. In this chapter we have considered it as independent from the
4D metric $g_{\mu\nu}$, although it interacts with them through the field equations. This is 
 quite similar to the Brans-Dicke (BD) scalar field in the BD theory, where the scalar field represents a degree of freedom
 of gravity,  is independent of the metric, and its effects to the spacetime are only through the field equations ~\cite{BD}. 
 On the contrary, in~\cite{HMT} the authors considered the gauge field $A$ as a part of the lapse function,
 $g_{tt} \simeq - (N - A)^2$ in the IR limit.

 We have also investigated anisotropic fluids with heat flow, and found   perfect fluid solutions, given 
 by Eq.~(\ref{Ch3-5.7a}). By properly choosing the gauge field $A$, the solutions can be free of
spacetime singularities at the center. This is in contrast to other versions of the HL theory~\cite{IM} due to 
the coupling of the fluid with the gauge field. We then have considered two particular cases in 
which  the pressure is a constant,  quite similar to the Schwarzschild perfect fluid solution. In all these cases, the
spacetimes are free of singularities at the center. 
 
 For a compact object, the spacetime outside of it  is vacuum and matching conditions  are needed across the surface of the star. With the minimal requirement that the junctions be mathematically meaningful,  we have worked out the general matching conditions, given by Eqs.~(\ref{7.8}) and (\ref{7.11})--(\ref{7.13b}),   in which a thin matter shell in general appears on the surface of the star. Applying them to the perfect   fluids, where the spacetime outside is described by the vacuum solutions (\ref{Ch3-23}), we have found   the matching conditions in terms of the free parameters of the solutions. When the thin shell is removed, these   conditions can also be satisfied by properly choosing the free parameters.
  
Finally, we note that da Silva argued, in the HMT setup, that the coupling constant $\lambda$ can still be different from one~\cite{Silva}. If this is indeed the case,  then one might be concerned with  the  strong coupling problem found in other versions of the  HL theory~\cite{Blas:2009yd,WWb,SC}. However, since the spin-0 graviton is eliminated completely here, as shown explicitly in~\cite{HMT, Silva, Wang:2010wi},    this question is automatically solved in the HMT setup even with $\lambda \not= 1$. It should be noted that    da Silva considered only perturbations of the case with the detailed balance condition, and found that the spin-0 mode is not propagating.     It is not clear if it is also true for the case without detailed balance. The problem certainly deserves further investigations.

\chapter{Gravitational Collapse}
\renewcommand{\theequation}{4.\arabic{equation}} \setcounter{equation}{0}

\begin{center}
\begin{singlespace}
\noindent This chapter published as:   J.~Greenwald, J.~Lenells, V.~H.~Satheeshkumar and A.~Wang,
  ``Gravitational collapse in Ho\v{r}ava-Lifshitz Theory,''  Phys.\ Rev.\ D {88}, 024044 (2013).
\end{singlespace}
\end{center}

The study of gravitational collapse provides useful insights into the {final} fate of a massive star~\cite{Joshi}. Within the framework of General Relativity, the dynamical collapse of a 
homogeneous spherical dust cloud under its own gravity was first considered by Datt~\cite{Datt} and Oppenheimer and Snyder~\cite{OppSnyder}. It was shown that it always leads to the formation of singularities. However, in a theory of quantum gravity,  it is expected that the formation of singularities in a gravitational collapse is prevented by short-distance quantum effects. 

In this chapter, we study gravitational collapse of a spherical star with a finite radius in the HL theory with the projectability condition,  an arbitrary coupling constant $\lambda$, and the extra $U(1)$ symmetry. In General Relativity, there are two common approaches for such studies. One approach relies on Israel's junction conditions~\cite{Israel}, which are essentially obtained by using the Gauss and Codazzi equations. An advantage of this method is that it can be applied to the case where the coordinate systems inside and outside a collapsing body are different. {Although Israel's method was initially developed only for non-null hypersurfaces, it was later generalized to the null hypersurface case~\cite{BI}. For a recent review of this method, we refer to~\cite{WS} and references therein.}
The other approach is originally due to Taub~\cite{Taub} and relies on distribution theory. In this approach, although the coordinate systems inside and outside the collapsing stars 
are taken to be the same, the null-hypersurface case can be easily included. Taub's approach was widely used to study colliding gravitational waves and other related issues
in General Relativity~\cite{Wang90s}. 

We follow Taub's approach, as it turns out to be more convenient when dealing with higher-order derivatives. Moreover, in contrast to the case of General Relativity, the foliation structure of  the HL theory implies that the coordinate systems inside and outside of the collapsing star are unique. Thus, {also} from a technical point of view, Taub's method seems a natural choice for the study of a collapsing star with a finite radius in the HL theory. 

\section{ Spherical Spacetimes Filled with a fluid}

Spherically symmetric static spacetimes in the framework of the HL theory with $U(1)$ symmetry   with or without  the projectability condition are studied systematically in~\cite{GSW,AP,LMW,LW,GLLSW,BLW,Alexandre:2013ura}.  In particular, the
 ADM variables  for   spherically symmetric spacetimes    with the projectability condition take the  forms
\bqn
\lb{3.1b}
N &=& 1,\nb\\
N^i &=& \delta^{i}_{r} e^{\mu(r, t) - \nu(r, t)},\nb\\
g_{ij}dx^idx^j &=&  e^{2\nu(r, t)} dr^2   + r^{2}d\Omega^2,
\eqn
in the spherical coordinates $x^{i} = (r, \theta, \phi)$, where $d\Omega^2 \equiv
d\theta^{2}  + \sin^{2}\theta\, d\phi^{2}$. The diagonal case $N^i = 0$ corresponds to $\mu(t, r) = -\infty$. {On the other hand, using}  the $U(1)$ gauge freedom, without loss of generality, we set
\bq
\lb{gauge}
\varphi = 0,
\eq
which  uniquely fixes the gauge. Then, we  find that  
 \bqn
\lb{3.3a}
{\cal{L}}_{\varphi}  &=& 0 = {\cal{L}}_{\lambda},\;\;\; F^{ij}_{\varphi} = 0,\nb\\
K_{ij} &=& e^{\mu+\nu}\Big((\mu'-\dot{\nu} e^{-\mu+\nu}) \delta^{r}_{i}\delta^{r}_{j} + re^{-2\nu}\Omega_{ij}\Big),\nb\\
R_{ij} &=&  \frac{2\nu'}{r}\delta^{r}_{i}\delta^{r}_{j} + e^{-2\nu}\Big[r\nu' - \big(1-e^{2\nu}\big)\Big]\Omega_{ij},\nb\\
{\cal{L}}_{K} &=& (1 - \lambda) \Bigg[\dot{\nu}^2 - 2\dot{\nu} \mu' e^{\mu-\nu} + \left( {\mu'}^2 + \frac{2}{r^{2}} \right) e^{2(\mu-\nu)} \Bigg]\nb\\
&& + \lambda \left[ \frac{4}{r} \dot{\nu} e^{\mu-\nu} - \frac{2}{r^{2}} e^{2(\mu-\nu)}\left(2r\mu' + 1\right)\right]  \nb\\ 
{\cal{L}}_{A} &=&  \frac{2A}{r^2} \Big[e^{-2\nu}\left(1 - 2r \nu'\right) + \Lambda_{g} r^2 - 1\Big],\nb\\
{\cal{L}}_{V} &=& \sum_{s=0}^{3}{{\cal{L}}_{V}^{(s)}},
\eqn
where  a prime denotes the partial derivative with respect to $r$, 
  $\Omega_{ij} \equiv \delta^{\theta}_{i}\delta^{\theta}_{j}  + \sin^{2}\theta\delta^{\phi}_{i}\delta^{\phi}_{j}$,
  and ${\cal{L}}_{V}^{(s)}$'s are given by Eq.~(\ref{LVs}). 
The Hamiltonian constraint (\ref{eq1}) reads
 \bq 
 \lb{3.3b}
\int{\left( {\cal{L}}_{K} + {\cal{L}}_{V}  - 8 \pi G J^{t} \right) e^{\nu} r^{2} dr}
= 0,
 \eq 
 while the momentum constraint (\ref{eq2}) yields
 \bqn
 \lb{3.3c}
  && (1-\lambda)\Big\{e^{\mu - \nu}\left[r^2 (\mu'' + \mu'^2 - \mu' \nu') + 2 (\mu' r - 1)\right] 
 - \dot{\nu}' r^2\Big\}  + 2 r \left(\lambda \nu' e^{\mu - \nu} - \dot{\nu} \right)  \nb\\
  && ~~~~~~~~~~~  =  - 8 \pi G r^2 e^{-\mu + \nu}  v,
 \eqn
 where 
  \bqn 
\nb
 J^{i} \equiv e^{-(\mu + \nu)}\big(v, 0, 0\big).
 \eqn
 It can also be shown that Eqs.~(\ref{eq4a}) and (\ref{eq4b}) now read
 \bqn
 \lb{3.3e}
& &  \Big[e^{2\nu}\left(\Lambda_{g}r^2 -1\right) + 1\Big] \Big( e^{\mu + \nu}\mu' - e^{2 \nu} \dot{\nu} \Big)\nb
 -2\Big(\nu' - \Lambda_{g}re^{2\nu}\Big) e^{\mu + \nu} \nb\\
&& + (1-\lambda)\Big\{e^{2\nu}\left( -r^2 \dot{\nu}'' + r^2 \dot{\nu}' \nu' - 2 r \dot{\nu}' \right) \nb\\
&& + e^{\mu + \nu} \left[ r^2 ({\mu}''' + 3 \mu' {\mu}'' - {\mu}' \nu'' - 3 {\mu}'' \nu' + \mu'^3 - 3 \nu' \mu'^2 \right. \nb\\ 
&&\left. + 2 \mu' \nu'^2 ) + 2r(2 \mu'' -\nu'' + 2 (\nu' -\mu')^2 ) + 2 \nu' \right]\Big\}\nb\\ 
&&= 8\pi G r^{2} e^{4\nu} J_{\varphi}, ~~~~~~~~ \\
\lb{3.3f}
& & 2 r \nu'  - \Big[e^{2\nu}\left(\Lambda_{g}r^2 - 1\right) +  1\Big] =   4\pi G r^2 e^{2\nu}  J_{A}.
\eqn
 The dynamical equations (\ref{eq3}), on the other hand, yield
 \bqn
 \lb{3.3g}
 && (1-\lambda)r \left[ e^{\nu+\mu} \big( \dot{\mu} \mu' + \dot{\mu}' -\dot{\nu}' \big) - e^{2\nu} \big( \ddot{\nu} + \frac{1}{2}{\dot{\nu}}^{2}\big) 
  + e^{2\mu} \big( \mu'' + \frac{1}{2}\mu'^{2} - \mu' \nu' \big) \right] \nb \\ 
& & + \left[2 \big(\mu' + \lambda \nu'\big)+ (4 \lambda - 3)\frac{1}{r}\right]e^{2\mu}  -2e^{\nu+\mu} \big(\lambda \dot{\mu} + \dot{\nu} \big) \nb\\
 & & + \frac{1}{2} r e^{2\nu} {\cal{L}}_{A} = - r\Big(F_{rr}  + F^{A}_{rr} + 8\pi G e^{2\nu}p_{r}\Big), \\
 \lb{3.3h}
  & &  \left[\lambda r \Big( \mu'' - \mu' \nu' \Big) + (2\lambda - 1) \big(2 \mu' - \nu' \big) 
+ \frac{1}{2} (3\lambda + 1) r \mu'^2 \right]e^{2\mu}  + \frac{1}{2} r e^{2\nu} {\cal{L}}_{A}  \nb\\
  && ~ + \Big( \lambda  \ddot{\nu} + \frac{1}{2} (\lambda + 1)  \dot{\nu}^2 \Big) r e^{2\nu}  
- \left[(2\lambda - 1) \dot{\mu} + r \mu' \big( \dot{\nu} + \lambda \dot{\mu} \big) + \lambda r \big( \dot{\nu}'+ \dot{\mu}' \big) \right] e^{\nu + \mu} \nb\\
 & &  = -\frac{e^{ 2\nu}}{r}\left(F_{\theta\theta} + F^{A}_{\theta\theta} + 8\pi G r^{2}p_{\theta}\right),
 \eqn
 where
 \bqn
 \tau_{ij} &=& e^{2\nu}p_{r}\delta^{r}_{i}\delta^{r}_{j} + r^{2}p_{\theta}\Omega_{ij},\nb\\ 
F^{A}_{ij} &=&\frac{ 2}{r}\big(A' + A\nu'\big) \delta^{r}_{i}\delta^{r}_{j}  +  e^{-2\nu}\Big[r^{2}\big(A'' - \nu'A'\big)\nb
+ r\big(A' + A\nu'\big)  - A\Big(1 - e^{2\nu}\Big)\Big]\Omega_{ij}, 
\eqn
 and $F_{ij}$   is given by 

\bqn
\lb{A.4}
(F_0)_{ij} & = & -\frac{e^{2 \nu }}{2} \delta_i^r\delta_j^r
-\frac{r^2}{2} \Omega_{ij},\nb\\
(F_1)_{ij} & = & \frac{1-e^{2 \nu }}{r^2} \delta_i^r\delta_j^r
-e^{-2 \nu } r \nu ' \Omega_{ij},\nb
\eqn
\bqn
(F_2)_{ij} & = & -\frac{2 e^{-2 \nu } }{r^4} \Big[6 e^{2 \nu }+e^{4 \nu }-8 r^2 \nu ''\nb
+12 r^2  \left(\nu '\right)^2-7\Big] \delta_i^r\delta_j^r
   	\nb\\
&& +\frac{2 e^{-4 \nu }}{r^2} \Big[6 e^{2 \nu }+e^{4 \nu }+4 \nu ^{(3)} r^3+24
   r^3 \left(\nu '\right)^3\nb\\
   &&  -2 r \nu ' \left(-3 e^{2 \nu }+14 r^2 \nu
   ''+7\right)-7\Big]\Omega_{ij},
	\nb\\
%
%
(F_3)_{ij} & = & -\frac{e^{-2 \nu }}{r^4} \Big[4 e^{2 \nu }+e^{4 \nu }-6 r^2 \nu ''\nb
+9 r^2    \left(\nu '\right)^2-5\Big]  \delta_i^r\delta_j^r
   	\nb\\
&& + \frac{e^{-4 \nu }}{r^2} \Big[4 e^{2 \nu }+e^{4 \nu }+3 \nu ^{(3)} r^3+18 r^3
   \left(\nu '\right)^3\nb\\
   && -r \nu ' \left(-4 e^{2 \nu }+21 r^2 \nu
   ''+10\right)-5\Big] \Omega_{ij},
	\nb\\
%
%
(F_4)_{ij} & = & -\frac{4 e^{-4 \nu } }{r^6} \left(e^{2 \nu }+2 r \nu '-1\right) \Big[22 e^{2
   \nu }+e^{4 \nu }-24 r^2 \nu ''\nb\\
   && +40 r^2 \left(\nu '\right)^2-2
   \left(e^{2 \nu }-1\right) r \nu '-23\Big] \delta_i^r\delta_j^r
   	\nb\\
&& + \frac{4 e^{-6 \nu } }{r^4} \biggl\{ 240 r^4 \left(\nu '\right)^4+4 \left(18 e^{2
   \nu }-17\right) r^3 \left(\nu '\right)^3\nb\\
   && -12 r^2 \left(\nu '\right)^2
   \left(-11 e^{2 \nu }+22 r^2 \nu ''+15\right)
   	\nb\\
&&  +3 r \nu ' \Big[-6 e^{2
   \nu }+7 e^{4 \nu }+8 \nu ^{(3)} r^3\nb
 -28 \left(e^{2 \nu }-1\right) r^2  \nu ''-1\Big]
   	\nb\\
&&    +2 \Big[12 r^4 \left(\nu ''\right)^2-24 \left(e^{2 \nu }-1\right) r^2 \nu''\nb\\
&& +\left(e^{2 \nu
   }-1\right) \left(22 e^{2 \nu }+e^{4 \nu }+6 \nu ^{(3)}
   r^3-23\right)\Big]\biggr\} \Omega_{ij},
	\nb\\
%
%
(F_5)_{ij} & = & -\frac{2 e^{-4 \nu }}{r^6}\biggl\{60 r^3 \left(\nu '\right)^3\nb
 +\left(e^{2 \nu   }-1\right) \left(16 e^{2 \nu }+e^{4 \nu }-14 r^2 \nu''-17\right)\nb\\
   && +\left(21 e^{2 \nu }-17\right) r^2 \left(\nu
   '\right)^2
   	\nb
 -4 r \nu ' \left(-7 e^{2 \nu }+9 r^2 \nu
   ''+7\right)\biggr\} \delta_i^r\delta_j^r
   	\nb\\
&& + \frac{2 e^{-6 \nu }}{r^4} \biggl\{18 r^4 \left(\nu ''\right)^2+180 r^4
   \left(\nu '\right)^4\nb
 +\left(e^{2 \nu }-1\right) \left(32 e^{2 \nu }+2
   e^{4 \nu }+7 \nu ^{(3)} r^3-34\right)
   	\nb\\
&&   +21 \left(2 e^{2 \nu }-1\right)
   r^3 \left(\nu '\right)^3-28 \left(e^{2 \nu }-1\right) r^2 \nu ''\nb
 -r^2   \left(\nu '\right)^2 \left(-77 e^{2 \nu }+198 r^2 \nu
   ''+101\right)
   	\nb\\
&&   +r \nu ' \Big[3 \left(-8 e^{2 \nu }+5 e^{4 \nu }+6 \nu
   ^{(3)} r^3+3\right)\nb
 -\left(49 e^{2 \nu }-41\right) r^2 \nu
   ''\Big]\biggr\} \Omega_{ij},
	\nb
\eqn
\bqn
(F_6)_{ij} & = & \frac{e^{-4 \nu } }{r^6} \biggl\{ -50 r^3 \left(\nu '\right)^3\nb
-\left(e^{2 \nu   }-1\right) \left(13 e^{2 \nu }+e^{4 \nu }-6 r^2 \nu ''-14\right)\nb\\
&&   -9    e^{2 \nu } r^2 \left(\nu '\right)^2
	 +6 r \nu ' \left(-2 e^{2 \nu }+5
   r^2 \nu ''+2\right)\biggr\} \delta_i^r\delta_j^r
   	\nb\\
&& + \frac{e^{-6 \nu }}{r^4}\biggl\{15 r^4 \left(\nu ''\right)^2+150 r^4 \left(\nu'\right)^4\nb
 +\left(e^{2 \nu }-1\right) \left(26 e^{2 \nu }+2 e^{4 \nu   }+3 \nu ^{(3)} r^3-28\right)
   	\nb\\
&&    +\left(18 e^{2 \nu }+25\right) r^3
   \left(\nu '\right)^3-12 \left(e^{2 \nu }-1\right) r^2 \nu ''\nb
-3 r^2   \left(\nu '\right)^2 \left(-11 e^{2 \nu }+55 r^2 \nu ''+12\right)
   	\nb\\
&&   +3
   r \nu ' \Big[-12 e^{2 \nu }+4 e^{4 \nu }+5 \nu ^{(3)} r^3\nb
   -\left(7   e^{2 \nu }-1\right) r^2 \nu ''+8\Big]\biggr\} \Omega_{ij},
	\nb
\eqn
\bqn
(F_7)_{ij} & = & \frac{8 e^{-4 \nu }}{r^6} \biggl\{6 e^{2 \nu }+e^{4 \nu }-2 \nu ^{(4)} r^4+15
   r^4 \left(\nu ''\right)^2\nb
 +40 r^4 \left(\nu '\right)^4 +4 r \nu'   \left(2 e^{2 \nu }+5 \nu ^{(3)} r^3-6\right)   	\nb\\
&& 
  -2 r^2 \left(\nu '\right)^2 \left(-3 e^{2 \nu   }+41 r^2 \nu ''+15\right)\nb
 -4 \left(e^{2 \nu  }-3\right) r^2 \nu '' -7 \biggr\} \delta_i^r\delta_j^r   	\nb\\
&&-\frac{8 e^{-6 \nu }}{r^4} \biggl\{12 e^{2 \nu }+2 e^{4 \nu }+\nu ^{(5)}   r^5\nb
 +120 r^5 \left(\nu '\right)^5+2 e^{2 \nu } \nu ^{(3)} r^3-6 \nu^{(3)} r^3   	\nb\\
&&   +r \nu ' \Big[24 e^{2 \nu }+e^{4 \nu }-16 \nu ^{(4)}   r^4+127 r^4 \left(\nu ''\right)^2\nb
 -2 \left(7 e^{2 \nu }-33\right) r^2   \nu ''-57\Big]   	\nb\\
&&  -r^2 \nu '' \left(8 e^{2 \nu }+25 \nu ^{(3)}   r^3-24\right)\nb
+r^2 \left(\nu '\right)^2 \left(22 e^{2 \nu }+101 \nu ^{(3)} r^3-102\right)   	\nb\\
&&    -2 r^3 \left(\nu '\right)^3 \left(-6 e^{2 \nu   }+163 r^2 \nu ''+45\right)-14\biggr\}\Omega_{ij},
	\nb
\eqn
\bqn
(F_8)_{ij} & = &\frac{e^{-4 \nu }}{r^6} \biggl\{6 \left(2 e^{2 \nu }+e^{4 \nu }-\nu ^{(4)}  r^4-3\right)\nb
 +45 r^4 \left(\nu ''\right)^2+120 r^4 \left(\nu'\right)^4+10 r^3 \left(\nu '\right)^3   	\nb\\
&&    -8 \left(e^{2 \nu }-4\right)   r^2 \nu ''\nb
 -r^2 \left(\nu '\right)^2 \left(-12 e^{2 \nu }+246 r^2 \nu''+77\right)   	\nb\\
&&    +2 r \nu ' \left(8 e^{2 \nu }+30 \nu ^{(3)} r^3-3 r^2
   \nu ''-32\right)\biggr\} \delta_i^r\delta_j^r   	\nb
 + \frac{e^{-6 \nu }}{r^4} \biggl\{-24 e^{2 \nu }-12 e^{4 \nu }-3 \nu ^{(5)}   r^5\nb\\
   && -360 r^5 \left(\nu '\right)^5-3 r^4 \left(\nu ''\right)^2-30 r^4   \left(\nu '\right)^4  
 -4 e^{2 \nu } \nu ^{(3)} r^3+16 \nu ^{(3)}   r^3\nb\\
   && +r^2 \nu '' \left(16 e^{2 \nu }+75 \nu ^{(3)} r^3-64\right)
  +2 r^3   \left(\nu '\right)^3 \left(-12 e^{2 \nu }+489 r^2 \nu
   ''+113\right)\nb\\
   && -r^2 \left(\nu '\right)^2 \left(44 e^{2 \nu }+303 \nu^{(3)} r^3-33 r^2 \nu ''-269\right)  
   -r \nu ' \Big[381 r^4 \left(\nu''\right)^2-2 \left(14 e^{2 \nu }-85\right) r^2 \nu ''\nb\\
   && +6    \left(8 e^{2 \nu }+e^{4 \nu }-8 \nu ^{(4)} r^4-25\right)
 +3\nu ^{(3)}   r^3\Big]+36\biggr\} \Omega_{ij}.
\eqn
where $\nu' = \partial \nu/\partial r$ and $\Omega_{ij} = \delta_i^\theta \delta_j^\theta + \sin^2\theta \delta_i^\phi \delta_j^\phi$.

We define a fluid with $p_{r} = p_{\theta}$ as a perfect fluid, which in general allows  energy flow along a radial direction, i.e.,  $v$ does not not necessarily vanish~\cite{Santos85}. 
 
The energy conservation law (\ref{eq5a}) now reads
\bqn
\lb{3.3ja}
\int dr \,\, e^\nu r^2 \Big[ \dot{\rho}_H + \left( \rho_H + 4 p_r\right)\dot{\nu}  
 + 4 \left( \dot{v} - v\dot{\mu} \right) 
 - 2\left( \dot{J}_A + \dot{\nu} J_A\right) \Big] = 0, \qquad
\eqn
while the momentum conservation law (\ref{eq5b}) yields
\bqn
\lb{3.3jb}
&& v\mu' - \big(v' - p_{r}'\big) - \frac{2}{r}\big(v - p_{r} + p_{\theta}\big)  - \frac{1}{2}J_{A} A' 
 - e^{\nu-\mu} \Big[\dot{v} + v \big(2\dot{\nu} - \dot{\mu}\big)\Big] = 0.
\eqn

 To relate the quantities $J^{t},\; J^{i}$ and $\tau_{ij}$   to the ones often used in General Relativity, 
 in addition to the normal vector $n_{\mu}$ defined in Eq.~(\ref{4Dnorm}), we also introduce  
the spacelike unit vectors $\chi_{\mu}, \; \theta_{\mu}$ and $\phi_{\mu}$ by
\bqn
 n_{\mu} &=& \delta^{t}_{\mu}, \;\;\; n^{\mu} = - \delta_{t}^{\mu} + e^{\mu-\nu}\delta_{r}^{\mu},\nb\\
 \chi^{\mu} &=& e^{-\nu}\delta^{\mu}_{r} , \;\;\; \chi_{\mu} = e^{\mu}\delta^{t}_{\mu} + e^{\nu} \delta^{r}_{\mu},\nb\\
 \theta_{\mu} &=& r\delta^{\theta}_{\mu},\;\;\; \phi_{\mu} = r\sin\theta \delta^{\phi}_{\mu}.
\eqn
In terms of these four unit  vectors,    
the energy-momentum tensor for an anisotropic fluid can be written as
\bqn
\lb{emt2}
T_{\mu\nu} &=& \rho_{H}n_{\mu}n_{\nu} + q \big(n_{\mu} \chi_{\nu} + n_{\nu} \chi_{\mu} \big)
 + p_{r}\chi_{\mu} \chi_{\nu}  + p_{\theta}\big(\theta_{\mu}\theta_{\nu} + \phi_{\mu}\phi_{\nu}\big),
\eqn
where $\rho_{H}, \; q,\; p_{r}$ and $p_{\theta}$ denote, respectively, the energy density, the heat flow 
along the radial direction and the radial and tangential pressures, measured by the observer with the four-velocity
$n_{\mu}$. 
This decomposition is consistent with  the quantities $J^{t}$ and $J^{i}$ defined by
 \bq \lb{rhoHv}
 \rho_{H} = -\frac{1}{2} J^{t},\qquad v= e^{\mu} q. 
\eq
  It should be noted that the definitions of the energy density $\rho_H$, 
 the radial pressure $p_r$ and the heat flow $q$ are different from the ones defined in a comoving frame 
 in General Relativity. 

\section{ Junction Conditions across the Surface of a Collapsing Sphere }

\begin{figure}[tbp]
\centering
\includegraphics[width=6cm]{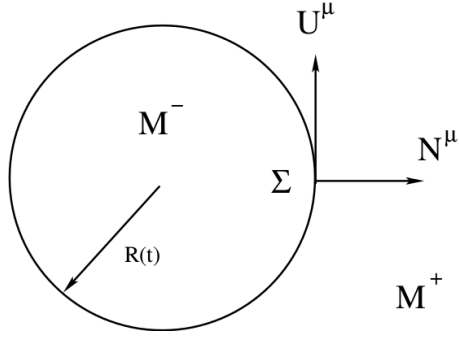} 
\caption{The spacetime is divided into two regions, the internal $M^{-}$ and external $M^{+}$, where 
$M^{-} = \left\{x^{\mu}: r < {\cal{R}} (t)\right\}$, and $M^{+} = \left\{x^{\mu}: r > {\cal{R}}(t)\right\}$. The surface $r = {\cal{R}}(t)$ is denoted by $\Sigma$.}
\label{fig0}
\vspace{0.75cm}
\end{figure}

The surface $\Sigma$ of a spherically symmetric collapsing star naturally divides the spacetime $M$ into 
two regions, the internal and the external regions, denoted by
$M^-$ and $M^+$ respectively, as shown schematically in Fig. \ref{fig0}. 
The surface $\Sigma = \partial M^- = -\partial M^+$ is described by
\bq
\Phi(t, r) = 0,
\eq
where $\Phi(t,r) \equiv r - {\cal{R}}(t)$. The spherical symmetry implies that the ADM variables  on $M$ take the form (\ref{3.1b}).

\subsection{Preliminaries}
We assume that the normal vector $\nabla \Phi$ to the hypersurface $\Sigma$ with components
\bqn \label{gradPhicomponents}
&& \Phi_{,\lambda} = \delta_{\lambda}^{r} - \dot{\cal{R}} \delta^{t}_{\lambda}, 
 	\\ \nb
&& \Phi^{,\lambda} = e^{-2\nu}(1 - e^{2\mu} - \dot{\mathcal{R}} e^{\mu + \nu}) \delta^{\lambda}_{r} + (e^{\mu - \nu} + \dot{\mathcal{R}}) \delta_{t}^{\lambda},
\eqn
is everywhere spacelike, i.e.
\bq
\lb{5.3a}
  \Phi^{,\lambda}\Phi_{,\lambda} = e^{-2\nu}\bigl[1 - (e^\mu + e^\nu \dot{\mathcal{R}})^2\bigr] > 0.
\eq
This is the case if $\dot{\mathcal{R}}$ is small enough. We may then define the vector field $N = \nabla \Phi/\|\nabla \Phi\|_g$ in a neighborhood of $\Sigma$. The vector field $N$ should not be confused with the lapse function which in the present case is set to one, see Eq.~(\ref{3.1b}).    $N$ has length one, i.e. $N_{\lambda}N^\lambda = 1$, and the restriction of $N$ to $\Sigma$ is the 
outward pointing unit normal vector field on $\Sigma$.

Let $H(\Phi)$ denote the Heaviside function defined by
\bqn
\lb{5.3}
H(\Phi) = 
\begin{cases}
1, & \Phi > 0,\\
\frac{1}{2}, & \Phi = 0,\\
0, &  \Phi < 0,
\end{cases} 
\eqn
and let $\delta(\Phi)$ denote the delta distribution with support on $\Sigma$.
By definition, $\delta(\Phi)$ acts on a smooth test function $\varphi \in C^\infty(M)$ of compact support by
\bq
(\delta(\Phi), \varphi) = \int_\Sigma \varphi d\Sigma,
\eq
where $d\Sigma = \iota_N \text{Vol}_g$ is the volume three-form induced by $g$ on $\Sigma$ and $\iota_N$ denotes interior multiplication by $N$. 
The derivatives $\delta^{(n)}(\Phi)$, $n \geq1$, of $\delta(\Phi)$ are defined in a standard way and the following relations are valid~\cite{GS1964}:
\bqn \nb
 \frac{\partial H(\Phi)}{\partial x^\lambda}
& = & \frac{\partial \Phi}{\partial x^\lambda} \delta(\Phi),
	\\ \nb
 \frac{\partial}{\partial x^\lambda} \delta^{(n)}(\Phi) & = & \frac{\partial \Phi}{\partial x^\lambda} \delta^{(n+1)}(\Phi), \qquad n = 0, 1,2, \dots,
 	\\ \label{deltarecursion}
\Phi \delta^{(n)}(\Phi) & =& -n \delta^{(n-1)}(\Phi), \qquad n = 1, 2, \dots.
\eqn

If $f$ is a function defined in a neighborhood of $\Sigma$, we define the distribution $f \delta^{(n)}(\Phi)$ by letting it act on a test function $\varphi$ by
\bq
(f \delta^{(n)}(\Phi), \varphi) = (\delta^{(n)}(\Phi), f \varphi).
\eq
The product $f\delta(\Phi)$ is well defined whenever $f$ is $C^0$ and it depends only on the restriction $f|_\Sigma$ of $f$ to $\Sigma$. 
More generally, the product $f\delta^{(n)}(\Phi)$ is well defined provided that $f$ is $C^n$ and it depends only on the values of $f$ and its 
partial derivatives of order $\leq n$ evaluated on $\Sigma$.

Let $F$ be a distribution on $M$ of the form
\bq\label{Fdef}
F = F^+ H(\Phi) + F^- [1-H(\Phi)] + \sum_{k = 0}^n F^{Im(k)} \delta^{(k)}(\Phi),
\eq
where the $F_n$'s are functions defined in a neighborhood of $\Sigma$ while $F^+$ and $F^-$ are sufficiently smooth functions defined on 
$M^+$ and $M^-$ respectively. We define the function $F^D$ on $M$ by
\bq
F^D = F^+ H(\Phi) + F^- [1-H(\Phi)],
\eq
and we define the jump $[F]^-$ of $F$ across $\Sigma$ by
\bq
[F]^-(x) = F^+(x) - F^-(x), \qquad x \in \Sigma.
\eq
We will also need the fact that the equation $F = 0$ is equivalent to the equations
\bqn\label{Fpm0}
F^\pm(x) = 0, \qquad x \in M^\pm,
\eqn
and
\bqn \nb
&& \sum_{k = 0}^j (-1)^k \frac{(n-k)!j!}{(j-k)!} \frac{\partial^{j-k}}{\partial \Phi^{j-k}} F^{Im(n-k)}\bigg|_{\Sigma} = 0,
	\\ \label{Fdeltaeqs}
&&\hspace{4cm} 0 \leq j \leq n,
\eqn
where $\frac{\partial}{\partial \Phi}$ acts on a function� $f$ by
\bqn\label{dfdPhi}
\frac{\partial f}{\partial \Phi} = \frac{1}{\|\nabla \Phi\|_g} df\cdot N,
\eqn
and, more generally, for any $j \geq 1$,
\bqn\label{partialPhij}
\frac{\partial^j f}{\partial \Phi^j} = \biggl(\frac{1}{\|\nabla \Phi\|_g} \iota_N d \biggr)^j f.
\eqn
A proof of this fact is given at the end of the chapter.

For $n = 3$, the conditions in (\ref{Fdeltaeqs}) are
\bqn \nb
&& F^{Im(3)}|_\Sigma = 0,
	\\ \nb
&&\biggl(3\frac{\partial F^{Im(3)}}{\partial \Phi} - F^{Im(2)}\biggr)\bigg|_\Sigma = 0,
	\\ \label{Fdeltaeqs2}
&&\biggl(3\frac{\partial^2 F^{Im(3)}}{\partial \Phi^2} - 2 \frac{\partial F^{Im(2)}}{\partial \Phi} 
+ F^{Im(1)}\biggr)\bigg|_\Sigma = 0,
	\\ \nb
&&\biggl(\frac{\partial^3 F^{Im(3)}}{\partial \Phi^3} - \frac{\partial^2 F^{Im(2)}}{\partial \Phi^2}  +\frac{\partial F^{Im(1)}}{\partial \Phi}
- F^{Im(0)}\biggr)\bigg|_\Sigma = 0. ~~
\eqn	
	
\subsubsection{Proof of Eqs.~(\ref{Fpm0})--(\ref{Fdeltaeqs}).}
  Let $F$ be given by (\ref{Fdef}). We will show that the equation $F = 0$ is equivalent to the conditions (\ref{Fpm0}) and (\ref{Fdeltaeqs}).
It is clear that the equation $F= 0$ is equivalent to (\ref{Fpm0}) together with the condition
\bqn\label{sumdelta0}
  \sum_{k = 0}^n F^{Im(k)} \delta^{(k)}(\Phi) = 0.
\eqn  
It remains to show that (\ref{sumdelta0}) is equivalent to (\ref{Fdeltaeqs}). 

Suppose first that (\ref{sumdelta0}) holds. Then, multiplying (\ref{sumdelta0}) by $\Phi^{n-j}$ and using the recursion relation (\ref{deltarecursion}) repeatedly, we find
\bq\label{GdeltaPhi}
  G \delta^{(j)}(\Phi) = 0, \qquad 0 \leq j \leq n,
\eq  
where the function $G$ is defined in a neighborhood of $\Sigma$ by
\bq\label{Gdef}
G(x) = \sum_{k = 0}^n (-1)^{k} k! F^{Im(k)}(x) \Phi^{n-k}(x).
\eq
Equation (\ref{GdeltaPhi}) with $j = 0$ implies that the restriction of $G$ to $\Sigma$ vanishes, i.e. $G|_\Sigma = 0$.
Equation (\ref{GdeltaPhi}) with $j = 1$ then gives
$$0 = G \delta'(\Phi) = \frac{G}{\Phi} \Phi \delta'(\Phi)
= -\frac{G}{\Phi} \delta(\Phi) \quad \text{i.e.} \quad \frac{G}{\Phi}\Big|_\Sigma = 0.$$
In terms of local coordinates $\{u^j\}$ such that $u^1 = \Phi$ while the remaining coordinates $\{u^j\}_{j \geq 2}$ parametrize the level surfaces of $\Phi$, we have
$$0 = \frac{G}{\Phi}\Big|_\Sigma = \frac{\partial G}{\partial \Phi}\bigg|_{\Phi = 0}.$$
Thus, $G$ vanishes to the first order on� $\Sigma$. Repeating the above procedure $n$ times, we infer that $G$ vanishes to the $n$th order on $\Sigma$:
\bq\label{partialjGzero}
\frac{G}{\Phi^j}\bigg|_{\Sigma} = 0 \quad \text{i.e.}\quad 
\frac{\partial^j G}{\partial \Phi^j}\bigg|_{\Phi = 0} = 0, \qquad 0 \leq j \leq n.
\eq
The partial derivatives denoted in the local coordinates $(u^j)$ by $\frac{\partial^j}{\partial \Phi^j}$ can be expressed invariantly as in (\ref{partialPhij}). 
Substituting the expression (\ref{Gdef}) for $G$ into (\ref{partialjGzero}), we find
\bqn\nb
0 &=& \sum_{k = 0}^n (-1)^{k} k! \sum_{r=0}^j {j \choose r} \frac{\partial^{j-r} F^{Im(k)}}{\partial\Phi^{j-r}} \frac{\partial^{r}\Phi^{n-k}}{\partial \Phi^r}\bigg|_{\Phi = 0}
	\\ \nb
&=& \sum_{k = n-j}^n (-1)^{k} k! (n-k)! {j \choose n-k} \frac{\partial^{j-(n-k)} F^{Im(k)}}{\partial\Phi^{j-(n-k)}}\bigg|_{\Phi = 0},
	\\ \nb
&&\hspace{4cm} \qquad 0 \leq j \leq n.
\eqn	
Replacing $k$ by $n-k$, we find (\ref{Fdeltaeqs}).

Conversely, if (\ref{Fdeltaeqs}) holds, then tracing the above steps backwards, we infer that (\ref{GdeltaPhi}), and hence also (\ref{sumdelta0}), holds.

\subsection{Distributional Metric Functions}

The field equations (\ref{eq1})--(\ref{eq3}) involve second-order derivatives of the metric coefficients with respect to $t$ and sixth-order derivatives with respect to $x^i$. Thus, one might require that
the metric coefficients be $C^{1}$ with respect to $t$ and $C^{5}$ with respect to $x^{i}$, where $C^{n}$ indicates that the first $n$ derivatives exist and are continuous across the hypersurface
$\Phi = 0$.  However, this assumption eliminates the important case of an infinitely thin shell of matter supported on $\Sigma$. Therefore, we will instead make weaker assumptions, so that a thin shell
located on the hypersurface $\Phi = 0$ is in general allowed, and consider the case without a thin shell only as a particular case of our general treatment to be provided below. In fact, we shall impose the minimal requirement that {\em the corresponding problem be mathematically meaningful in terms of distribution theory}.
Then, in review of Eqs.~(\ref{3.3b})--(\ref{3.3jb}), we find that the cases $\lambda = 1$ and $\lambda \not=1$ have different dependencies on the derivatives of $\mu$. In particular, the term $\mu'\mu''$ appears when $\lambda \not=1$. Thus, in the following we consider the two cases separately. 

\subsubsection{$\lambda = 1.$} In this case, we assume that: (a) $\mu$ and $\nu$ are $C^5$ in each of the regions $M^+$ and $M^-$ up to the boundary $\Sigma$; (b) $\mu$ is $C^{0}$ across $\Sigma$; (c) $\nu$ is $C^{0}$ with respect to $t$ and $C^{2}$ with respect to $r$ across $\Sigma$.

The above regularity assumptions ensure that the mathematically ill-defined products $\delta(\Phi)^2$ and $\delta(\Phi)H(\Phi)$ do not appear in the field equations. Indeed, the terms in the field equations (\ref{3.3b})--(\ref{3.3jb}) that could lead to products of this type are
\bq
\lb{5.5}
{\mu'}^{2},\;\; \dot{\mu}\mu',\;\; \dot{\nu}^2, \;\; {\nu''}^{2},\;\; \nu''\nu'''.
\eq
Our assumptions imply that these terms may contain $H(\Phi)^2$ but not $\delta(\Phi)^2$ or $\delta(\Phi)H(\Phi)$.

In order to compute the derivatives of $\mu$ and $\nu$, we note that
\bqn 
\mu & =& \mu^D = \mu^+ H(\Phi) + \mu^-[1-H(\Phi)], 
	 \nb\\
\nu &=& \nu^D = \nu^+ H(\Phi) + \nu^-[1-H(\Phi)],
\eqn
where the functions $\mu^+$ and $\nu^+$ are $C^5$ on $M^+$, while the functions $\mu^-$ and $\nu^-$ are $C^5$ on $M^-$. Let $V_\Sigma$ denote an open neighborhood of $\Sigma$. Let $\tilde{\mu}^+$ and $\tilde{\nu}^+$ denote $C^5$-extensions of $\mu^+$ and $\nu^+$ to $M^+ \cup V_\Sigma$. Let $\tilde{\mu}^-$ and $\tilde{\nu}^-$ denote $C^5$-extensions of $\mu^-$ and $\nu^-$ to $M^- \cup V_\Sigma$. Then the functions
\bqn\label{muhatnuhatdef}
\hat{\mu} \equiv \tilde{\mu}^+ - \tilde{\mu}^-,	 \qquad
\hat{\nu} \equiv \tilde{\nu}^+ - \tilde{\nu}^-,
\eqn
are defined on $V_\Sigma$ and the following relations are valid on $\Sigma$ whenever $\alpha + \beta \leq 5$:
\bqn
\lb{4.19}
\hat{\mu} = [\mu]^-, \quad 
\frac{\partial^{\alpha + \beta} }{\partial t^{\alpha} \partial r^{\beta}}\hat{\mu} = \biggl[\frac{\partial^{\alpha + \beta} }{\partial t^{\alpha} \partial r^{\beta}} \mu\biggr]^-, 
	\\ \label{muhatnuhat}
\hat{\nu} = [\nu]^-, \quad 
\frac{\partial^{\alpha + \beta} }{\partial t^{\alpha} \partial r^{\beta}}\hat{\nu} = \biggl[\frac{\partial^{\alpha + \beta} }{\partial t^{\alpha} \partial r^{\beta}} \nu\biggr]^-.
\eqn

Since $\mu$ is $C^0$ across $\Sigma$, we find
\bqn \nb
\mu_{,t} &=& (\mu_{,t})^D, 
	\\ \nb
 \mu_{,r} & = & (\mu_{,r})^D, 
	\\ \nb
\mu_{,tr} &=&  (\mu_{,tr})^{D} + \hat{\mu}_{,t} \delta(\Phi),
	\\ \nb
\mu_{,rt} &=&  (\mu_{,rt})^{D} - \dot{\cal{R}} \hat{\mu}_{,r}\delta(\Phi),
	\\ \nb
\mu_{,rr} &=&  (\mu_{,rr})^{D} + \hat{\mu}_{,r} \delta(\Phi).
	\\ \label{muderivatives}
\mu_{,rrr} &=&  (\mu_{,rrr})^{D} + 2\hat{\mu}_{,rr} \delta(\Phi) +  \hat{\mu}_{,r} \delta'(\Phi).
\eqn
Since $\mu$ is $C^0$ across $\Sigma$, the derivatives of $\mu^+$ and $\mu^-$ in any direction tangential to $\Sigma$ must 
coincide when evaluated on $\Sigma$. In particular, since the vector $U$ defined by
\bq
U^{\lambda} \equiv \delta^{\lambda}_{t} +  \dot{\cal{R}} \delta^{\lambda}_{r},
\eq
is tangential to $\Sigma$ (i.e. $U^{\lambda} N_{\lambda} = 0$), we obtain
$$
U^\lambda [\mu_{,\lambda}]^- = [\mu_{,t}]^- + \dot{\cal{R}} [\mu_{,r}]^- = 0,
$$
that is,
\bq
\hat{\mu}_{,t}  = -\dot{\cal{R}} \hat{\mu}_{,r},  
\eq
after Eq.~(\ref{4.19}) is taken into account. Then,   from Eq.~(\ref{muderivatives}) one finds  $\mu_{,tr} = \mu_{rt}$, as it is expected. 

Similarly, since $\nu$ is $C^0$ across $\Sigma$,
we also have
\bq
0 = U^\lambda [\nu_{,\lambda}]^- = [\nu_{,t}]^- + \dot{\cal{R}} [\nu_{,r}]^-.
\eq
But $[\nu_{,r}]^- = 0$, because $\nu$ is assumed to be $C^2$ with respect to $r$. Thus $[\nu_{,t}]^- = 0$. Therefore,  $\nu$ is in 
fact $C^1$ across $\Sigma$. The same argument applied to $\nu_{,t}$  and $\nu_{,r}$ now implies that $\nu$ is in fact $C^2$ across $\Sigma$.
We find 
\bqn \nb
\nu_{,t} &=& (\nu_{,t})^D, 
	\\ \nb
\nu_{,r} &=& (\nu_{,r})^{D}, 
	\\ \nb
\nu_{,rr} &=& (\nu_{,rr})^{D},
	\\ \nb
\nu^{(3)} &=& (\nu^{(3)})^D, 
	\\ \nb
\nu^{(4)} &=& (\nu^{(4)})^D + \hat{\nu}^{(3)} \delta(\Phi),
	\\ \label{nuderivatives}
\nu^{(5)} &=& (\nu^{(5)})^D + 2\hat{\nu}^{(4)} \delta(\Phi)
	 + \hat{\nu}^{(3)} \delta'(\Phi),
\eqn
where $\nu^{(n)} \equiv \partial^{n}\nu/\partial r^{n}$. 
We emphasize that the expressions on the right-hand sides of (\ref{muderivatives}) and (\ref{nuderivatives}) are independent of the extensions used to define $\hat{\mu}$ and $\hat{\nu}$ in (\ref{muhatnuhatdef}), because the values of $\hat{\mu}$, $\hat{\nu}$, and their partial derivatives of order $\leq 5$ are uniquely prescribed on $\Sigma$ in view of (\ref{muhatnuhat}).

\subsubsection{The Junction Conditions.} We will find the junction conditions across $\Sigma$ by substituting the expressions (\ref{muderivatives}) and (\ref{nuderivatives}) for the derivatives of $\mu$ and $\nu$ into the field equations (\ref{3.3b})--(\ref{3.3jb}).

\begin{table*}[htdp]
\lb{table1}
\begin{center}
\label{default}
\caption{A list of all field equations for $\lambda = 1$.}
\begin{tabular}{c p{3.5cm} c c c}
\hline
\hline
Variation &  & General & Spherically & Junction \\ 
w. r. t. & Name of equation &  version & symmetric case &  condition \\ 
\hline
lapse $N(t)$ & Hamiltonian ~~~~ constraint & (\ref{eq1}) & (\ref{3.3b}) & (\ref{distHamiltonianconstraint}) \\
shift $N^i$ & Momentum ~~~~~~ constraint & (\ref{eq2}) & (\ref{3.3c}) & (\ref{vDJDJD}) \\
$\varphi$ & - & (\ref{eq4a}) & (\ref{3.3e}) & (\ref{vDJDJD}) \\
 &  &  &  &  \\
gauge field $A$ & - & (\ref{eq4b}) & (\ref{3.3f}) & (\ref{vDJDJD}) \\
 &  &  &  &  \\
metric $g_{ij}$ & Dynamical ~~~~~~ equations & (\ref{eq3}) & (\ref{3.3g}) and (\ref{3.3h}) & (\ref{distdynamic1}) and (\ref{distdynamic2}) \\
- & Energy ~~~~~~~~~~~~~~~~~~    conservation law & (\ref{eq5a}) & (\ref{3.3ja}) & (\ref{distenergyconservation}) \\
- & Momentum ~~~~~~  conservation law & (\ref{eq5b}) & (\ref{3.3jb}) & (\ref{distmomentumconservation}) \\
\hline
\hline
\end{tabular}
\end{center}
\end{table*}%

Suppose that the energy density $\rho_H = -2 J^{t}$ has the form
\bq
\rho_H = (\rho_H)^D + \sum_{n=0}^\infty \rho_H^{Im(n)} \delta^{(n)}(\Phi),
\eq
where it is understood that only finitely many of the $\rho_H^{Im(n)}$'s are nonzero.
Since, by (\ref{3.3a}),
$$\mathcal{L}_K = (\mathcal{L}_K)^D, \qquad \mathcal{L}_V = (\mathcal{L}_V)^D,$$
the Hamiltonian constraint (\ref{3.3b}) reads
\bqn \nb
&& \int_{r < \mathcal{R}(t)} \left( \mathcal{L}_K^- + \mathcal{L}_V^-  + 4 \pi G \rho_H^- \right) e^{\nu} r^2 dr
	\\ \label{distHamiltonianconstraint}
&& + \int_{r > \mathcal{R}(t)} \left( {\cal{L}}_{K}^+ + {\cal{L}}_{V}^+  + 4 \pi G \rho_H^+ \right) e^{\nu} r^{2} dr
	\\ \nb 
&& + 4\pi G \sum_{n=0}^\infty (-1)^n \frac{\partial^n}{\partial r^n}\bigg|_{r = \mathcal{R}(t)}\bigl(\rho_H^{Im(n)}e^{\nu} r^2\bigr)
= 0.
\eqn

The left-hand sides of Eqs.~(\ref{3.3c}), (\ref{3.3e}) and (\ref{3.3f}) have no supports on the hypersurface  $r = {\cal{R}} (t)$. 
Thus, these equations remain unchanged in the regions $M^+$ and $M^-$, while on the hypersurface $\Sigma$ they yield
\bq \label{vDJDJD}
v = v^D, \qquad J_{\varphi} = (J_\varphi)^D, \qquad J_{A} = (J_A)^D.
\eq
In fact, in order to avoid that the ill-defined product $H(\Phi) \delta(\Phi)$ arises from the term $J_A A'$ in (\ref{3.3jb}), we will assume that $J_A$ is $C^0$.

The gauge field $A$ has dimension $[A] = 4$, so the action cannot contain terms like $A^n$ with $n \ge 2$, that is, it must be linear in $A$.
We therefore assume that $A$ has the form
\bqn\nb 
 A(t, r) &=& A^D + \sum_{n =0}^{\infty} A^{Im(n)}\delta^{(n)}(\Phi).
 \eqn
It follows that
 \bqn \nb
 A_{,r} &=& (A_{,r})^D + \bigl[\hat{A} + A_{,r}^{Im(0)}\bigr] \delta(\Phi)
 + \sum_{n=1}^\infty \bigl[ A_{,r}^{Im(n)} + A^{Im(n-1)}\bigr] \delta^{(n)}(\Phi),	\\ \nb
 A_{,rr} &=& (A_{,rr})^D + \bigl[2\hat{A}_{,r} + A_{,rr}^{Im(0)}\bigr] \delta(\Phi)
 + \bigl[\hat{A} + 2A_{,r}^{Im(0)}+  A_{,rr}^{Im(1)}\bigr] \delta'(\Phi)	\\ \nb
&& + \sum_{n=2}^\infty \bigl[ A_{,rr}^{Im(n)} + 2A_{,r}^{Im(n-1)} + A^{Im(n-2)}\bigr] \delta^{(n)}(\Phi).
\eqn
Thus,
\bqn 
F_{rr}^{A} &=& \frac{2}{r}\biggl\{(A_{,r})^D + \nu_{,r}A^D 
+ \bigl[\hat{A} + A_{,r}^{Im(0)} + \nu_{,r} A^{Im(0)}\bigr]\delta(\Phi) 
	\nb\\
&& + \sum_{n=1}^\infty \Bigl[\bigl(A_{,r}^{Im(n)} + A^{Im(n-1)}\bigr) \delta^{(n)}(\Phi)
+ \nu_{,r}  A^{Im(n)}\delta^{(n)}(\Phi)\Bigr] \biggr\},\nb\\
F_{\theta\theta}^{A} &=& (F_{\theta\theta}^A)^D + \sum_{n=0}^\infty F_{\theta\theta}^{A, Im (n)} \delta^{(n)}(\Phi), \nb
\eqn
where
\bqn \nb
(F_{\theta\theta}^A)^D &=& e^{-2\nu}\bigl[r^2(A_{,rr})^D  - \nu_{,r}r^2(A_{,r})^D +r(A_{,r})^D 
 + r\nu_{,r} A^D -(1 - e^{2\nu})A^D\bigr],	\\ \nb  
 F_{\theta\theta}^{A, Im (0)} &=& e^{-2\nu}\bigl[r^2(2\hat{A}_{,r} + A_{,rr}^{Im(0)}) 
 - r^2\nu_{,r}(\hat{A} + A_{,r}^{Im(0)})  	\\ \nb
&& + r(\hat{A} + A_{,r}^{Im(0)})  + r\nu_{,r} A^{Im(0)} 	
 - (1 - e^{2\nu})A^{Im(0)}\bigr],	\\ \nb
  F_{\theta\theta}^{A, Im (1)} &=& e^{-2\nu}\bigl[r^2(\hat{A} + 2 A_{,r}^{Im(0)} + A_{,rr}^{Im(1)}) 
 - r^2 \nu_{,r}(A^{Im(0)} + A_{,r}^{Im(1)}) 	\\ \nb
&&  + r(A^{Im(0)} + A_{,r}^{Im(1)}) 
 + r \nu_{,r} A^{Im(1)}   - (1 - e^{2\nu})A^{Im(1)}\bigr],  	\\ \nb
   F_{\theta\theta}^{A, Im (n)} &=& e^{-2\nu}\bigl[r^2\big(A^{Im(n-2)} + 2 A_{,r}^{Im(n-1)} 
 + A_{,rr}^{Im(n)}\big)  - r^2 \nu_{,r}(A^{Im(n-1)} + A_{,r}^{Im(n)}) 	\\ \nb
&& + r(A^{Im(n-1)} + A_{,r}^{Im(n)}) 
 + r \nu_{,r} A^{Im(n)}  - (1 - e^{2\nu})A^{Im(n)}\bigr], \; n \geq 2.
\eqn

From Eq.~(\ref{A.4}) we find that the functions $\{F_n\}_{n=1}^6$ contain no delta functions whereas
\bqn \nb
  (F_7)_{rr} &=& (F_7)_{rr}^D - \frac{16e^{-4\nu}}{r^2}\hat{\nu}^{(3)}\delta(\Phi),  	\\ \nb
  (F_8)_{rr} &=& (F_8)_{rr}^D - \frac{6e^{-4\nu}}{r^2} \hat{\nu}^{(3)}\delta(\Phi),	\\ \nb
  (F_7)_{\theta\theta} &=& (F_7)_{\theta\theta}^D - 8re^{-6\nu}\bigl[(2\hat{\nu}^{(4)} - 16\nu_{,r}\hat{\nu}^{(3)}) \delta(\Phi)
  + \hat{\nu}^{(3)} \delta'(\Phi)\bigr],  	\\ \nb
  (F_8)_{\theta\theta} &=& (F_8)_{\theta\theta}^D - 3 r e^{-6\nu}\bigl[(2\hat{\nu}^{(4)} - 16 \nu_{,r} \hat{\nu}^{(3)})\delta(\Phi) 
 + \hat{\nu}^{(3)}\delta'(\Phi)\bigr].
\eqn
Thus, (\ref{eq3a}) gives
\bqn \nb
F_{rr} &=& (F_{rr})^D - (16 g_7 + 6 g_8) \frac{e^{-4\nu}}{r^2 \zeta^4} \hat{\nu}^{(3)}\delta(\Phi),
	\\ \nb
F_{\theta\theta} &=& (F_{\theta\theta})^D - (8 g_7 + 3 g_8) \frac{r e^{-6\nu}}{\zeta^4}
 \bigl[(2\hat{\nu}^{(4)} - 16 \nu_{,r} \hat{\nu}^{(3)})\delta(\Phi) + \hat{\nu}^{(3)}\delta'(\Phi)\bigr].
\eqn

Writing $p_r$ in the form 
\bqn\nb
p_r(t, r) = p_r^D + \sum_{n =0}^{\infty} p_r^{Im(n)}\delta^{(n)}(\Phi),
\eqn
we find that Eq.~(\ref{3.3g}) remains unchanged in the regions $M^+$ and $M^-$, while on the hypersurface $\Sigma$ it yields
\bqn 
\label{distdynamic1}
\sum_{n=0}^\infty && \biggl\{\frac{e^{2\nu}}{r} [e^{-2\nu}(1 - 2r\nu') + \Lambda_g r^2 - 1]   A^{Im(n)} 
	\nb\\
&& + r\Bigl[F_{rr}^{Im(n)} + F_{rr}^{A Im(n)} + 8\pi G e^{2\nu} p_r^{Im(n)}\Bigr]\biggr\} \nb\\
&& ~~~~~~~~~ \times \delta^{(n)}(\Phi) = 0.
\eqn
Using (\ref{Fdeltaeqs}), Eq.~(\ref{distdynamic1}) can be rewritten as a hierarchy of scalar equations on $\Sigma$.

Similarly, Eq.~(\ref{3.3h}) remains unchanged in the regions $M^+$ and $M^-$, while on the hypersurface $\Sigma$ it yields 
\bqn 
 \label{distdynamic2}
&&  r (\hat{\mu}_{,r} e^{2\mu} + \dot{\mathcal{R}} \hat{\mu}_{,r} e^{\nu + \mu})\delta(\Phi)
	\nb\\
&& + \sum_{n=0}^\infty \biggl\{\frac{e^{2\nu}}{r} [e^{-2\nu}(1 - 2r\nu') + \Lambda_g r^2 - 1]   A^{Im(n)} 
   	 \nb\\
 &&  + \frac{e^{2\nu}}{r}\Bigl[ F_{\theta\theta}^{Im(n)} + F_{\theta\theta}^{AIm(n)} + 8\pi G r^2 p_\theta^{Im(n)}\Bigr]\biggr\} \delta^{(n)}(\Phi) = 0.\nb\\
 \eqn

Note that
\bqn \nb
\rho_{H,t} &=& (\rho_{H,t})^D + \bigl[\rho_{H,t}^{Im(0)} - \dot{\mathcal{R}} \hat{\rho}_{H} \bigr]\delta(\Phi) 
+ \sum_{n=1}^\infty \bigl(\rho_{H,t}^{Im(n)} - \dot{\mathcal{R}} \rho_H^{Im(n-1)} \bigr) \delta^{(n)}(\Phi),
\eqn
and, by (\ref{vDJDJD}),
\bqn \nb
v_{,t} &=& (v_{,t})^D - \dot{\mathcal{R}} \hat{v} \delta(\Phi).
\eqn
Thus, in view of (\ref{vDJDJD}), the energy conservation law (\ref{3.3ja}) takes the form
\bqn \nb
 \int dr \, e^\nu r^2 \Bigl\{&&(\rho_{H,t})^D +\bigl[\rho_{H,t}^{Im(0)} - \dot{\mathcal{R}} \hat{\rho}_{H} \bigr]\delta(\Phi) 	
 	\\ \nb
&& + \sum_{n=1}^\infty \bigl(\rho_{H,t}^{Im(n)}  - \dot{\mathcal{R}} \rho_H^{Im(n-1)} \bigr) \delta^{(n)}(\Phi)
	\\ \nb
&& + \biggl(  (\rho_H)^D +  \sum_{n=0}^\infty \rho_H^{Im(n)} \delta^{(n)}(\Phi)
 	\\ \nb
&& + 4  (p_r)^D + 4 \sum_{n=0}^\infty p_r^{Im(n)} \delta^{(n)}(\Phi)\biggr)\nu_{,t}
	\\ \nb
 && + 4 ( (v_{,t})^D - \dot{\mathcal{R}} \hat{v} \delta(\Phi) - v^D \mu_{,t} ) 
 	\\ \nb
&& - 2\bigl( (J_{A,t})^D + \nu_{,t} (J_A)^D\bigr) \Bigr\} = 0,
\eqn
that is,
\bqn \nb
&& \biggl(\int_{r < \mathcal{R}(t)} + \int_{r > \mathcal{R}(t)}\biggr) e^\mu r^2 \bigl(\rho_{H,t} 
+ \nu_{,t}(\rho_H + 4 p_r)
 	\\ \nb
&&  + 4 v_{,t} - 4 v \mu_{,t}  
- 2(J_{A,t}+ \nu_{,t} J_A) \bigr) dr
 + \Bigl[e^\nu r^2\Bigl(\rho_{H,t}^{Im(0)} 
 	\\\nb
&& - \dot{\mathcal{R}} \hat{\rho}_{H} + \nu_{,t}\bigl(\rho_H^{Im(0)} + 4p_r^{Im(0)}\bigr)
 - 4\dot{\mathcal{R}} \hat{v}\Bigr)\Bigr]\Big|_{r = \mathcal{R}(t)}
	\\ \nb
&& + \sum_{n=1}^\infty (-1)^n \frac{\partial^n}{\partial r^n}\bigg|_{r = \mathcal{R}(t)}\Bigl[e^\nu r^2\Bigl(\rho_{H,t}^{Im(n)} - \dot{\mathcal{R}} \rho_H^{Im(n-1)} 
	\\  \label{distenergyconservation}
&& + \nu_{,t}\bigl(\rho_H^{Im(n)} + 4p_r^{Im(n)}\bigr)\Bigr)\Bigr] = 0.
 \eqn

The momentum conservation law (\ref{3.3jb}) remains unchanged in $M^+$ and $M^-$ while on the hypersurface $\Sigma$ it yields
\bqn \nb
&& - \hat{v} \delta(\Phi) + \hat{p}_r \delta(\Phi) 
 + \sum_{n=0}^\infty \bigl[(p_r^{Im(n)})_{,r} \delta^{(n)}(\Phi) + p_r^{Im(n)} \delta^{(n+1)}(\Phi)\bigr] 	\\ \nb
&& + \frac{2}{r}\sum_{n=0}^\infty ( p_r^{Im(n)} - p_\theta^{Im(n)}) \delta^{(n)}(\Phi)	
 - \frac{1}{2}J_{A} \biggl[\bigl(\hat{A} + A_{,r}^{Im(0)}\bigr) \delta(\Phi)
 	\\ \nb
&&  + \sum_{n=1}^\infty \bigl(A_{,r}^{Im(n)} + A^{Im(n-1)}\bigr) \delta^{(n)}(\Phi)\biggr] 
 + e^{\nu-\mu} \dot{\mathcal{R}} \hat{v} \delta(\Phi)  = 0,
	 \label{distmomentumconservation}
\eqn
where we have used that
\bqn \nb
p_{r}' = \hat{p}_r \delta(\Phi)  + \sum_{n=0}^\infty \bigl[(p_r^{Im(n)})_{,r} \delta^{(n)}(\Phi) + p_r^{Im(n)} \delta^{(n+1)}(\Phi)\bigr]. 
\eqn
This completes the general description of the junction conditions for the case $\lambda =1$, which are summarized in Table 1.   

\subsubsection{$\lambda \not= 1.$} In this case,  the nonlinear terms
\bq
\lb{5.5b}
{\mu'}^{2},\;\; \dot{\mu}\mu',\;\; \mu'\mu'',\;\; \dot{\nu}^2, \;\; {\nu''}^{2},\;\; \nu''\nu''',
\eq
appear in the field equations (\ref{3.3b})--(\ref{3.3jb}). Thus, to ensure these field equations are well-defined, we assume that:
(a) $\mu$ and $\nu$ are $C^5$ in each of the regions $M^+$ and $M^-$ up to the boundary $\Sigma$; (b) $\mu$ is 
$C^{0}$ with respect to $t$ and $C^{1}$ with respect to $r$ across $\Sigma$;  
(c) $\nu$ is $C^{0}$ with respect to $t$ and $C^{2}$ with respect to $r$ across $\Sigma$.

The same argument as above shows that $\nu$ is $C^2$ and that $\mu$ is $C^1$ across $\Sigma$. Equations (\ref{muderivatives}) and (\ref{nuderivatives}) for the derivatives of $\mu$ and $\nu$ are still valid, but since $\mu$ now is $C^1$, we have $\hat{\mu}_{,t} = \hat{\mu}_{,r} = 0$.
It follows that all the junction conditions (\ref{distHamiltonianconstraint})--(\ref{distmomentumconservation}) remain unchanged, except that the presence of the term $\mu'''$ in (\ref{3.3e}) implies that the expression for $J_\varphi$ now may include a delta function:  
\bqn \label{Jvarphidelta}
&& J_{\varphi} = (J_\varphi)^D + (1-\lambda)\frac{e^{\mu - 3\nu}}{4 \pi G} \hat{\mu}_{,rr} \delta(\Phi).
\eqn

In what follows, we will consider some 
specific models of gravitational collapse for which the spacetime inside the collapsing sphere is described by the  Friedman-Lemaitre-Robertson-Walker (FLRW)
universe. 

\section{Gravitational Collapse of Homogeneous and Isotropic Perfect Fluid}

In this section, we consider the gravitational collapse of a spherical cloud consisting of a homogeneous and isotropic  perfect fluid described by  the  FLRW universe. Gravitational collapse of a homogeneous and isotropic dust fluid filled in the whole space-time was  considered in~\cite{TS}, using a method proposed in~\cite{WW09}:
\bqn
ds^2 = -d\bar{t}^2 + a^2(\bar{t})\bigg(\frac{d\bar{r}^2}{1 - k\bar{r}^2} + \bar{r}^2 d^2\Omega\biggr),\nb
\eqn
where $k =0, \pm1$. Letting  $r = a(\bar{t})\bar{r}, \;  t = \bar{t}$, 
the corresponding ADM variables  take the form (\ref{3.1b}) with $N^- = 1$, and 
\bqn
\nu^{-}(t,r) &=& -\frac{1}{2} \ln\biggl(1 - k \frac{r^2}{a^2(t)}\biggr),\nb\\
\mu^{-}(t,r) &=&  \ln\biggl(\frac{-\dot{a}(t) r}{\sqrt{a^2(t) - k r^2}}\biggr),
\eqn
where $\dot{a} \le 0$ for a collapsing cloud. For a perfect fluid, we assume that 
\bq
\lb{PF}
p_\theta^{-} = p_r^{-} = p^-(t), \;\;\;  v = 0.  
\eq
We anticipate that the junction condition for $\nu$ requires  $k = 0$. Then, we find that
\bqn\label{numuflrwk0}
  \nu^{-}(t,r) = 0, \qquad \mu^{-}(t,r) = \ln\big({- rH}\big), \quad (k = 0), \quad
\eqn  
 where $ H \equiv \dot{a}(t)/a(t)$, and that
\bqn\nb
&&
\mathcal{L}_K^{-} = {3 (1-3 \lambda ) H^2}, \;\;\;  \mathcal{L}_V^{-} = 2\Lambda,
	\\  \label{FLRWk0Ls}
&& \mathcal{L}_\varphi^{-} =   \mathcal{L}_\lambda^{-} = 0,\;\;\; \mathcal{L}_A^{-} = 2 \Lambda_g A^{-}.
\eqn
It is easy to verify that the momentum constraint (\ref{3.3c}) is satisfied, whereas the equations (\ref{3.3e}) and (\ref{3.3f}) 
obtained by variation with respect to $\varphi$ and $A$ respectively, reduce to
\bqn
\lb{JJ}
  {3 \Lambda_g H}  +8 \pi  G J_\varphi^{-} &=& 0,
  	\\ \nb
  4\pi G J_A^{-} + \Lambda_g &=& 0.
\eqn 
Since $\nu^{-} = 0$, we have $F_{ij}^{-} = -\Lambda g_{ij}^{-}$, and the first dynamical equation (\ref{3.3g}) reduces to the condition
\bqn \nb
&& \frac{4}{r} a^2 A^{-}_{, r} + 2 a^2 \Lambda_g {A^{-}} +
2 (3 \lambda -1) a \ddot{a} + (3 \lambda -1) \dot{a}^2
 + 2 a^2 (8 \pi  G p^- - \Lambda )  = 0.
 \eqn
If this condition is satisfied the second dynamical equation (\ref{3.3g}) also holds provided that
$A^{-}_{,r} - r A^{-}_{,rr} = 0.$
On the other hand, the momentum conservation law (\ref{3.3jb}) reduces to
$J_A^{-} A^{-}_{,r} = 0.$
We conclude that the general solution when $k = 0$ is given by
\bqn\label{flrwsolution1}
&&   J_\varphi^{-} = -\frac{3 \Lambda_g H}{8 \pi  G}, \quad
J_A^{-} = -\frac{\Lambda_g}{4\pi G},
\eqn
with $A^{-}= A^{-}(t)$ being given by 
\bqn
\lb{AA}
&&  \Lambda_g A^{-} +
 (3 \lambda -1)\left(\frac{\ddot{a}}{a} + \frac{H^2}{2}\right)
	   -     \Lambda = - 8 \pi  G p^-. ~~~~
\eqn

In the rest of this section, we consider only the case where $\Lambda_g =0$. Then, Eq.~(\ref{flrwsolution1}) yields
\bqn\label{flrwsolution2}
&& J_A^{-} = J_\varphi^{-} = 0, 
\eqn
for which Eq.~(\ref{AA}) shows that now $A^{-}(t)$    is an arbitrary function  of $t$, and $a(t)$ is given by
\bqn  
 \label{aeq}
&&  (3 \lambda -1)\left(\frac{\ddot{a}}{a} + \frac{H^2}{2}\right)
	   -     \Lambda = - 8 \pi  G p^-. 
\eqn

It is interesting to note that, since the Hamiltonian constraint is global, there is no analog of the Friedman equation in the current situation. This is in contrast 
to the case of HL cosmology~\cite{HW}, where a Friedman-like equation still exists, because of the homogeneity and isotropy of the whole universe. Although there is no analog of the Birkhoff theorem in HL theory, so that the spacetime outside the collapsing cloud can be either static or dynamical, 
we assume in this chapter that the exterior solution is a static spherically symmetric vacuum spacetime. 
We also assume that the value of $\Lambda_g$ is the same in the exterior and interior regions, i.e. 
\bq
\lb{lambdag}
\Lambda_g^+ = \Lambda_g^- = 0.
\eq
It is convenient to consider the cases $\lambda = 1$ and $\lambda \not=1$ separately. 
\vspace{0.75cm}

\subsection{Gravitational Collapse with $\lambda = 1$}

We first consider the case of $\lambda = 1$. In this case, the static spherically symmetric exterior vacuum solution has the form ~\cite{GSW}
\bqn
\lb{Outside}
&& \mu^+ = \mu^+(r)   = \frac{1}{2}\ln\biggl(\frac{2m^+}{r} + \frac{1}{3}\Lambda r^2 - 2A^+(r) 
 + \frac{2}{r} \int_{r_0}^r A^+(r')dr'\biggr),\nb\\
&& \nu^+ = 0, 
\eqn
for which we find that
\bqn
 \label{exteriorsolution}
&& \mathcal{L}_K^+ = \frac{4}{r} A^+_{,r} - 2\Lambda, \quad \mathcal{L}_V^+ = 2\Lambda,   \quad \mathcal{L}_A^+ = 0,\nb\\
&& v^+ =   J_A^+ =   J_\varphi^+ =   \rho_H^+ = 0,
\eqn
where $m^+$, $r_0$ are constants and $A^+ = A^+(r)$ is a function  of  $r$ only, yet to be determined.  

As mentioned previously, the condition that $\nu$ be continuous across $\Sigma$ implies that $k = 0$.
We let the interior solution be of the form (\ref{numuflrwk0}), 
and assume that the thin shell of matter separating the interior and exterior solutions is such that
\bqn
\nb
&& p = p_{r}^{-}, \quad v = 0, \quad J_\varphi = J_\varphi^{Im(0)} \delta(\Phi),
	\\\nb
&& p_\theta = p_{\theta}^{-} + p_\theta^{Im(0)} \delta(\Phi), \quad \rho_H = \rho_H^- + \rho_H^{Im(0)}\delta(\Phi),\quad
	\\\label{shellassumptions}
&& A = A^D + A^{Im(0)}\delta(\Phi), \quad J_A = 0,  
\eqn
where 
$\rho^{+}_{H} =  J_\varphi^{\pm} = p_\theta^+ = p_r^+ = 0$.

\begin{proposition}\label{lambda1prop}
For the spacetime defined by (\ref{numuflrwk0}), (\ref{aeq}), (\ref{Outside}), the six 
junction conditions (\ref{distHamiltonianconstraint})--(\ref{distmomentumconservation}) reduce to the following six conditions:
\bqn 
\label{JuncHamiltonianconstraint} 
&&
 \left(-6 H^2 + 2\Lambda + 4 \pi G \rho_H^-(t) \right) \frac{\mathcal{R}(t)^3}{3} 
 + 4\int_{\mathcal{R}(t)}^\infty  A_{,r}^+ r dr
 + 4\pi G \rho_H^{Im(0)} r^2 \Big|_{r = \mathcal{R}(t)} = 0,\\
\label{JuncvDJDJD}
&&
 J_\varphi^{Im(0)} = 0,	\\
\label{Juncdynamic1}
&& 
\text{$A(t,r)$ is continuous across $\Sigma$},\\	
\label{Juncdynamic2}
&& 
A_{,t}^- = \mathcal{R} \biggl( \frac{\Lambda}{2} - H^2 \biggr)(\dot{\mathcal{R}} - H \mathcal{R}) - 8\pi G p_\theta^{Im(0)} H \mathcal{R},\\
\label{Juncenergyconservation}
&&
 \rho_{H}^{Im(0)}(t, \mathcal{R}(t)) = e^{-\int_0^t \frac{2\dot{\mathcal{R}}(\tau)}{\mathcal{R}(\tau)}d\tau}\biggl[\rho_{H}^{Im(0)}(0, \mathcal{R}(0)) 
 + \int_0^t e^{\int_0^s \frac{2\dot{\mathcal{R}}(\tau)}{\mathcal{R}(\tau)}d\tau} \biggl(\frac{1}{4} H(s) \mathcal{R}(s)^2 \rho_{H,t}^-(s) \nb\\
&& ~~~~~~~~~~~~~~~~~~~~~~~
 - \dot{\mathcal{R}}(s) \rho_H^-(s)\biggr) ds\biggr],
	\\ \label{Juncenergyconservation}
	\label{Juncmomentumconservation}
&&
 \text{$r p + 2p_\theta^{Im(0)}  = 0$ on $\Sigma$}.
\eqn
Moreover, the condition that $\mu$ be continuous across $\Sigma$ implies that
\bqn \label{mucontinuouscondition}
A^-_{,t} = \frac{\Lambda  - 3H^2}{2} \mathcal{R} \dot{\mathcal{R}}- HH_{,t} \mathcal{R}^2.
\eqn
\end{proposition}
\proofbegin
For the spacetime defined by (\ref{exteriorsolution})--(\ref{shellassumptions}), condition (\ref{distHamiltonianconstraint}) reduces to
\bqn\nb
 \left(-6H(t)^2 + 2\Lambda + 4 \pi G \rho_H^-(t) \right) \int_0^{\mathcal{R}(t)}  r^2 dr
  + 4\int_{\mathcal{R}(t)}^\infty  A_{,r}^+ r dr + 4\pi G \rho_H^{Im(0)} r^2 \Big|_{r = \mathcal{R}(t)}
= 0,
\eqn
which yields (\ref{JuncHamiltonianconstraint}). 
Moreover, condition (\ref{vDJDJD}) reduces immediately to (\ref{JuncvDJDJD}).

Conditions (\ref{distdynamic1}) and (\ref{distdynamic2}) reduce to
\bqn \label{distdynamic12}
  F_{rr}^{A Im(0)} \delta(\Phi) + F_{rr}^{A Im(1)} \delta'(\Phi) = 0,
\eqn
and
\bqn \nb
 r (\hat{\mu}_{,r} e^{2\mu} + \dot{\mathcal{R}} \hat{\mu}_{,r} e^{\mu})\delta(\Phi)
	\\ \nb
+ \frac{1}{r} F_{\theta\theta}^{AIm(0)} \delta(\Phi) 
+ \frac{1}{r} F_{\theta\theta}^{AIm(1)} \delta'(\Phi) 
	\\ \label{distdynamic22}
 + \frac{1}{r} F_{\theta\theta}^{AIm(2)}  \delta''(\Phi) + 8\pi G r p_\theta^{Im(0)} \delta(\Phi) = 0,
\eqn
respectively, where we have used that 
\bqn\label{FrrFthetatheta}
  F_{rr} = (F_{rr})^D, \qquad F_{\theta\theta} = (F_{\theta\theta})^D.
\eqn  
Now
\bqn \nb
F_{rr}^{A} &=& \frac{2}{r}\biggl\{(A_{,r})^D + \bigl[\hat{A} + A_{,r}^{Im(0)} \bigr]\delta(\Phi) 
	 + A^{Im(0)} \delta'(\Phi) \biggr\},
	 	\\ \label{FArrFAthetatheta}
  F_{\theta\theta}^{A} &=& (F_{\theta\theta}^A)^D + \sum_{n=0}^2 F_{\theta\theta}^{A, Im (n)} \delta^{(n)}(\Phi),
\eqn
where
\bqn \nb
(F_{\theta\theta}^A)^D &=& r^2(A_{,rr})^D +r(A_{,r})^D,
	\\ \nb  
 F_{\theta\theta}^{A, Im (0)} &=& r^2(2\hat{A}_{,r} + A_{,rr}^{Im(0)}) 
  + r(\hat{A} + A_{,r}^{Im(0)}),
	\\ \nb
  F_{\theta\theta}^{A, Im (1)} &=& r^2(\hat{A} + 2 A_{,r}^{Im(0)}) + rA^{Im(0)},
  	\\ \nb
   F_{\theta\theta}^{A, Im (2)} &=& r^2 A^{Im(0)}.
\eqn
Thus, equation (\ref{distdynamic12}) can be written as
\bqn\label{hatAArIm0}
\bigl[\hat{A} + A_{,r}^{Im(0)} \bigr]\delta(\Phi) + A^{Im(0)} \delta'(\Phi) = 0.
\eqn
{Thus, by}  (\ref{Fdeltaeqs2}), $A^{Im(0)}|_\Sigma = 0$. Hence, $A^{Im(0)}\delta(\Phi) = 0$ which gives 
$$0=(A^{Im(0)}\delta(\Phi))_{,r} = A_{,r}^{Im(0)} \delta(\Phi) + A^{Im(0)} \delta'(\Phi).$$
Equation (\ref{hatAArIm0}) then gives $\hat{A}|_\Sigma = 0$ so that in fact $A$ is continuous across $\Sigma$, which proves (\ref{Juncdynamic1}).
Equation (\ref{distdynamic22}) can now be written as
\bqn \nb
&& \Bigl[\hat{\mu}_{,r}  (e^{2\mu} + \dot{\mathcal{R}} e^{\mu})
+ 2\hat{A}_{,r} + 8\pi G p_\theta^{Im(0)} \Bigr]\delta(\Phi)  
 + \hat{A} \delta'(\Phi) = 0. 
\eqn
In view of (\ref{Fdeltaeqs2}) this yields
\bqn \label{hatmure2mu2}
\hat{\mu}_{,r}  (e^{2\mu} + \dot{\mathcal{R}} e^{\mu})
+ 2\hat{A}_{,r} + 8\pi G p_\theta^{Im(0)} = \frac{\partial \hat{A}}{\partial \Phi} \; \; \text{�on $\Sigma$}. \qquad
\eqn

Now observe that if a function $f(t,r)$ is $C^0$ across $\Sigma$, then
\bqn\label{dfhatdPhi}
  \frac{\partial \hat{f}}{\partial \Phi} = \frac{\partial \hat{f}}{\partial r} \; \text{ on $\Sigma$}.
\eqn  
Indeed, the continuity of $f$ implies that the derivative of $\hat{f}$ in any direction tangential to $\Sigma$ 
must vanish when evaluated on $\Sigma$; thus $\hat{f}_{,t} + \dot{\cal{R}} \hat{f}_{,r} = 0$ on $\Sigma$.
A computation using (\ref{gradPhicomponents}), (\ref{5.3a}), and (\ref{dfdPhi}) now gives (\ref{dfhatdPhi}).

On the other hand, since
$$\mu_{,r}^+ =  \frac{1}{2}(\Lambda r - 2A_{,r}^+) e^{-2\mu^+}, \qquad \mu_{,r}^- = \frac{1}{r},$$
we find
\bqn\label{hatmur}
\hat{\mu}_{,r} = \frac{1}{2}(\Lambda r - 2A_{,r}^+) e^{-2\mu^+} - \frac{1}{r}.
\eqn

Inserting the equations (\ref{dfhatdPhi}) and (\ref{hatmur}) into (\ref{hatmure2mu2}), we find
\bqn \nb
&& \biggl(\frac{1}{2}(\Lambda r - 2 A_{,r}^+)e^{-2\mu} - \frac{1}{r}\biggr)(e^{2\mu} + \dot{\mathcal{R}}e^\mu)  + \hat{A}_{,r} 
	\\ \nb
&&\qquad + 8\pi G p_\theta^{Im(0)} = 0 \text{  on $\Sigma$.}
\eqn
Since $ \hat{A}_{,r} = A_{,r}^+ = A_{,t}^- \dot{\mathcal{R}}^{-1}$, simplification yields (\ref{Juncdynamic2}).

Condition (\ref{distenergyconservation}) reduces to
\bqn \nb
&&\int_0^{\mathcal{R}(t)} e^\mu r^2 \rho_{H,t}^- dr + r^2 \Bigl[ \rho_{H,t}^{Im(0)} - \dot{\mathcal{R}} \hat{\rho}_H \Bigr]\Big|_{r = \mathcal{R}(t)} 
 + \frac{\partial}{\partial r}\bigg|_{r = \mathcal{R}(t)} \Bigl[r^2 \dot{\mathcal{R}} \rho_H^{Im(0)}\Bigr] = 0.
 \eqn
That is,
 \bqn \nb
&&- \frac{\dot{a}(t)\rho_{H,t}^-(t)}{a(t)}   \int_0^{\mathcal{R}(t)} r^3  dr 
 + \mathcal{R}(t)^2 \Bigl[\rho_{H,t}^{Im(0)}(t, \mathcal{R}(t)) + \dot{\mathcal{R}}(t) \rho_H^-(t)\Bigr]
	\\ \nb
&& 
 + 2\mathcal{R}(t) \dot{\mathcal{R}}(t) \rho_H^{Im(0)} (t, \mathcal{R}(t))
 + \mathcal{R}(t)^2 \dot{\mathcal{R}}(t)\rho_{H,r}^{Im(0)}(t, \mathcal{R}(t)) = 0.
 \eqn
Consequently, 
\bqn 
 \label{preJuncenergyconservation}
-\frac{H(t) \rho_{H,t}^-(t) \mathcal{R}(t)^2}{4}  + \frac{d}{dt}\Bigl[\rho_{H}^{Im(0)}(t, \mathcal{R}(t))\Bigr]
  +  2 \frac{\dot{\mathcal{R}}(t)}{\mathcal{R}(t)} \rho_H^{Im(0)} (t, \mathcal{R}(t)) + \dot{\mathcal{R}}(t) \rho_H^-(t) = 0. \nb\\
\eqn
Solving this differential equation for $\rho_H^{Im(0)}$, we find (\ref{Juncenergyconservation}).

Condition (\ref{distmomentumconservation}) reduces to 
 \bqn \nb
&&  \left(p   + \frac{2}{r} p_\theta^{Im(0)}\right) \delta(\Phi) = 0.
\eqn
This yields (\ref{Juncmomentumconservation}). 

Finally, the condition that $\mu$ be continuous across $\Sigma$ can be written as
\bqn 
\label{premucondition}
\frac{2m^+}{\mathcal{R}(t)} + \frac{1}{3}\Lambda \mathcal{R}(t)^2 - 2A^+(\mathcal{R}(t)) 
+ \frac{2}{\mathcal{R}(t)} \int_{r_0}^{\mathcal{R}(t)} A^+(r')dr'
= H^2\mathcal{R}^2.
\eqn
Since $A$ is continuous across $\Sigma$, we have $A^+(\mathcal{R}(t)) = A^-(t)$. Hence, multiplying 
(\ref{premucondition}) by $\mathcal{R}$ and then differentiating with respect to $t$, we find
\bqn \nb
&&\Lambda \mathcal{R}^2 \dot{\mathcal{R}} 
- 2A^-_{,t} \mathcal{R} = 2HH_{,t} \mathcal{R}^3 + 3H^2\mathcal{R}^2\dot{\mathcal{R}}.
\eqn
Solving this equation for $A^-_{,t}$, we find (\ref{mucontinuouscondition}).
\proofend

The conditions (\ref{Juncdynamic2}) and (\ref{mucontinuouscondition}) imply that
\bqn 
 \biggl( \frac{\Lambda}{2} - H^2 \biggr)(\dot{\mathcal{R}} - H \mathcal{R}) - 8\pi G p_\theta^{Im(0)} H 
 = \frac{\Lambda  - 3H^2}{2} \dot{\mathcal{R}}- HH_{,t} \mathcal{R},
\eqn
i.e.
\bqn \nb
&& H \dot{\mathcal{R}} + (2H^2 + 2H_{,t} - \Lambda ) \mathcal{R}
 - 16\pi G p_\theta^{Im(0)}  = 0.
\eqn
Solving this equation for $\mathcal{R}(t)$ we find the following equation which expresses $\mathcal{R}(t)$ in terms of $H(t)$ and the pressure $p_\theta^{Im(0)}$ on the shell:
\bqn 
\label{Rsolvedfor}
\mathcal{R}(t) & = & e^{-\int_0^t I(s) ds}\biggl\{\mathcal{R}(0) 
 + 16\pi G \int_0^t e^{\int_0^s I(\tau) d\tau}\frac{p_\theta^{Im(0)}(s, \mathcal{R}(s))}{H(s)} ds\biggr\}, \qquad \quad
\eqn
where $I(t)$ is defined by
\bqn\label{Idef}
 I = 2H + \frac{2H_{,t}}{H} - \frac{\Lambda}{H}.
\eqn

\subsection{Dust Collapse with $\lambda = 1$}

Suppose now that the perfect fluid in the interior region consists of dust, i.e. 
\bqn\label{przero}
p_r^-  = p^{-}_{\theta} = 0.
\eqn 
Then, the condition (\ref{Juncmomentumconservation}) implies that
\bqn
  p_\theta^{Im(0)} = 0.
\eqn  
Solving equation (\ref{aeq}) for $a(t)$ we find
\bqn \label{acases}
a(t) = 
\begin{cases}
a_0 \cosh^{\frac{2}{3}}\biggl(\frac{\sqrt{3\Lambda}}{2} (t - t_0)  \biggr), & \Lambda \not= 0,\\
a_0   (t_0 - t)^{2/3}, & \Lambda = 0,\\
\end{cases}
\eqn
where $a_0$ and $t_0$ are constants. 
In the following, let us consider the cases $\Lambda \not= 0$ and $\Lambda = 0$, separately.

\subsubsection{$\Lambda > 0.$} In this case, substituting the expression for $a(t)$ into (\ref{Idef}) we obtain
\bqn\nb
  I(t) =  \sqrt{\frac{\Lambda}{3}}\tanh\biggl(\frac{\sqrt{3\Lambda}}{2} (t_0  - t)  \biggr),
\eqn
and then (\ref{Rsolvedfor}) yields
\bqn\label{Rdustformula}
  \mathcal{R}(t) = \mathcal{R}_0 \cosh^{\frac{2}{3}}\biggl(\frac{\sqrt{3\Lambda}}{2} (t_0  - t)\biggr), 
\eqn
where $\mathcal{R}_0$ is a constant.
Condition (\ref{mucontinuouscondition}) now implies that $A^-_{,t} = 0$, i.e. $A^-(t) = A_0$ for some constant $A_0$. 
Then, by (\ref{Juncdynamic1}), $A^+(\mathcal{R}(t)) = A_0$. That is, $A^+(r) = A_0$ for all $r$ such that $r = \mathcal{R}(t)$ for some $t$. 
Hence, the form of (\ref{Rdustformula}) implies that $A^+ = A_0$ for all $(t,r)$ in the exterior region. This gives
\bqn  
  A(t,r) = A_0.
\eqn
Condition (\ref{JuncHamiltonianconstraint}) now implies
\bqn  \nb
 &&  \rho_H^{Im(0)}(t, \mathcal{R}(t)) = -\frac{(-6H^2 + 2\Lambda + 4\pi G \rho_H^-(t))\mathcal{R}(t)}{12\pi G}
  	\\  \label{rhoHIm0dust}
&& = - \mathcal{R}_0 \frac{\Lambda + \pi G [1 + \cosh(\sqrt{3\Lambda}(t_0  - t))]\rho_H^-(t)}{6\pi G \cosh^{\frac{4}{3}}(\frac{\sqrt{3\Lambda}}{2}(t_0  - t))}.
\eqn
Substituting this into condition (\ref{Juncenergyconservation}), or its equivalent form (\ref{preJuncenergyconservation}), we infer that $\rho_H^-(t)$ satisfies:
\bqn \nb
&& -\frac{\mathcal{R}_0}{12}\cosh^{\frac{1}{3}}\biggl(\frac{\sqrt{3\Lambda}}{2} (t_0  - t)  \biggr)
\biggl\{4 \cosh^{\frac{1}{3}}\biggl(\frac{\sqrt{3\Lambda}}{2} (t_0  - t)  \biggr)
	\\ \nb
&& - \mathcal{R}_0\sqrt{3\Lambda} \sinh\biggl(\frac{\sqrt{3\Lambda}}{2} (t_0  - t)  \biggr)\biggr\}
\rho_{H,t}^-(t)  = 0,
\eqn
i.e.
\bqn
\rho_H^-(t) = \rho_{H}^{(0)}, 
\eqn
where $\rho_{H}^{(0)}$ is a constant.
All the conditions of Proposition \ref{lambda1prop} are now satisfied. It only remains to consider the condition that $\mu$ be continuous across $\Sigma$. This condition reduces to
\bqn \nb
0 &=& \frac{2m^+}{\mathcal{R}} + \frac{1}{3}\Lambda \mathcal{R}^2 - 2A_0 + \frac{2}{\mathcal{R}} A_0(\mathcal{R} - r_0)
-  H^2 \mathcal{R}^2
	\\ \nb
&=& \frac{6 m^+ - 6A_0 r_0 + \mathcal{R}_0^3 \Lambda}{3\mathcal{R}_0 \cosh^{2/3}(\frac{\sqrt{3\Lambda}}{2} (t_0  - t))}.
\eqn
That is, the parameter $r_0$ is fixed by
\bqn 
  r_0 = \frac{6 m^+  + \mathcal{R}_0^3 \Lambda}{6A_0}.
\eqn
This implies that
\bqn
\mu^+ = \frac{1}{2}\ln\biggl(\frac{\Lambda r^2}{3} - \frac{\Lambda \mathcal{R}_0^3}{3r}\biggr).
\eqn
Since all the field equations and junction conditions are now satisfied we have proved the following result.

\begin{proposition}
Ho\v{r}ava-Lifshitz gravity admits the following explicit solution when $\lambda = 1$ and $\Lambda > 0$:
\bqn 
\lb{PropV.2}
&& \mu^+ = \frac{1}{2}\ln\biggl(\frac{\Lambda r^2}{3} - \frac{{\cal{R}}_{0}^3\Lambda}{3r}\biggr), \quad
\mu^- = \ln\big(-H(t)r\big),
	 \nb\\
&&  \nu = 0, \quad H(t) = - \sqrt{\frac{\Lambda}{3}}\tanh\biggl(\frac{\sqrt{3\Lambda}}{2} (t_0 - t)  \biggr),
	 \nb\\
&& \mathcal{R}(t) = \mathcal{R}_0 \cosh^{\frac{2}{3}}\biggl(\frac{\sqrt{3\Lambda}}{2} (t_0 - t)  \biggr),
	 \\
&& p_r = p_\theta = 0, \quad \rho_H^-(t) = \rho_{H}^{(0)}, \quad A(t,r) = A_0,
	 \nb\\
&& \text{$\rho_H^{Im(0)}$ is given by (\ref{rhoHIm0dust})},\nb
\eqn
where $t_0$, $\mathcal{R}_0$, $A_0$, and $\rho_{H}^{(0)}$ are constants
and $M_{+} \equiv -\Lambda {\cal{R}}_{0}^3/6$.
\end{proposition}

\begin{figure}[tbp]
\centering
\includegraphics[width=8cm]{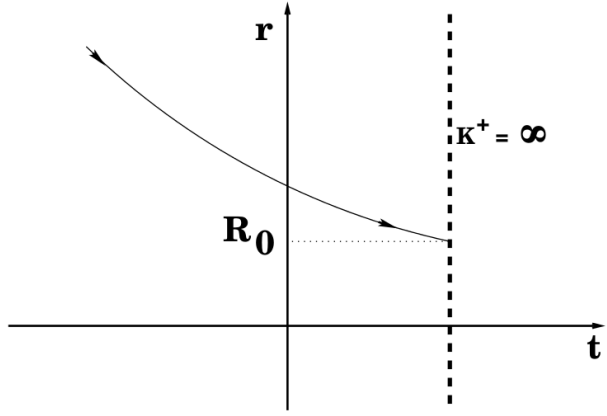} 
\caption{The evolution of the surface of the collapsing star for $\lambda = 1$ and $\Lambda > 0$, given by Eq.~(\ref{PropV.2}). At the moment
$t = t_0$,  the star collapses to its minimal radius   ${\cal{R}}(t_0) =  {\cal{R}}_0$, at which the extrinsic curvature $K^{+}$ becomes unbounded,
while the four-dimensional Ricci scalar remains finite.}
\label{hlfig1}
\vspace{0.75cm}
\end{figure}

For $t < t_0$ the dust cloud is contracting.
As $t \rightarrow t_0$, the radius of the dust sphere approaches its minimal value of $\mathcal{R} = \mathcal{R}_0$ at $t = t_{0}$, and the function
 $e^{\mu^+}$  approaches zero:
$$
\lim_{t \rightarrow t_0} \mathcal{R}(t) = \mathcal{R}_0, \qquad
\lim_{t \rightarrow t_0} e^{\mu^+(t)} = 0,
$$
as shown schematically in Fig. \ref{hlfig1}.   After the star collapses to this point, it is not clear how spacetime {evolutes}, because $\mu_{+}$ 
becomes unbounded as one can see from Eq.~(\ref{PropV.2}), for which the extrinsic scalar $K^+$,
\bq
\lb{K}
K^+(r) =    e^{\mu^+(r)}\left({\mu^+_{,r}(r)} + \frac{2}{r}\right),
\eq
also becomes unbounded, which indicates the existence of a scalar singularity at this point~\cite{CW}.  However, such a singularity is weak. In particular,   the corresponding four-dimensional 
Ricci scalar remains finite, ${}^{(4)}R = 4 \Lambda$.  Thus, it is not clear whether the spacetime across this point is extendible or not. 

In addition,  Eq.~(\ref{rhoHIm0dust}) shows that $\rho_H^{Im(0)}$ and $\rho_H^-$ cannot both be positive. To understand this,  
letting $M = - \Lambda \mathcal{R}_0^3/6$ we can write $\mu^{+}$ in the form
\bq
\mu^+ = \frac{1}{2}\ln\biggl(\frac{2M}{r} + \frac{\Lambda r^2}{3}\biggr).
\eq
However,  this is nothing but the Schwarzschild-de Sitter solution with mass $M$ and a cosmological constant $\Lambda$, where
$M$ is  negative.

\subsubsection{$\Lambda < 0.$} In this case, substituting the expression for $a(t)$ into (\ref{Idef}) we obtain
\bqn\nb
  I(t) = \sqrt{\frac{|\Lambda|}{3}}\tan\biggl(\frac{\sqrt{3|\Lambda|}}{2} (t - t_0)  \biggr),
\eqn
and then (\ref{Rsolvedfor}) yields
\bqn
  \mathcal{R}(t) = \mathcal{R}_0 \cos^{\frac{2}{3}}\biggl(\frac{\sqrt{3|\Lambda|}}{2} (t - t_0)\biggr), 
\eqn
where $\mathcal{R}_0$ is another constant.
Condition (\ref{mucontinuouscondition}) now implies that $A^-_{,t} = 0$, i.e. $A^-(t) = A_0$ for some constant $A_0$. 
Then, by (\ref{Juncdynamic1}), $A^+(\mathcal{R}(t)) = A_0$. That is, $A^+(r) = A_0$ for all $r$ such that $r = \mathcal{R}(t)$ for some $t$. 
We will assume that $A^+ = A_0$ for all $(t,r)$ in the exterior region, i.e.
$
  A(t,r) = A_0.
$
Condition (\ref{JuncHamiltonianconstraint}) now implies
\bqn  \nb
 &&  \rho_H^{Im(0)}(t, \mathcal{R}(t)) = -\frac{(-6H^2 + 2\Lambda + 4\pi G \rho_H^-(t))\mathcal{R}(t)}{12\pi G}
  	\\  \label{rhoHIm0dust2}
&& = \mathcal{R}_0 \frac{|\Lambda| - \pi G [1 + \cos(\sqrt{3|\Lambda|}(t-t_0))]\rho_H^-(t)}{6\pi G \cos^{\frac{4}{3}}(\frac{\sqrt{3|\Lambda|}}{2}(t-t_0))}.
\eqn
Substituting this into condition (\ref{Juncenergyconservation}), or its equivalent form (\ref{preJuncenergyconservation}), we infer that $\rho_H^-(t)$ satisfies
\bqn
\rho_H^-(t) = \rho_{H}^{(0)}, 
\eqn
where $\rho_{H}^{(0)}$ is a constant.
All the conditions of Proposition \ref{lambda1prop} are now satisfied, while  the condition that $\mu$ be continuous across $\Sigma$   reduces to
\bqn \nb
0 &=& \frac{2m^+}{\mathcal{R}} + \frac{1}{3}\Lambda \mathcal{R}^2 - 2A_0 + \frac{2}{\mathcal{R}} A_0(\mathcal{R} - r_0)
-  H^2 \mathcal{R}^2
	\\ \nb
&=& \frac{6 m^+ - 6A_0 r_0 + \mathcal{R}_0^3 \Lambda}{3\mathcal{R}_0 \cos^{2/3}(\frac{\sqrt{3|\Lambda|}}{2} (t - t_0))}.
\eqn
Thus, the parameter $r_0$ is fixed by
\bqn 
  r_0 = \frac{6 m^+  + \mathcal{R}_0^3 \Lambda}{6A_0}.
\eqn
This implies that
\bqn
\mu^+ = \frac{1}{2}\ln\biggl(\frac{2M}{r} -\frac{|\Lambda|}{3}  r^2\biggr),
\eqn
where $M \equiv |\Lambda| \mathcal{R}_0^3/6$. Clearly, this corresponds to the Schwarzschild-anti-de Sitter solution. For $\mu^{+}$ to be real, we must assume that
$r \le {\cal{R}}_0$. Similar to the last case, the extrinsic curvature $K^+$ at $r =  {\cal{R}}_0$ becomes unbounded, while the four-dimensional Ricci scalar ${}^{(4)}R$ remains
constant. Thus, in this case it is also not clear whether or not the spacetime is extendible cross $r =  {\cal{R}}_0$. 

In any case,  all the field equations and junction conditions are now satisfied for $r \le {\cal{R}}_0$, and  we have proved the following result.

\begin{proposition}
Ho\v{r}ava-Lifshitz gravity admits the following explicit solution when $\lambda = 1$ and $\Lambda < 0$:
\bqn 
\lb{PropV.3}
&& \mu^+ = \frac{1}{2}\ln\biggl(\frac{|\Lambda|}{3 r}(\mathcal{R}_0^3 - r^3)\biggr), \quad
\mu^- = \ln(-H(t)r),
	 \nb\\
&&  \nu = 0, \quad H(t) = -\sqrt{\frac{|\Lambda|}{3}}\tan\biggl(\frac{\sqrt{3|\Lambda|}}{2} (t - t_0)  \biggr),
	\nb\\ 
&& \mathcal{R}(t) = \mathcal{R}_0 \cos^{\frac{2}{3}}\biggl(\frac{\sqrt{3|\Lambda|}}{2} (t - t_0)  \biggr),
	\\ 
&& p_r = p_\theta = 0, \quad \rho_H^-(t) = \rho_{H}^{(0)}, \quad A(t,r) = A_0,
	 \nb\\
&& \text{$\rho_H^{Im(0)}$ is given by (\ref{rhoHIm0dust2})},\nb
\eqn
where $t_0$, $\mathcal{R}_0$, $A_0$, and $\rho_{H}^{(0)}$ are constants.  
\end{proposition}

\begin{figure}[tbp]
\centering
\includegraphics[width=8cm]{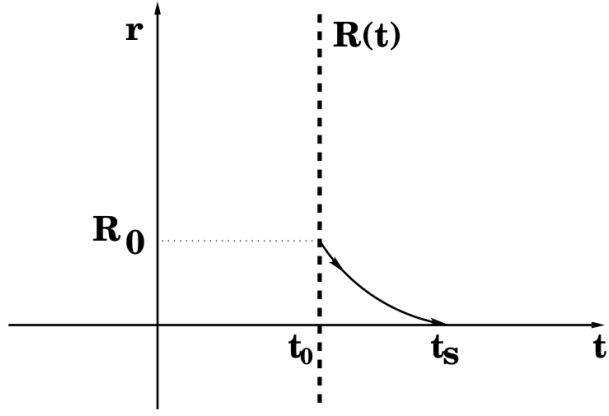} 
\caption{The evolution of the surface of the collapsing star for $\lambda = 1$ and $\Lambda < 0$, given by Eq.~(\ref{PropV.3}). The star starts to collapse at a time $t = t_i \ge t_0$. At the later time $t = t_s$, at which ${\cal{R}}(t_s) =  0$, the star collapses and a central singularity is formed.}
\label{hlfig2}
\vspace{0.75cm}
\end{figure}

The evolution of the surface of the collapsing star is illustrated in Fig. \ref{hlfig2}. The collapse starts at an initial time $t_i \le t_0$, and at time $t = t_s$, the star collapses to a central singularity at which we have $\mathcal{R}(t_s) = 0$, where $t_s \equiv t_0 + \pi/\sqrt{3|\Lambda|}$.
Equation (\ref{rhoHIm0dust2}) shows that now both $\rho_H^{Im(0)}$ and $\rho_H^{-}$ can be positive, provided that $|\Lambda| > 2 \pi G \rho_H^{(0)}.$


\subsubsection{$\Lambda = 0.$} In this case, substituting the expression (\ref{acases}) for $a(t)$ into (\ref{Idef}) we obtain
\bqn\nb
  I(t) = \frac{2}{3(t_0 -t)},
\eqn
and then (\ref{Rsolvedfor}) yields
\bqn\label{Rdustformula2}
  \mathcal{R}(t) = \mathcal{R}_0 (t_0 -t)^{\frac{2}{3}}, 
\eqn
where $\mathcal{R}_0$ is a constant.
Condition (\ref{mucontinuouscondition}) now implies that $A^-_{,t} = 0$, i.e. $A^-(t) = A_0$ for some constant $A_0$. 
Then, by (\ref{Juncdynamic1}), $A^+(\mathcal{R}(t)) = A_0$. That is, $A^+(r) = A_0$ for all $r$ such that $r = \mathcal{R}(t)$ for some $t$. 
Hence (\ref{Rdustformula2}) implies that $A^+ = A_0$ for all $(t,r)$ in the exterior region. Thus, in the present case we also have  
$  A(t,r) = A_0.
$
Condition (\ref{JuncHamiltonianconstraint}) now implies
\bqn  \nb
 &&  \rho_H^{Im(0)}(t, \mathcal{R}(t)) = -\frac{(-6H^2 + 4\pi G \rho_H^-(t))\mathcal{R}(t)}{12\pi G}
  	\\  \nb
&& = \mathcal{R}_0 \frac{2 - 3G\pi(t_0-t)^2 \rho_H^-(t)}{9G\pi(t_0 -t)^{4/3}}.
\eqn
Substituting this into condition (\ref{Juncenergyconservation}), or its equivalent form (\ref{preJuncenergyconservation}), we infer that 
\bqn
\rho_H^-(t) = \rho_{H}^{(0)}, 
\eqn
where $\rho_{H}^{(0)}$ is a constant.
All the conditions of Proposition \ref{lambda1prop} are now satisfied, and  the condition that $\mu$ be continuous across $\Sigma$ becomes
\bqn \nb
0 &=& \frac{2m^+}{\mathcal{R}} - 2A_0 + \frac{2}{\mathcal{R}} A_0(\mathcal{R} - r_0)
-  H^2 \mathcal{R}^2
	\\ \nb
&=& 2\frac{9m^+ - 9A_0 r_0 - 2\mathcal{R}_0^3}{9\mathcal{R}(t)}.
\eqn
Hence,  the parameter $r_0$ is fixed to
\bqn 
  r_0 = \frac{9 m^+ - 2\mathcal{R}_0^3}{9A_0},
\eqn
which implies that
\bqn
  \mu^+ = \frac{1}{2}\ln\biggl(\frac{r_g}{r}\biggr),\;\; \nu^{+} = 0,\;\; N^{+} = 1, 
\eqn
where $r_g \equiv 4\mathcal{R}_0^3/9$. This is nothing but is the Schwarzschild solution written in the Painlev\'e-Gullstrand coordinates~\cite{GP}. 
All the field equations and junction conditions are satisfied, so we have proved the following result.

\begin{proposition}
Ho\v{r}ava-Lifshitz gravity admits the following explicit solution when $\lambda = 1$ and $\Lambda = 0$:
\bqn 
\lb{PropV.4}
&& \mu^+ =  \frac{1}{2}\ln\biggl(\frac{r_g}{r}\biggr), \quad
\mu^- = \ln\big(-H(t)r\big),
	 \nb\\
&&  \nu = 0, \quad H(t) = -\frac{2}{3(t_0 -t)},
	\nb\\
&& \mathcal{R}(t) = \mathcal{R}_0 (t_0 -t)^{\frac{2}{3}},
	\\ 
&& p_r = p_\theta = 0, \quad \rho_H^-(t) = \rho_{H}^{(0)}, \quad A(t,r) = A_0,
	 \nb\\
&& \rho_H^{Im(0)} = \mathcal{R}_0 \frac{2 - 3G\pi(t_0-t)^2 \rho_{H}^{(0)}}{9G\pi(t_0 -t)^{4/3}},\nb
\eqn
where $t_0$, $\mathcal{R}_0$, $A_0$, and $\rho_{H}^{(0)}$ are constants.  
\end{proposition}

\begin{figure}[tbp]
\centering
\includegraphics[width=8cm]{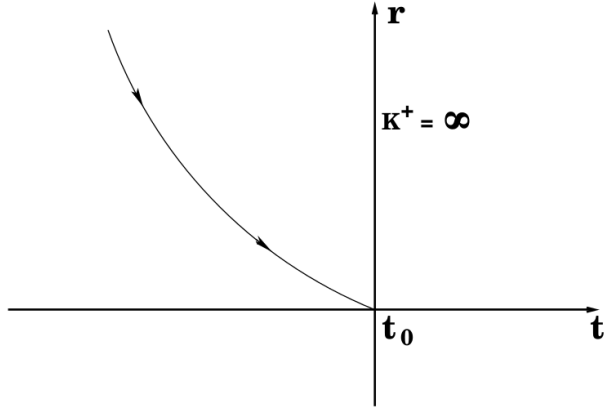} 
\caption{The evolution of the surface of the collapsing star for $\lambda = 1$ and $\Lambda = 0$, given by Eq.~(\ref{PropV.4}). At the moment
$t = t_i \le t_0$,  the star starts to collapse until the moment $t = t_0$, at which we have    ${\cal{R}}(t_0) =  0$, whereby a central singularity is formed.}
\label{fig4}
\vspace{0.75cm}
\end{figure}

The evolution of the surface of the collapsing star is shown in Fig. \ref{fig4}. The star begins to collapse at the moment $t_i$ with a radius 
$\mathcal{R}_i [\equiv \mathcal{R}(t_i)]$ until the moment $t = t_0$, at which we have ${\cal{R}}(t_0) =  0$ and a central singularity is formed. 
The spacetime outside of the star is given by the Schwarzschild solution.   
Thus, as in GR, the Schwarzschild spacetime can be formed by the collapse of a homogeneous and isotropic dust perfect fluid 
\cite{Joshi}. We note that $\rho_H^{Im(0)} > 0$ for  
$$t_0 - \sqrt{\frac{2}{3G\pi \rho_H^{(0)}}} < t < t_0.$$

\subsection{Gravitational Collapse with $\lambda \not= 1$}

We now consider the case of $\lambda \neq 1$. For an exterior static spherically symmetric vacuum  spacetime with $\lambda \neq 1$ 
and $\Lambda_g = 0$, equation (\ref{3.3f}) implies that 
\bq
\nu^+ = -\frac{1}{2}\ln\biggl(1 - \frac{2B}{r} \biggr),
\eq
where $B$ is a constant. On the other hand, for the interior FLRW region, we have
\bq
\nu^- = -\frac{1}{2}\ln\biggl(1 - k\frac{r^2}{a^2(t)}\biggr).
\eq
Hence, the condition $\nu_{,t}^+ = \nu_{,t}^-$ on $\Sigma$ implies that
$0 = k\mathcal{R}(t)^2.$
Consequently, in order for a solution with $\mathcal{R}(t) \neq 0$ to exist, we must have $k = 0$. 
The conditions that $\nu$ and $\nu_{,r}$ be continuous across $\Sigma$ then reduce to
${2B}/{\mathcal{R}(t)}  = 0$.
Thus, in order for a nontrivial solution to exist we must have 
$k  = B = 0$.
Thus,  
we have
\bqn
\label{nunumu}
  \nu^- = \nu^+ = 0, \qquad \mu^- = \ln\big({- rH}\big).
\eqn  
On the other hand, since $\lambda \neq 1$, the momentum constraint (\ref{3.3c}) yields
\bqn
\lb{mu2}
   \mu^+(r) = \ln\biggl(C_1 r + \frac{C_2}{r^2}\biggr),
\eqn
where $C_1$ and $C_2$ are constants.
The field equations (\ref{3.3e})--(\ref{3.3h}) are then satisfied provided that
\bqn
  A^+(r) = A_0^+ - \frac{3C_2^2}{8r^4} + \frac{3(1-3\lambda)C_1^2 + 2\Lambda}{8}r^2,
\eqn
where $A_0^+$ is a constant. It is interesting to note that this class of solutions was first found in~\cite{LMW} in the IR limit. However,  
{since the restriction of the spacetime to the leaves $t = $ constant is flat, we have $R_{ij} = 0$, and the higher-order derivative terms {of $R_{ij}$ 
vanish identically, so} they are also solutions of the full theory.} 
Moreover, since
$$\mu_{,r}^+ = \frac{C_1 r^3 - 2C_2}{C_1 r^4 + C_2 r}, \qquad \mu_{,r}^- = \frac{1}{r},$$
we find
\bqn\label{hatmur2}
\hat{\mu}_{,r} = \frac{- 3C_2}{C_1 r^4 + C_2 r}.
\eqn
Thus, the requirement that $\mu$ is $C^1$ implies that $C_2 = 0$. The continuity of $\mu$ then requires that $H(t) = - C_1$ is a constant and so
$$a(t) = a_0 e^{-C_1 t}.$$
It follows that $\mu$ is smooth across $\Sigma$. 
Note also that
the asymptotical-flatness condition requires $C_1 = 0$. However, in the following we leave the possibility of $C_1 \neq 0$ open. 

We find that
\bqn \nb
&& \mathcal{L}_K^+ = 3C_1^2 (1-3 \lambda ),
\quad 
\mathcal{L}_V^+ = 2\Lambda, \quad \mathcal{L}_A^+ = 0,
	\\
&& v^+ = 0, \quad J_A^+ = 0, \quad J_\varphi^+ = 0, \quad \rho_H^+ = 0.	
\eqn
In order for the integral over the exterior region in the Hamiltonian constraint (\ref{distHamiltonianconstraint}) to converge, we also need to assume that
\bqn\label{C1Lambdaconstraint}
  3C_1^2(1 - 3\lambda) + 2\Lambda = 0.
\eqn
Thus, $A^+(r) = A_0^+$ is a constant and equation (\ref{aeq}) implies that $p^-(t) = 0$, that is, the perfect fluid in the interior region consists of dust.

Similar to the case with $\lambda = 1$,  the interior solution is still  of the form (\ref{flrwsolution2}), i.e.
\bqn\nb
&&J_A^- = J_\varphi^- = 0, \quad A^- = A^-(t).
\eqn
In view of (\ref{FLRWk0Ls}), we have
\bqn\nb
&&\mathcal{L}_\varphi^- = 0, \qquad \mathcal{L}_\lambda^- = 0, \qquad
\mathcal{L}_K^- = 3(1-3\lambda)H^2, 
	\\  
&& \mathcal{L}_V^- = 2\Lambda, \qquad \mathcal{L}_A^- = 0, \qquad v^- = 0.	
\eqn
We assume that the thin shell of matter separating the interior and exterior solutions is such that
\bqn\nb
&& p_r = 0, \quad v = 0, \quad J_\varphi = J_\varphi^{Im(0)} \delta(\Phi),
	\\\nb
&& p_\theta = p_\theta^{Im(0)} \delta(\Phi), \quad \rho_H = \rho_H^- + \rho_H^{Im(0)}\delta(\Phi), \qquad
	\\\label{shellassumptions2}
&& A = A^D + A^{Im(0)}\delta(\Phi), \quad J_A = 0, \quad \mathcal{L}_A = \mathcal{L}_A^-,
\eqn
with $\rho_H^- = \rho_H^-(t)$. 

\begin{proposition} 
For the spacetime defined by (\ref{nunumu})--(\ref{shellassumptions2}), the six junction conditions
 (\ref{distHamiltonianconstraint})--(\ref{distmomentumconservation}) reduce to the following six conditions:
\bqn  \label{Junc2Hamiltonianconstraint}
&& \rho_H^-(t) \frac{\mathcal{R}(t)}{3} + \rho_H^{Im(0)}(t, \mathcal{R}(t)) = 0,
	\\ \label{Junc2vDJDJD}
&& J_\varphi^{Im(0)} = 0,	
	\\ \lb{AAa}
&& \text{$A(t,r) = A_0$ is a constant},	
	\\ \label{Junc2dynamic2}
&&  p_\theta^{Im(0)} = 0 \text{  on $\Sigma$,} 
	\\ \nb
&& \frac{d}{dt}\Bigl[\rho_{H}^{Im(0)}(t, \mathcal{R}(t))\Bigr] +  2 \frac{\dot{\mathcal{R}}(t)}{\mathcal{R}(t)} \rho_H^{Im(0)} (t, \mathcal{R}(t))
	\\ \label{Junc2energyconservation}
&&   + \dot{\mathcal{R}}(t) \rho_H^-(t) + \frac{C_1 \rho_{H,t}^-(t) \mathcal{R}(t)^2}{4} = 0.
\eqn
\end{proposition}
\proofbegin
For the spacetime defined by (\ref{nunumu})--(\ref{shellassumptions2}), condition (\ref{distHamiltonianconstraint}) reduces to
\bqn\nb
&& \left( 3(1-3\lambda)C_1^2 + 2\Lambda + 4 \pi G \rho_H^-(t) \right) \int_0^{\mathcal{R}(t)}  r^2 dr
	\\  \nb
&& + \int_{\mathcal{R}(t)}^\infty 
\biggl(3C_1^2(1 - 3\lambda) + 2\Lambda\biggr) r^2 dr 
	\\ \nb
&& + 4\pi G \rho_H^{Im(0)} r^2 \Big|_{r = \mathcal{R}(t)}
= 0,
\eqn
which, in view of (\ref{C1Lambdaconstraint}), yields (\ref{Junc2Hamiltonianconstraint}). Moreover, equation (\ref{Jvarphidelta}) reduces to (\ref{Junc2vDJDJD}).

The functions $F$ and $F^A$ are given by (\ref{FrrFthetatheta})--(\ref{FArrFAthetatheta}) also for $\lambda \neq 1$. 
Hence, condition (\ref{distdynamic1}) implies that $A$ is continuous across $\Sigma$ just like in the case of $\lambda = 1$. Since $A^+ = A_0^+$ is constant and $A^-(t)$ is independent of $r$, this gives (\ref{AAa}).
Condition (\ref{distdynamic2}) then reduces to
\bqn \nb
&& \frac{1}{r} F_{\theta\theta}^{AIm(0)} \delta(\Phi) 
+ \frac{1}{r} F_{\theta\theta}^{AIm(1)} \delta'(\Phi) 
	\\ \nb
&& + \frac{1}{r} F_{\theta\theta}^{AIm(2)}  \delta''(\Phi) + 8\pi G r p_\theta^{Im(0)} \delta(\Phi) = 0.
\eqn
Since $A$ is a constant, this yields (\ref{Junc2dynamic2}).

Conditions (\ref{distenergyconservation}) and (\ref{distmomentumconservation}) reduce to (\ref{Junc2energyconservation}) and (\ref{Junc2dynamic2}).
\proofend
   
Conditions (\ref{Junc2Hamiltonianconstraint}) and (\ref{Junc2energyconservation}) imply that
$$\dot{\rho}_{H}^-(t) \mathcal{R}(t) \bigg(\frac{C_1}{4}\mathcal{R}(t) - \frac{1}{3}\bigg) = 0.$$
Excluding the case of no collapse where $\mathcal{R}(t)$ is a constant, it follows that $\rho_H^-$ must be a constant. If this is the case, then the junction conditions are satisfied provided that $ \rho_H^{Im(0)}(t, \mathcal{R}(t)) = -\rho_H^- \mathcal{R}(t)/3$.

In summary, in the case $\lambda \not= 1$ a static spherical spacetime can be produced by gravitational collapse of a  homogeneous and isotropic
dust fluid. However, the space-time outside of such a fluid is not asymptotically flat, as one can see from
Eqs.~(\ref{nunumu}) and (\ref{mu2}) with $C_2 = 0$.

\section{Conclusions}

In this chapter, we have studied gravitational collapse of  a spherical cloud of fluid  with a finite radius  in the framework of the nonrelativistic 
general covariant theory of HL gravity with the projectability condition and an arbitrary coupling constant $\lambda$.  
Using distribution theory, we have developed the general junction conditions for such a collapsing spherical body, under the minimal 
requirement that {\em the junctions should be mathematically meaningful in the sense  of generalized functions.}
The general junction conditions have been  summarized in Table I.  

As one of the simplest applications, we have studied  a collapsing star that is made of a homogeneous and isotropic perfect fluid, while the external  {region} is described 
by a stationary  spacetime. We have found that the problem reduces to the matching of six independent conditions that the Arnowitt-Deser-Misner  
variables ($N, N^i, g_{ij}$) and the gauge field $A$ and Newtonian prepotential $\varphi$ must satisfy. 

For the case of  a homogeneous and isotropic dust fluid (a perfect fluid with vanishing pressure), we have found explicitly the space-time outside of the collapsing
sphere. In particular,   in the case $\lambda  = 1$,  the external spacetimes are described by the Schwarzschild (anti-) de Sitter solutions, written in Painlev\'e-Gullstrand
coordinates~\cite{GP}.  It is remarkable that the collapse of a homogeneous and isotropic dust to a 
 Schwarzschild black hole, studied by Oppenheimer and Snyder in General Relativity more than 80 years ago~\cite{OppSnyder}, is a particular case. However, there are 
 fundamental differences. First, in General Relativity a thin shell does not necessarily appear on the surface of the collapsing sphere~\cite{OppSnyder}, while in the current case we have shown that
 such a thin shell must exist, in General Relativity, because of the local conservation of  energy of the collapsing body, the energy density of the dust fluid is inversely proportional to the cube of the radius of the fluid, while in the current case it remains a constant, as the conservation law is global [cf. Eq.~(\ref{eq1})] and the energy of the collapsing star is not 
 necessarily conserved locally.
 
  In the case $\lambda \not= 1$, the space-time outside of the homogeneous and isotropic dust fluid is described by Eqs.~(\ref{nunumu}) and (\ref{mu2}) with $C_2 = 0$. 
 It is clear that such a space-time is not asymptotically flat. Therefore, in this case to obtain an asymptotically flat space-time outside of a collapsing dust fluid,  it must not be
 homogeneous and/or isotropic. 
 
 From the above simple examples, one can already see the significant differences between the HL theory and General Relativity in the strong gravitational field regime. Therefore, it is very interesting to study 
 gravitational collapse of more general fluids, such as perfect fluids with different equations of state, or anisotropic fluids with or without heat flows. Particular attention should be paid to the roles that the equation of state and heat flows might play. It would be extremely interesting to study  the implications for black hole physics, or more generally for (observational) astrophysics and cosmology~\cite{Joshi}. Since the general formulas have been already laid down in this chapter, we expect that such studies  can be carried out easily.   
 
As emphasized previously, our treatment of the junction conditions of a collapsing star presented in this chapter can be easily generalized to other versions of Ho\v{r}ava-Lifshitz gravity or, more generally, 
to any model of a higher-order derivative gravity theory.

\chapter{Global Structure of Spacetime}
\renewcommand{\theequation}{5.\arabic{equation}} \setcounter{equation}{0}

\begin{center}
\begin{singlespace}
\noindent This chapter published as:   J.~Greenwald, J.~Lenells, J.~X.~Lu, V.~H.~Satheeshkumar and A.~Wang,
  ``Black holes and global structures of spherical spacetimes in Ho\v{r}ava-Lifshitz Theory,''
  Phys.\ Rev.\ D {84}, 084040 (2011).
\end{singlespace}  
\end{center}

\section{Introduction}    

\begin{figure}[tbp]
\centering
\includegraphics[width=8cm]{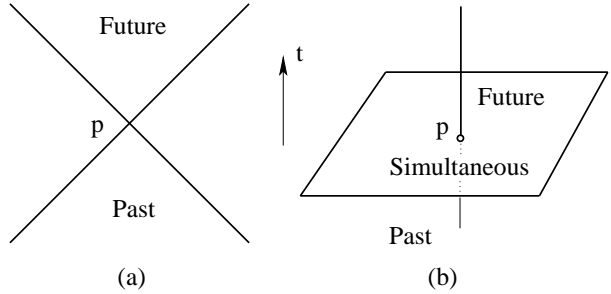}
\caption{ (a) The light cone of the event $p$ in special relativity. (b) The causal structure of the point $p$  in Newtonian theory. }
\label{HLcone}
\vspace{0.75cm}
\end{figure}

In the HL theory, due to the breaking of the general covariance, the dispersion relations of particles  usually contain high order momentum terms
\cite{Mukc,Sotiriou,Padilla,Hreview,Visserb},
\bq
\lb{1.4}
\omega_{k}^{2} = m^{2} + k^{2}\left(1 + \sum^{z-1}_{n=1}{\lambda_{n}\left(\frac{k}{M_{n}}\right)^{2n}}\right),
\eq
for which the group velocity is given by~\cite{CH}
\bq
\lb{1.4b}
v_{k} = \frac{k}{\omega}\left(1 + \sum^{z-1}_{n=1}{(n+1)\lambda_{n}\left(\frac{k}{M_{n}}\right)^{2n}}\right).
\eq
As an immediate result,  the speed of light becomes unbounded in the UV. This makes the causal
structure of the spacetimes quite different from that given in GR,  where the light cone of a given point $p$ plays a fundamental 
role in determining the causal relationship of $p$ to other events [cf. Fig. \ref{HLcone}]. 
However, once the general covariance is broken, the causal structure will be dramatically changed. For example, in the Newtonian theory,  time is absolute and the speeds of signals are not limited. Then, the causal structure of  a given point $p$ is uniquely determined by the time difference, $\Delta{t} \equiv t_{p} - t_{q}$, between the two events.  
 In particular, if $\Delta{t} > 0$, the event $q$ is to the past of $p$; if $\Delta{t} < 0$, it  is to the future; and if $\Delta{t} = 0$, the two events are
simultaneous.  

Another  consequence of the breaking of the general covariance is that  free particles now do not follow geodesics. This immediately makes all the  definitions of black holes given in General Relativity invalid~\cite{HE73,Tip77,Hay94,Wang}.  To provide a proper definition of black holes,  anisotropic conformal boundaries~\cite{HMT2} and  kinematics of particles ~\cite{KM} have been studied   within the HL framework. In this chapter, we shall adopt the  approach of Kiritsis and Kofinas (KK)~\cite{KKb}, where a horizon is defined as the infinitely redshifted 2-dimensional (closed) surface of massless test particles.  Clearly, such a definition reduces to that given in General Relativity when the dispersion relation is relativistic [Where  $\lambda_{n} = 0$, as shown in Eq.~(\ref{1.4}).]. 

It should be noted that black holes in the HL theory with or without the projectability condition have been extensively studied, mainly using the definition borrowed directly from GR. In this chapter, we shall show explicitly how these definitions are changed by considering some  particular examples, found in the HMT set up with $\lambda = 1$. 
 
 Another  interesting approach is the equivalence  between   the HL theory (without the projectability
 condition) and the  Einstein-aether theory    in the IR ~\cite{Jaco}, where the former is equivalent to the latter for the case where the aether vector field  $u_{\mu}$ is hypersurface-orthogonal. In the spherically symmetric case, this is not a restriction as the aether field  $u_{\mu}$ now is always hypersurface-orthogonal. From such studies one already sees the difficulties to define black holes, because of the fact that  different modes may have different velocities even in the IR. In~\cite{Jaco}, black holes are defined to possess both a metric horizon and a spin-0 mode horizon.   Since the equivalence holds only in the IR, it is still unclear how to extend such definitions  to high energy scales, where high order  curvature   terms become important.

\section{Black Holes in HL Theory }

Kristis and Kofinas (KK) considered a scalar field with a given dispersion relation $F(\zeta)$  ~\cite{KKb}. In the geometrical optical  approximations,  $\zeta$ is given by $\zeta = g_{ij}k^{i}k^{j}$, where $k_{i}$ denotes the 3-momentum of the corresponding spin-0 particle. With this  approximation, the trajectory of a test particle is given by
\bqn
\lb{eqq1}
S_{p} &\equiv& \int_{0}^{1}{{\cal{L}}_{p} d\tau} \nb\\
&=& \frac{1}{2} \int_{0}^{1}{ d\tau\Bigg\{\frac{c^{2}N^{2}}{e} \dot{t}^{2} + e \Big[F(\zeta) - 2 \zeta F'(\zeta)\Big]\Bigg\}}, ~~~~
\eqn
where $e$ is a one-dimensional einbein, and   $\zeta$ is now considered as a functional of $t, x^{i}, \dot{t}, \dot{x}^{i}$ and $e$,  given by the relation,
\bq
\lb{eqq2}
\zeta\;  [F'(\zeta)]^{2} = \frac{1}{e^{2}} g_{ij}\big(\dot{x}^{i} + N^{i} \dot{t}\big) \big(\dot{x}^{j} + N^{j}  \dot{t}\big),
\eq
with $\dot{t} \equiv dt/d\tau$, etc.  For detail, we refer readers to~\cite{KKb}.

It should be noted that KK obtained the above action starting from a scalar field. So, strictly speaking, it is valid only for spin-0 test particles. However, what is really important in their derivations is the dispersion relationship $F(\zeta)$. As shown in~\cite{Wangb}, a spin-2 particle  has a similar dispersion relation. It is expected that a spin-1 test particle, such as photons,  should have a similar dispersion relation too~\cite{CH,KKb}. Therefore, in the rest of this chapter and without proof, we simply consider the action (\ref{eqq1})  to describe all massless test particles. 

Spherically symmetric static spacetimes in the framework of the HMT setup were studied systematically in~\cite{AP,GSW}, and  the metric for  static spherically symmetric spacetimes that preserve the form of Eq.~(\ref{1.2}) with the projectability condition can be cast in the form~\cite{GPW} given below. Note the slight difference between the $g_{tr}$ term defined here and the one defined in~\cite{GPW,GSW}.   
\bq
\lb{3.1b}
ds^{2} = - c^{2}dt^{2} + e^{2\nu} \left(dr + e^{\mu - \nu} cdt\right)^{2}  + r^{2}d^2\Omega,  
\eq
  where  $d^2\Omega = d\theta^{2}  + \sin^{2}\theta d\phi^{2}$,
and 
\bq
\lb{3.1}
 \mu = \mu(r),\;\;\; \nu = \nu(r),\;\;\; N^{i} = \left\{ce^{\mu - \nu}, 0, 0\right\}.
 \eq
The corresponding timelike Killing vector is  $\xi = \partial_{t}$, and the diagonal case $N^{r} = 0$ corresponds to $\mu = -\infty$.

However, to study black hole solutions in a more general case, in this (and only in this) section, we also consider the case without projectability condition, and write   the metric as,
\bq
\lb{eqq3}
ds^{2} = - N^{2}c^{2}dt^{2} + \frac{1}{f} \left(dr + N^{r} c dt\right)^{2}  + r^{2}d^2\Omega. 
\eq
where $N,\; f$ and $N^{r}$ are all functions of $r$.  
Without loss of generality, in the rest of the chapter we shall set $c = 1$, 
which is equivalent to the coordinate transformations
$x_{0} = c t,\; \bar{N}^{r} = N^{r}/c$. Taking
\bq
\lb{F-function}
F(\zeta) = \zeta^{n}, \; (n = 1, 2, ...),
\eq
Eq.~(\ref{eqq2}) yields,
\bq
\lb{eqq4}
\zeta = \left(\frac{\dot{r} + N^{r}\dot{t}}{n e \sqrt{f}}\right)^{{2}/{(2n - 1)}} \equiv \left(\frac{{\cal{D}}}{e^{2}}\right)^{{1}/{(2n - 1)}}.
\eq
Inserting this into Eq.~(\ref{eqq1}), we find that, for  radially moving particles, ${\cal{L}}_{p}$ is given by
\bq
\lb{eqq5}
{\cal{L}}_{p} = \frac{N^{2}}{2e} \dot{t}^{2} + \frac{1}{2} \big(1 - 2n\big)e^{1/(1-2n)} {\cal{D}}^{{n}/{(2n - 1)}}.
\eq
Then, from the equation $\delta{\cal{L}}_{p}/\delta{e} = 0$ we obtain
\bq
\lb{eqq6}
N^{2}\dot{t}^{2} - e^{2(n-1)/(2n-1)} {\cal{D}}^{{n}/{(2n - 1)}} = 0.
\eq
On the other hand, since $\delta{\cal{L}}_{p}/\delta{t} = 0$, the Euler-Lagrange equation,
$$
\frac{\delta{\cal{L}}_{p}}{\delta{t}} - \frac{1}{d\tau}\left(\frac{\delta{\cal{L}}_{p}}{\delta{\dot{t}}}\right) = 0,
$$ 
yields
\bq
\lb{eqq8}
N^{2}\dot{t} - e^{2(n-1)/(2n-1)}\frac{N^{r}}{\sqrt{f}} {\cal{D}}^{{1}/{[2(2n - 1)]}} = e E,
\eq
where $E$ is an integration constant, representing the total energy of the test particle. 

To solve Eqs.~(\ref{eqq6}) and (\ref{eqq8}),   we first consider the case $n = 1$, which  corresponds to the relativistic dispersion relation. 
From such considerations, we shall see how to generalize the definition of black holes given in General Relativity to the HL theory where $n$ is generically 
different from 1, as required by the renormalizability condition in  the UV.

\subsection{$ n = 1$}

In this  case,  Eqs.~(\ref{eqq6}) and (\ref{eqq8}) reduce, respectively, to,
\bqn
\lb{eqq9a}
N^{2}\dot{t}^{2} - {\cal{D}} &=& 0,\\
\lb{eqq9b}
N^{2}\dot{t}  -  {N^{r}}\sqrt{\frac{\cal{D}}{f}}  &=& e E.
\eqn
Eq.~(\ref{eqq9a}) simply tells us that now the particle moves along   null  geodesics. The above equations  can be
easily solved according to whether $N^{r}$ vanishes or not. 

 \subsubsection{$N^{r} = 0.$} When $N^{r} = 0$, from Eq.~(\ref{eqq9a}) we find
 \bq
 \lb{eqq10}
 dt = \pm \frac{dr}{N \sqrt{f}},
 \eq
where ``+" (``$-$") corresponds to out-going (in-going) light rays. If $f$ has an a-th order zero and $N^{2}$ a b-th order zero at a surface, say, $r = r_{g}$, that is,
\bq
\lb{fN}
f = f_{0}(r)(r - r_{g})^{a},\;\;\;
N = N_{0}(r) (r - r_{g})^{b/2}, 
\eq
where  $N_{0}(r_{g}) \not= 0$ and $f_{0}(r_{g}) \not= 0$, then from the above
equations we find that in the neighborhood of $r = r_{g}$, 
\bq
\lb{eqq11}
t \simeq t_{0} \pm \frac{1}{N_{0}\sqrt{f_{0}}}
\begin{cases}
\frac{2}{2 - (a+b)}(r-r_{g})^{1 - (a+b)/2}, & a+b \not= 2,\\
\ln\left|r-r_{g}\right|, & a +b  =2.\\
\end{cases}
\eq
Therefore, when
\bq
\lb{eqq12}
a + b \ge 2, \; (n = 1),
\eq
$t$  becomes unbounded, 
as $r \rightarrow r_{g}$, 
at which 
the light rays are infinitely redshifted. 
This indicates that an event  horizon might exist at $r = r_{g}$, provided that the spacetime has no curvature singularity there. 
A simple example is the Schwarzschild solution, $N^{2} = f = (r - r_{g})/r$, which is also a solution of the HL theory without the projectability condition, but with
the  detailed balance condition  softly broken~\cite{KK},
and for which we have $a = b = 1$. Clearly, it satisfies  the above condition with the equality, so $r = r_{g}$ indeed defines a horizon. 

 \subsubsection{$N^{r} \not = 0.$} When $N^{r} \not = 0$, Eq.~(\ref{eqq9a}) yields
 \bq
 \lb{eqq13}
 t = t_{0} +  \int{\frac{\epsilon dr}{N\sqrt{f} - \epsilon N^{r}}},
 \eq
where $\epsilon = +1\; (\epsilon = - 1)$ corresponds to out-going (in-going) light rays. If 
\bq
\lb{eqq13aaa}
H(r) \equiv N\sqrt{f} - \epsilon N^{r},
\eq
 has $\delta$-th order zero at $r_{g}$, 
\bq
\lb{H-functin}
H(r) = H_{0}(r)(r - r_{g})^{\delta},
\eq
with $H_{0}(r_{g}) \not= 0$, 
we find that in the neighborhood $r = r_{g}$ Eq.~(\ref{eqq13}) yields
\bq
\lb{eqq14}
t = t_{0} +\frac{\epsilon}{H_{0}(r_{g})}
\begin{cases}
\frac{1}{1-\delta}(r-r_{g})^{1-\delta}, & \delta \not= 1,\\
\ln(r-r_{g}), & \delta = 1.\\
\end{cases}
\eq
Clearly, when 
\bq
\lb{eqq15}
\delta \ge 1, (n = 1), 
\eq
$|t|$ becomes unbounded as $r \rightarrow r_{g}$, and  an event horizon might exist. 

The Schwarzschild solution in the Painlev\'e-Gullstrand coordinates ~\cite{GP} is given by 
\bq
\lb{Sch}
N_{Sch}^{2} = f_{Sch}  = 1,\;\;\;
N_{Sch}^{r} = \epsilon_{1} \sqrt{\frac{r_{g}}{r}},
\eq
 where
$\epsilon_{1} = \pm 1$.  As shown in~\cite{AP,GSW}, this is also a vacuum solution of the HL theory in the HMT setup~\cite{HMT}. Then, we find that $H(r) = 1 - \epsilon_{1}\epsilon  \sqrt{r_{g}/r}$. 
Thus, for the solution with  $ \epsilon_{1} = +1$, the time
of  the out-going null rays, measured by asymptotically flat observers,   
becomes unbounded at $r_{g}$, and for  the solution with  $ \epsilon_{1} = -1$, the time
of  the in-going null rays   becomes unbounded. Therefore,  an event horizon is indicated to exist at $r = r_{g}$ in both cases. 

 In review of the above, KK generalized the notion of black holes defined in General Relativity to the case of a  non-standard dispersion relation~\cite{KKb}. 
 In summary, {\em a horizon is defined as a surface on which light rays are infinitely redshifted}. It should be noted that this redshift should be understood as measured by asymptotically 
 flat observers  at $N(r \gg r_{g}) \simeq 1$ and $N^{r}(r\gg r_{g}) \simeq 0$, with $r$ being the geometric radius, $r = \sqrt{A/4\pi}$, of the 2-sphere:  $t, r = $ constants,  where $A$
 denotes the area of the 2-sphere.

 
 \subsection{$n \ge 2$}
 
 In this case, eliminating $e$ from Eqs.~(\ref{eqq6}) and (\ref{eqq8}) we find that
 \bq
 \lb{eqq16}
 X^{n} - p(r) X - q(r, E) = 0,
 \eq
 where
 \bqn
 \lb{eqq17}
 X &\equiv& \left(\frac{\sqrt{\cal{D}}}{\dot{t}}\right)^{1/(n-1)} = \left(\frac{\left|r' + N^{r}\right|}{n \sqrt{f}}\right)^{1/(n-1)} ,\nb\\
 p(r) &\equiv& \frac{N^{r}}{\sqrt{f}},\;\;\;
 q(r, E) \equiv E N^{1/(n-1)}, 
 \eqn
 with $r' \equiv \dot{r}/\dot{t} = dr/dt$. To solve the above equation, again it is found convenient to consider the cases $N^{r} = 0$ and $N^{r} \not=0$
 separately. 
 
 \subsubsection{$N^{r} = 0.$} When $N^{r} = 0$, Eq.~(\ref{eqq16}) has the solution,
 \bq
 \lb{eqq18}
 t = t_{0} + \epsilon \int{\frac{dr}{n E^{(n-1)/n}\sqrt{f}N^{1/n}}},
 \eq
 where $\epsilon = +1$ corresponds to outgoing rays, and $\epsilon = -1$   to  ingoing rays.
 Thus, if  $f$ has an a-th order zero and $N^{2}$ a b-th order zero at $r = r_{g}$,  as given by Eq.~(\ref{fN}),
 we have $\sqrt{f}N^{1/n}\sim (r-r_{g})^{(a + b/n)/2}$. Then, from the above, we find that
  the time $t$, measured by asymptotically flat observers, becomes infinitely large
 at $r = r_{g}$, provided that ~\cite{KKb}
 \bq
 \lb{eqq19}
 a + \frac{b}{n} \ge 2.
 \eq
  For the solutions with the projectability condition ($N = 1,\; b = 0$), this is possible only when $a \ge 2$. 
  
  Considering  again 
  the Schwarzschild solution, $N^{2} = f = (r - r_{g})/r$, one finds  that this does not satisfy the condition (\ref{eqq19}) 
  with $n \ge 2$. Therefore, the  Schwarzschild black hole in General Relativity is no longer a black hole in the HL theory, because of the non-relativistic dispersion relations
  (\ref{1.4}).
  This is expected, since even in General Relativity when quantum effects are taken into account, such as the Hawking radiation, classical black holes are no longer
  black. 
  
\subsubsection{$N^{r} \not= 0.$} In this case,  let us consider an ingoing ray ${r}'< 0$. Suppose there is a horizon located at $r = r_{H}$. Then $r'(r) \simeq 0$ as we approach the horizon. Thus, if $N^r > 0$ and bounded away from zero,   $(r' + N^r)$ will also be positive, when the ray is sufficiently near the horizon. Conversely, if $N^r < 0$ and bounded away from zero, then $(r' + N^r)$ will  be negative sufficiently near the horizon.
Defining $H$ by $H(r, E) \equiv r'$, we find that for an ingoing ray near the horizon we have,
\bqn
\lb{4.18}
& t = t_0 + \int \frac{dr}{H(r,E)},
	\\ 
\lb{4.19}
 & H(r,E) = \epsilon n \sqrt{f}X^{n-1} - N^r,
\eqn
where
\bq
\epsilon = 
\begin{cases}
1, & N^r > 0, \\
-1, &  N^r < 0. \\
\end{cases}
\eq
Dividing (\ref{eqq16}) by $X$ and solving for $X^{n-1}$, we obtain
$$
X^{n-1} = \frac{N^r}{\sqrt{f}} + \frac{EN^{\frac{1}{n-1}}}{X}.
$$
Substituting  this   into (\ref{4.19}), we find,
\bq
\lb{*}
  H = (\epsilon n - 1) N^r + \epsilon n \sqrt{f}\frac{EN^{\frac{1}{n-1}}}{X}.
\eq
It follows that if $H$ has a zero at $r = r_{H}$, then
\bq
\lb{**}
  X|_{r = r_{H}} = - \frac{\epsilon n \sqrt{f} EN^{\frac{1}{n-1}}}{(\epsilon n - 1) N^r}.
\eq
The expression on the rhs is positive (negative) for $\epsilon = -1$ ($\epsilon = 1$). Thus, $H$ can have a zero only if $\epsilon = -1$.
Thus, we will henceforth consider only this case.
Differentiation of (\ref{*}) with respect to $r$ yields
\bqn
\lb{***}
H'(r) = - (n + 1)N^{r'} - n \sqrt{f}\frac{EN^{\frac{1}{n-1} - 1}}{(n-1)X}N' 
- n \frac{EN^{\frac{1}{n-1}}}{2 \sqrt{f}X}f'  +  n \sqrt{f}\frac{EN^{\frac{1}{n-1}}}{X^2}X'. ~~~~
\eqn
On the other hand, differentiation of (\ref{eqq16}) with respect to $r$ yields
\bqn
X'(r)  =  \frac{1}{nX^{n-1} - \frac{N^r}{\sqrt{f}}} \left[\left(\frac{1}{\sqrt{f}} \frac{dN^r}{dr} - \frac{N^r}{2f^{3/2}} \frac{df}{dr}\right)X 
 + \frac{EN^{\frac{1}{n-1} -1}}{n-1}\frac{dN}{dr}\right].
\eqn
Substituting the above into Eq.~(\ref{***}), we find that
\bqn
& &H'(r) \bigg|_{r = r_{H}} =
-\frac{n+1}{2} \Biggl(\frac{H_{1}}{H_{2}} -\frac{N^r f'}{f}
   +\frac{2 N^rN'}{N-n N}+2 N^{r'}\Biggr),
   \eqn
where
\bqn
& &  H_{1}\equiv 2 E (n+1) f N^r N' 
+ E (n-1) n N \left(N^r f' - 2 f N^{r'}\right),\nb\\
& &  H_{2} \equiv (n-1) n f N \Bigg[E + (n+1) N^{\frac{1}{1-n}} 
 \left(\frac{E n \sqrt{f}   N^{\frac{1}{n-1}}}{(-n-1) N^r}\right)^n\Bigg].
\eqn

If $H$ has a zero of order $\delta > 0$ at $r_{H}$, we can write it in the form, 
\bq
\lb{H-functionb}
H(r) = H_0(r_{H})(r - r_{H})^\delta + \cdots,  
\eq
as $r \to r_{H}$, where $H_0(r_{H}) \neq 0$.  
Therefore,
\bq
\lb{CT}
H'(r) \bigg|_{r = r_{H}} = 
\begin{cases}
 0, &  \delta > 1, \\
H_0(r_{H}), &  \delta  = 1, \\
\pm \infty, & 0 <\delta < 1.
\end{cases}
\eq
Now $t \to \infty$ as $r \rightarrow  r_{H}^{+}$ if and only if 
\bq
\lb{delta}
\delta \geq 1, 
\eq
which happens if and only if ${dH}/{dr} \big|_{r = r_{H}}$ is  finite. 
This gives an explicit condition on $f, N, N^r, E, n$ for the blow-up of $t$ at $r_{H}$.

It should be noted that $r_{H}$ usually depends on the energy $E$ of the test particles, as can be seen 
from the above and specific examples considered below. 

%

{Case $n=2$:} In this case, we have
\bq
H'(r)\bigg|_{r = r_{H}} =
\frac{H_{3}}{2 N \left[4 E f N+3 (N^r)^2\right]}, 
\eq
where
\bqn
  H_{3} \equiv  3 \left[4 E N^2 \left(N^r f'-2 f   N^{r'}\right)+8 E f N N^r N' \right. 
 -3 (N^r)^3 N'\Big], (n = 2).
\eqn
Again, for the Schwarzschild solution (\ref{Sch}),  
we have
\bqn
X(r) &=& \frac{1}{2} \left( - \sqrt{\frac{r_g}{r}} + \frac{\sqrt{4 E r+ r_g}}{\sqrt{r}}\right), \nb\\
H(r) &=& \frac{-3 \sqrt{r_g (4 E r+r_g)}+4 E r+3
   r_g}{\sqrt{r r_g}-\sqrt{r (4 E r+r_g)}}, 
  \eqn
 so that $H(r) = 0$ has the solution,
 \bq
 \lb{root} 
  r_{H} = \frac{3 r_g}{4E}, \; (n = 2),
  \eq
  at which we have
 \bq
  H'(r_{H}) = -\frac{2 E^{3/2}}{\sqrt{3} r_g}.
 \eq
 Then, according to Eq.~(\ref{CT}), we have $\delta = 1$,  i.e.,  $t$ diverges logarithmically as $r \rightarrow  r_{H}^{+}$. Therefore, in this case
there does exist a horizon. But, the location of it  depends on the energy $E$ of the test particle, and approaches zero   when $E \gg r_{g}$.  
This is understandable, as  the speed of light is unbounded in the UV, and in principle the  singularity located at $r = 0$ 
can be seen by asymptotically flat observers, as long as the  light rays sent by the observers have sufficiently high energies.  

{Case $n=3$:} In this case, we have
\bq
\lb{casen3}
X^{3} - p(r) X - q(r, E) = 0.
\eq
Assuming that $H(r) = 0$ has a real and positive root $r_{H}$, we find that
\bq
\lb{CT3}
H'(r) \bigg|_{r = r_{H}} = 
\frac{H_{4}}{3 N \left(27 E^2 f^{3/2} N-16 (N^r)^3\right)},
\eq
where
\bqn
 H_{4} \equiv 162 E^2 \sqrt{f} N^2 N^r f'  +32 (N^r)^4 N' 
 -162 E^2 f^{3/2} N \left(2 N N^{r'}-N^r N'\right). 
\eqn

For the Schwarzschild solution (\ref{Sch}),  
we have $p(r) = -\sqrt{r_g/r}$ and $q(r, E) = E$. Then,
we find that 
\bqn
\lb{casen3a}
& & X^{3} +\sqrt{\frac{r_g}{r}} X -E = 0,\\
\lb{casen3b}
&& H(r) = 4\sqrt{\frac{r_g}{r}} - \frac{3E}{X},
\eqn
from which  we find that $H (r) =0$ has a solution,  
\bq
\lb{rootb}
r_{H} = r_{g}\left(\frac{16}{27E^{2}}\right)^{2/3}, \; ( n = 3),
\eq
which  also depends on $E$, and approaches zero as $E \rightarrow \infty$. 
Substituting $r_{H}$  into Eq.~(\ref{CT3}),  we find
$H'(r_{H}) = - 27E^{2}/(16r_{g})$. That is, the hypersurface $r = r_{H}$ is also an observer-dependent  horizon in the case $n = 3$, and the radius of the horizon is
inversely proportional to the energy of the test particle. For $E \gg r_{g}$, we have $r_{H} \simeq 0$.

Another (simpler)  consideration  for the existence of the horizon is given as follows: First, from Eq.~(\ref{eqq17}) we find that
\bqn
\lb{JX1}
X = \left(\frac{|r' + N^r|}{n \sqrt{f}}\right)^{1/(n -
       1)} \simeq  \left(\frac{\epsilon N^r}{n\sqrt{f}}\right)^{1/(n - 1)} 
            \times \left(1 + \frac{H}{(n - 1) N^r}\right),  
\eqn
for $r \simeq r_{H}$. Inserting it into Eq.~(\ref{*}), we have, to leading order,
\bqn
\lb{JX2}
 \left(1 + \frac{E N^{1/(n - 1)}}{(n - 1)  \left(\frac{\epsilon N^r}{n \sqrt{f}}\right)^{n/(n -
       1)}}\right) H (r, E) = (\epsilon n - 1) N^r 
\left(1 + \frac{E N^{1/(n - 1)}}{(\epsilon n - 1) \left(\frac{\epsilon N^r}{n \sqrt{f}}\right)^{n/(n -
       1)}}\right).
\eqn
Then, we obtain
\bq
\lb{JX3}
\left.\frac{E N^{1/(n - 1)}}{( n + 1)
       \left(\frac{- N^r}{n \sqrt{f}}\right)^{n/(n -
       1)}}\right|_{r = r_H} = 1. 
\eq
 Given this, we can further simplify Eq.~(\ref{JX2}) to, 
 \bqn
 \lb{JX4}
   \frac{2 n}{n - 1} H (r, E) = - (n + 1) N^r (r_H) 
\left(1 - \frac{E N^{1/(n - 1)} (r)}{( n + 1)
       \left(\frac{- N^r (r)}{n \sqrt{f (r)}}\right)^{n/(n -
       1)}}\right). ~~~~~~~~~~
 \eqn
 Then, using Eq.~(\ref{H-functionb}), we have the following constraint for $N, N^r, f$ to
       satisfy so that a horizon can indeed exist, 
\bqn
\lb{CDb}       
 \frac{E N^{1/(n - 1)} (r) }{( n + 1)   \left(\frac{- N^r (r)}{n \sqrt{f (r)}}\right)^{n/(n -
       1)}} 
 = 1 + \frac{2 n H_0 (r_H)}{(n^2 - 1) N^r (r_H)} (r - r_H)^\delta + \cdots. ~~~~~~~~
\eqn
This equation can be first used to determine $r_H$   and then   $\delta$,  once $N, \; N^r$ and $ f$ are given.  
To illustrate how to use it,  let us consider  the Schwarzschild metric (\ref{Sch}). For $n = 2$, $r_H$
 can  be obtained simply from the above, and is given   exactly by Eq.~(\ref{root}), 
for which we have
\bqn
\lb{JX6}
&& \frac{E N^{1/(n - 1)} (r) }{( n + 1)
       \left(\frac{- N^r (r)}{n \sqrt{f (r)}}\right)^{n/(n -
       1)}} 
       \simeq 1 + \frac{r - r_H}{r_H}, ~~~~~
\eqn
that is, $\delta = 1$.

For $n = 3$, from Eq.~(\ref{JX3}) we find that $r_{H}$ is given by Eq.~(\ref{rootb}), and 
\bqn 
\frac{E N^{1/(n - 1)} (r) }{( n + 1) \left(\frac{- N^r (r)}{n \sqrt{f (r)}}\right)^{n/(n -
       1)}}  = \frac{3^{3/2} E r^{3/4}}{4 r_g^{3/4}}  
 \simeq 1 +  \frac{3}{4 r_H} (r - r_H) + \cdots.
\eqn
Therefore, in this case we have  $\delta = 1$ too.

 It should be noted that in the above analysis, we assumed that $F(\zeta) = \zeta^{n}$. In  more realistic models, the dispersion relation is a polynomial of
 $\zeta$, as shown by Eq.~(\ref{1.4}), or more specifically, 
 \bq
 \lb{Poly}
 F(\zeta) = \zeta + \frac{\zeta^{2}}{M^{2}_{A}}  +   \frac{\zeta^{4}}{M^{4}_{B}} + ...,
 \eq
 where $M_{A}$ and $M_{B}$ are the energy scales, which can be significantly different from the Planck one~\cite{BPS}. Therefore, for observers in low energy scales, where $\zeta \ll M_{A}, M_{B}$, the first term dominates, and some solutions, including the Schwarzschild solution, look like black holes, as shown in the case $n = 1$.
But,  for observers with high energies, those solutions may not be black holes any longer. Even if they are,  their horizons in general are observer-dependent,  as shown in the 
cases $n = 2$ and $n = 3$ explicitly for the Schwarzschild solution. To illustrate the main properties of the dispersion relation (\ref{Poly}),  we shall consider 
the case where only the first two terms are  important. 

\subsection{Trajectories of Test Particles with the Dispersion Relation $F(\zeta) = \zeta + {\zeta^{2}}/{M^{2}_{A}}$}

For the sake of simplicity, we  restrict ourselves to the  case $N^{r} = 0$. 
Substituting 
\bq
 \lb{Polya}
 F(\zeta) = \zeta + \frac{\zeta^{2}}{M^{2}_{A}},
 \eq
  into  Eq.~(\ref{eqq2}), we find 
\bq
\label{zetaeq}
  \zeta \left(1 + \frac{2\zeta}{M_A^2}\right)^2 = \frac{\dot{r}^2}{e^2f}.
\eq
Solving this equation directly for $\zeta$ yields a very complicated expression, and it is not clear how to proceed along this direction. 
Instead, we note that our goal is to find the analog of equation (\ref{eqq6}), i.e. of the equation $\delta \mathcal{L}_p/\delta e = 0$, where
\bqn
  \mathcal{L}_p &=& \frac{1}{2}\left(\frac{N^2}{e}\dot{t}^2 + e\Big[F(\zeta) - 2\zeta F'(\zeta)\Big]\right)\nb\\
	& =& \frac{1}{2}\left(\frac{N^2}{e}\dot{t}^2 - e\left[\zeta + \frac{3\zeta^2}{M_A^2}\right]\right).
\eqn
Thus, 
we will first calculate   $\delta \zeta/\delta e$,  implicitly by applying $\delta/\delta e$ to both sides of (\ref{zetaeq}),
which yields,
\bq
\frac{\delta \zeta}{\delta e} 
 = -\frac{2M_A^4\dot{r}^2}{e^3 f(M_A^4 + 8M_A^2 \zeta + 12 \zeta^2)}.
 \eq
Substituting this into the expression
$$
\frac{\delta  \mathcal{L}_p }{\delta e} = \frac{1}{2}\left(-\frac{N^2}{e^2}\dot{t}^2 - \left[\zeta + \frac{3\zeta^2}{M_A^2}\right]
- e\left[\frac{\delta \zeta}{\delta e} + \frac{6\zeta}{M_A^2}\frac{\delta \zeta}{\delta e}\right]\right),
$$
we find the following analog of equation (\ref{eqq6}),
\bqn
\label{4.6analog}  
 \zeta  \left(e^2 M_A^2+2 N^2 \dot{t}^2\right)+5 e^2 \zeta ^2+\frac{6 e^2 \zeta^3}{M_A^2} 
 + M_A^2   \left(N^2 \dot{t}^2- \frac{2 \dot{r}^2}{f}\right)  = 0,
\eqn
where $\zeta$ is given implicitly by Eq.~(\ref{zetaeq}). Note that in the limit $M_A \rightarrow \infty$,  the above equation   reduces 
precisely to  Eq.~(\ref{eqq6}) for $F(\zeta) = \zeta$ and $N^r = 0$,  as expected.

On the other hand, the analog of Eq.~(\ref{eqq8}) is simply
\bq
\label{4.7analog}  
  N^2 \dot{t} = eE.
\eq

Using Eqs.~(\ref{zetaeq}) and (\ref{4.7analog}) to eliminate $\dot{r}$ and $\dot{t}$ from Eq.~(\ref{4.6analog}), we find, 
$$
\frac{\delta  \mathcal{L}_p }{\delta e} = \frac{1}{2}\left(\zeta + \frac{\zeta^2}{M_A^2} - \frac{E^2}{N^2}\right) = 0.
$$
In the limit $M_A \to \infty$, this equation reduces to  $\zeta - \frac{E^2}{N^2} = 0,$ which is again consistent with the case $F(\zeta) = \zeta$. Solving this equation for $\zeta$, we infer that  
$$
\zeta = -\frac{M_A^2}{2} + \frac{M_A}{2N}\sqrt{4 E^2 + M_A^2 N^2}.
$$
Substitution of this expression    into Eq.~(\ref{4.6analog}) yields
\bqn
\label{4.6analog2} 
  \frac{M_A \left(N^2 \left(\frac{2\dot{r}^2}{e^2 f}+M_A^2\right)+4
   E^2\right)}{N \sqrt{4 E^2+M_A^2  N^2}}-\frac{N^2 \dot{t}^2}{e^2} 
 -\frac{3 E^2}{N^2}-M_A^2  = 0.
\eqn
Replacing   $e$ by $N^2 \dot{t}/E$ and then solving the resulting equation for $\dot{r}/\dot{t}$, we find  
$$\frac{\dot{r}^2}{\dot{t}^2} 
= \frac{f N \left(4 E^2+M_A^2 N^2\right) \left(\sqrt{4 E^2+M_A^2 N^2}-M_A
   N\right)}{2 E^2 M_A}.$$
In the limit $M_A \to \infty$, this equation becomes $\frac{\dot{r}^2}{\dot{t}^2} = fN^2,$ which is again consistent with the case $F(\zeta) = \zeta$.  Thus, the trajectory  is given by
\bq
t = t_0 + \int \frac{dr}{H(r,E)},
\eq
where
\bqn
H(r,E) &=& \sqrt{\frac{f N \left(4 E^2+M_A^2 N^2\right)}{2 E^2 M_A}} 
\sqrt{\sqrt{4 E^2+M_A^2 N^2}-M_A  N}.
 \eqn

As an example, let us consider the Schwarzschild solution,  $N^2 = f = 1 - r_g/{r}$, for which we find
$$
H = 2\sqrt{\frac{E}{M_A r_g^{3/2}}}(r - r_g)^{3/4} + {\cal{O}}\Big((r- r_g)^{5/4}\Big), 
$$
as $r \to r_g$, so that $t$ remains finite.  On the other hand, as $M_A \to \infty$,
$$
H = \frac{r - r_g}{r_g} + \frac{3E^2}{2M_A^2} +  {\cal{O}}\left(\frac{1}{M_A^4}\right).
$$
Thus, if we take the limit $M_A \to \infty$ before letting the trajectory  approach $r_g$, then $t$ will blow up logarithmically as $r \to r_g$.
As a result,  a horizon exists in this limit. 

More generally, if $f$ has an $a$th order zero and $N^2$ has a $b$th order zero at $r = r_g$, as given in Eq.~(\ref{fN}),
then, we find that
$$
H = 2\sqrt{\frac{Ef_0(r_g)N_0(r_g)}{M_A}} (r - r_g)^{\frac{a}{2} + \frac{b}{4}} +  {\cal{O}}\left((r-r_g)^{\frac{a}{2} + \frac{3b}{4}}\right), 
$$
as $r \to r_g$.  It follows that
\bqn
t &\simeq& t_0 + \frac{1}{2\sqrt{\frac{Ef_0(r_g)N_0(r_g)}{M_A}}} 
\begin{cases}
 \frac{(r-r_g)^{1 - \frac{a}{2} - \frac{b}{4}}}{1 - \frac{a}{2} - \frac{b}{4}},
 & \frac{a}{2} + \frac{b}{4} \neq 1, \\
\ln(r - r_g), 
& \frac{a}{2} + \frac{b}{4} = 1.\\
\end{cases}
\eqn
Therefore, $t$ blows up as $r \to r_g$,  if and only if
\bq
a + \frac{b}{2}  \geq 2,
\eq
which is exactly Eq.~(\ref{eqq19}) for $n = 2$, as expected.

 \section{Vacuum Solutions with $N^{r} = 0$ }

 When $N^{r} = 0$, the vacuum equations with  $J^t = v = p_{r} = p_{\theta} = J_{A} = J_{\varphi} = 0$
 yield the following most general solutions~\cite{GSW},  
 \bq
 \lb{4.1}
  f(r) = 1 + \frac{C}{r} -\frac{1}{3} \Lambda_g r^2, \;\;\; 
  N = 1,\;\;\; N^{r} = 0 = \varphi, 
 \eq
with the Hamiltonian constraint
  \bq 
 \lb{4.3c}
\int{{\cal{L}}_{V}  e^{\nu} r^{2} dr} = 0,
\eq
where ${\cal{L}}_{V} = {\cal{L}}_{V}(r, \Lambda_{g}, C, g_{s})$, as defined in Eq.~(\ref{2.5a}).


The gauge field  $A$ must satisfy the equations, 
\bqn
 \lb{4.3a}
& &  A' + A \nu' + \frac{1}{2} r F_{rr}= 0,\\
  \lb{4.3b}
& & r^{2}\big(A'' - \nu'A'\big) + r \big(A' + \nu' A\big) - A\big(1 - e^{2\nu}\big) \nb\\
& &~~~~~~~~~~~~~~~~~~~ +    e^{2\nu} F_{\theta \theta} = 0,
 \eqn
where $F_{ij}$ is  given by Eqs.~(\ref{eq3a}) and (\ref{A.2}). Then, from  Eq.~(\ref{4.3a}) we find that
\bq
\lb{4.4}
A = A_{0} e^{-\nu}  - \frac{1}{2}e^{-\nu}\int^{r}{r'e^{\nu(r')} F_{rr}(r') dr'},
\eq
where $A_{0}$ is an integration constant. The solutions with $\Lambda_{g} = 0$ were first studied in~\cite{HMT,AP}. 

Since now we have $N = 1$ and $ b = 0$,  Eq.~(\ref{eqq19}) shows that   a horizon exists only when $a \ge 2$.
It can be shown that for the solutions given by Eq.~(\ref{4.1}), this is impossible for any 
 chosen $C$ and $\Lambda_{g}$. Therefore, it is concluded that {\em the solutions given by Eq.~(\ref{4.1}) do not represent black holes}.
 
However, in some cases $f(r) = 0$ does have a real and positive root. So, there indeed exists some kind of coordinate singularities, and to obtain a maximally (geodesically) complete spacetime, some kind of extensions  are needed. Because of the breaking of the general covariance and the restricted diffeomorphism (\ref{1.2}), it is not clear if this requirement is still applicable  here in the HL theory. Even if it is not, some kind of extensions still seems needed. Such extensions are also needed in order to determine the range of $r$, from  which  the Hamiltonian constraint (\ref{4.3c}) can be carried out. Once this constraint is satisfied, one can integrate Eq.~(\ref{4.4}) to obtain the gauge field
$A$. To this end,  we divide the   solutions into the
cases:
$(i)\; C  = \Lambda_{g} = 0,\;
(ii) \; C \not= 0,\; \Lambda_{g} = 0, \; 
(iii) \; C = 0,\; \Lambda_{g} \not= 0$, and 
$(iv) \; C \not= 0,\;   \Lambda_{g} \not= 0$.
The first case is trivial, and it corresponds to the Minkowski spacetime with
$\nu = \Lambda = 0$ and $A = A_{0}$. Thus, in the following we shall consider only the last three cases.

\subsection{$C \not= 0,\;\;\; \Lambda_{g} = 0$}

In this case the metric takes the form
\bq
\lb{4.5a}
ds^{2} = - dt^{2} + \frac{dr^{2}}{1 + \frac{C}{r}} + r^{2}d^{2}\Omega,
\eq
from which we find that
\bqn
\lb{4.5}
{\cal{L}}_{V} &=& 2\Lambda + \frac{3g_{3}C^2}{2\zeta^2 r^{6}}  +  \frac{3g_{6}C^3}{4\zeta^4 r^{9}} 
+  \frac{45g_{8}C^2}{2\zeta^4 r^{8}}\left(1 + \frac{C}{r}\right),
\eqn
where $\Lambda = g_{0}\zeta^2/2$.
To consider the Hamiltonian constraint (\ref{4.3c}), we need to further distinguish the cases $C > 0$ and $C < 0$.

\subsubsection{$C > 0.$} When $ C > 0$, the metric (\ref{4.5a}) is singular only at $r = 0$, so the solution covers the whole spacetime
 $r \in (0, \infty)$. The singularity at the center  is a curvature one ~\cite{CW}, as it can be seen  from the  expressions,
 \bqn
 \lb{4.5b}
 R^{ij}R_{ij} &=& \frac{3C^{2}}{2r^{6}},\nb\\
R^{i}_{j}R^{j}_{k}R^{k}_{i} &=& - \frac{3C^{3}}{4r^{9}}, \nb\\
\left(\nabla_{i}R_{jk}\right) \left(\nabla^{i}R^{jk}\right) &= &   \frac{45 C^2}{ 2 r^{8}}\left(1 + \frac{C}{r}\right).
\eqn
 Since   event horizons do not exist for $C> 0$, this singularity is also  naked.  Inserting it into Eq.~(\ref{4.3c}), 
 we find that the Hamiltonian constraint is satisfied only when
\bq
\lb{4.6}
\Lambda = g_{3} = g_{6} = g_{8} = 0.
\eq
Considering Eq.~(\ref{A.2}), we find that $F_{ij}$ now has only two non-vanishing terms, given by
\bq
\lb{4.7}
F_{ij} = -\left(F_{1}\right)_{ij} + \frac{g_{5}}{\zeta^{4}}\left(F_{5}\right)_{ij}.
\eq
Substituting it into Eqs.~(\ref{4.3a}) and (\ref{4.3b}), we obtain 
\bq
\lb{4.8}
A = 1 + A_{0}\sqrt{1 + \frac{C}{r}}, \;\;\; g_{5} = 0.
\eq
It should be noted that the above solution holds not only   in the infrared (IR) regime but also in the UV.

\begin{figure}[tbp]
\centering
\includegraphics[width=8cm]{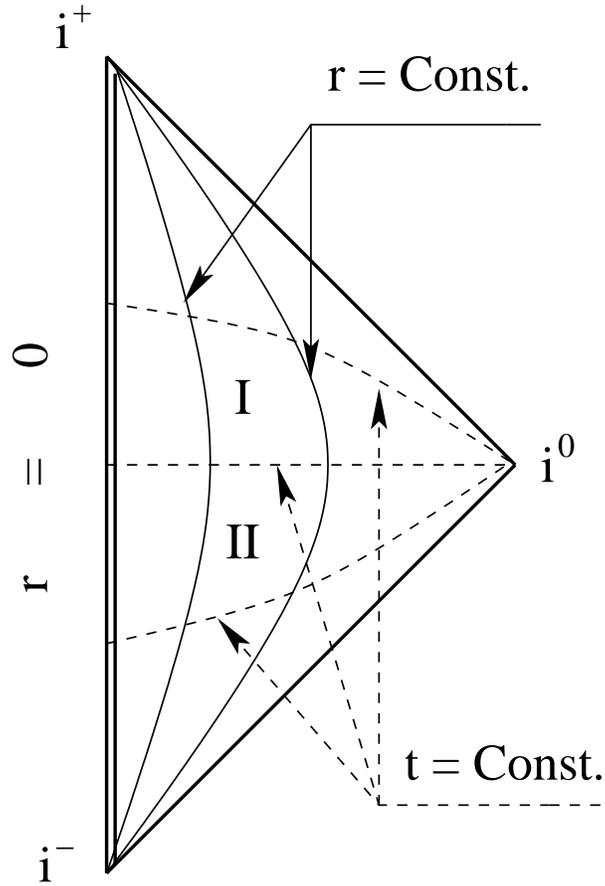}
\caption{The Penrose diagram for $N^{r} = 0,\; C > 0$ and $\Lambda_{g} = 0$. The double vertical solid lines represent the center ($r = 0$), at which the
spacetime is singular.  This singularity is clearly naked. Note that the restricted diffeomorphisms (\ref{1.2}) do not allow for the transformations
needed in order to draw Penrose diagrams. Therefore, these diagrams cannot be used to study the global structures of spacetimes in the 
HL theory but are included only for comparison.   }
\label{fig1}
\vspace{0.75cm}
\end{figure}

To study the global structure of the spacetime, let us first introduce a new radial coordinate $r^{*}$ via the relation
\bqn
\lb{4.8a}
r^{*} \equiv \int{\frac{dr}{\sqrt{1 + \frac{C}{r}}}}  =  - \frac{C}{2}\ln\frac{\left(\sqrt{r + C} + \sqrt{r}\right)^{2}}{C} 
 + \sqrt{r(r+C)}   = 
\begin{cases}
0, &  r = 0,\\
\infty, & r = \infty.
\end{cases}
\eqn
In terms of $r^{*}$ the metric takes the form,
\bq
\lb{4.8b}
ds^{2} = - dt^{2} + { dr^{*}}^{2} + r^{2}(r^{*})d^{2}\Omega.
\eq
Then, one might introduce the two double null coordinates $u$ and $v$ via the relations,
\bq
\lb{4.8c}
u = \tan^{-1}(t + r^{*}),\;\;\;
v = \tan^{-1}(t - r^{*}),
\eq
so that the metric finally takes the form, 
\bq
\lb{4.8d}
ds^{2} = -\frac{dudv}{\cos^{2}u\cos^{2}v} + r^{2}(u, v)d^{2}\Omega,
\eq
where $- \pi/2 \le u, v \le \pi/2$. The corresponding Penrose diagram is given by Fig. \ref{fig1}.

However, the coordinate transformations (\ref{4.8c}) are not allowed by the   foliation preserving diffeomorphisms Diff($M, \; {\cal{F}}$) of Eq.~(\ref{1.2}). So, 
in the HL theory the restricted  diffeomorphisms do not permit    Penrose diagrams.  In addition, due to the breaking of the general covariance, even if
one were allowed to do so, the causal structure of the spacetime cannot be studied in terms of it, as shown explicitly  in the previous sections for 
the Newtonian  theory.  

\begin{figure}[tbp]
\centering
\includegraphics[width=8cm]{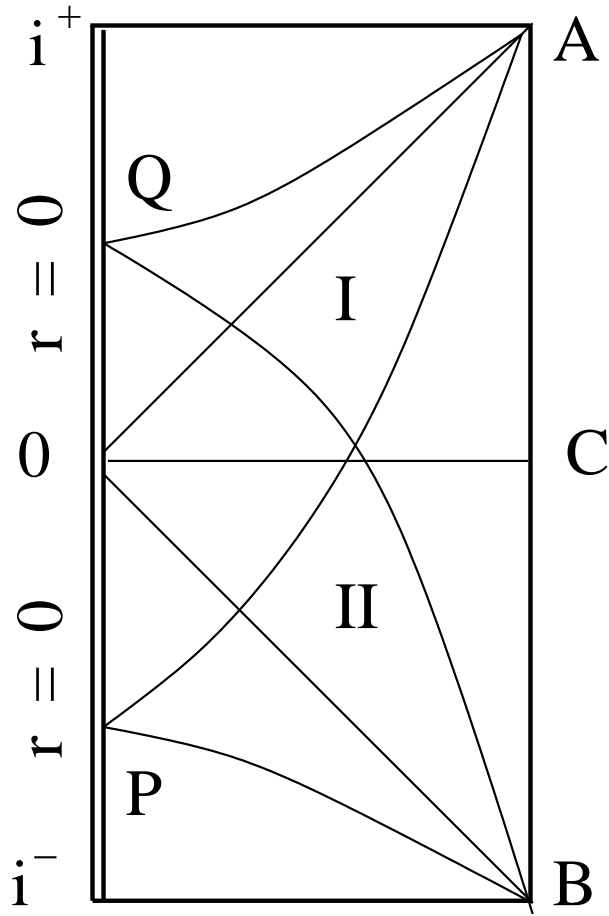}
\caption{The global structure of the spacetime in the ($\bar{t},\; \bar{r}^{*}$)-plane  for $N^{r} = 0,\; C > 0$ and $\Lambda_{g} = 0$. 
The double vertical solid lines represent the center ($r = 0$), at which the
spacetime is singular.   The vertical line $AB$ represents the spatial infinity  $r = \infty$, while the horizontal line $i^{+}A\; (i^{-}B)$ 
is the line where $t = \infty\; (t = -\infty)$. The lines $t = $ constant are the straight lines parallel to $OC$, while the ones $r = $ constant 
are the straight lines parallel to $i^{-}i^{+}$. The lines $BP, BO, BQ, PA, OA$ and $QA$ represent the radial null geodesics.}
\label{fig1a}
\vspace{0.75cm}
\end{figure}

Allowed are the coordinate transformations 
\bq
\lb{4.8e}
t = \tan{\bar{t}},\;\;\;
r^{*} = \tan{\bar{r}^{*}},
\eq
where  $- \pi/2 \le \bar{t} \le \pi/2$ and  $0 \le  \bar{r}^{*} \le \pi/2$. Then, the global structure of the spacetime is given by Fig. \ref{fig1a}.
 
\subsubsection{$ C < 0.$} In this case, setting $C = - 2M < 0$, the corresponding metric reads, 
\bq
\lb{4.8f}
ds^{2} = - dt^{2} + \left(1 - \frac{2M}{r}\right)^{-1}dr^{2} + r^{2}d^{2}\Omega.
\eq
This is the solution first found in~\cite{HMT}, in which it was
argued that the relativistic lapse function  should be ${\cal{N}} = N - A$  in the IR.  It is not clear how to then relate ${\cal{N}}$ to $N$ and $A$ in other regimes.   Instead, 
in this paper we shall simply take the point of view that $A$ and $\varphi$ are just gravitational gauge fields, and their effects on the 
spacetime itself occur only through  the field equations~\cite{GSW}. 
With the above arguments,  we  can consider the solution valid in any regimes, including   the IR and UV.  

Let us first note that
the metric (\ref{4.5a}) is asymptotically flat and singular at both $r = 0$ and $ r =  2M$. The singularity at $r = 0$ is a
curvature one,
as can be seen from Eq.~(\ref{4.5b}), but the one at $r =  2M$ is more peculiar. In particular,  in the region $r < 2M$ both $t$ and $r$ 
are timelike, in contrast to General Relativity where $t$ and $r$ exchange their roles across $r = 2M$.  
All the above indicate that the nature of the singularity at $r = 2M$ now is different. In fact, as to be shown explicitly below, the region $r < 2M$ actually is not part of the 
spacetime. 

To see this closely,    let us 
first consider the radial timelike geodesics. It can be shown that they are given by,
\bqn
\lb{4.9}
t &=&E \tau + t_{0},\nb\\
\tau  &=& \pm \frac{1}{\sqrt{E^{2} -1}}\Bigg\{M \ln\Big[(r -M) + \sqrt{r(r -2M)}\Big] 
 -  \sqrt{r(r -2M)}\Bigg\} +  \tau_{0},
\eqn
where $E$ is an integration constant, and $\tau$ denotes the proper time. 
The constant $\tau_{0}$ is chosen so that $\tau(r_{0}) = 0$ at the initial position of the test particle, $r = r_{0} > 2M$. 
The ``+" (``-") sign corresponds to the out-going (in-going) radial geodesics. It is clear that, starting at any given finite radius $r_{0}$,
 observers that follow the null geodesics will arrive at $r = 2M$ within a finite proper time. As shown in the last section, massless test  particles in the HL theory do not follow  null geodesics, because of the non-relativistic dispersion relations (\ref{1.4}). In other words,  in the HL theory  particles that follow the null geodesics are not massless and even may not be   test particles.  Setting
\bq
\lb{4.10}
e^{\alpha}_{(0)} \equiv \frac{dx^{\alpha}}{d\tau} = \left(E, - \sqrt{(E^{2} -1)f}, 0, 0\right),
\eq
where $f \equiv 1 - 2M/r$, we find that the  spacelike unit vectors, 
\bqn
\lb{4.11}
e^{\alpha}_{(1)}   &=& \left(\sqrt{E^{2} -1}, - E \sqrt{f}, 0, 0\right),\nb\\
e^{\alpha}_{(2)}   &=&  \frac{1}{r}\left(0, 0, 1, 0\right),\nb\\
e^{\alpha}_{(3)}   &=& \frac{1}{r\sin\theta} \left(0, 0, 0, 1\right),
\eqn
together with $e^{\alpha}_{(0)}$ form a freely-falling frame,
\bq
\lb{4.12}
e^{\alpha}_{(a)} e_{\alpha\; (b)}   = \eta_{ab}, \;\;\;
e^{\alpha}_{(0)}D_{\alpha} e^{\beta}_{(a)} = 0, 
\eq
where $D_{\alpha}$ denotes the 4D covariant derivatives, and $\eta_{ab}$ is the 4D Minkowski metric
with $a, b, = 0, ..., 3$. Then, from the geodesic deviations,
\bq
\lb{4.13}
\frac{D^{2}\eta^{a}}{D\tau^{2}} + {\cal{K}}^{a}_{b}\eta^{b} = 0,
\eq
where ${\cal{K}}_{ab} \equiv - R_{\sigma\alpha\beta\gamma} e^{\sigma}_{(a)}  e^{\alpha}_{(0)}e^{\beta}_{(0)}
e^{\gamma}_{(b)}$ denotes the tidal forces exerting on the observers, we find that  in the present case 
${\cal{K}}_{ab} $ is given by
\bq
\lb{4.14}
{\cal{K}}_{ab} = - \frac{(E^{2} -1)M}{r^{3}}\left(\delta^{2}_{a}\delta^{2}_{b} + \delta^{2}_{a}\delta^{3}_{b}\right).
\eq
Clearly, ${\cal{K}}_{ab}$ is finite at $r = 2M$. All the above considerations indicate that the singularity at $r = 2M$ is a coordinate
one, and to have a (geodesically) complete spacetime, extension beyond this surface is needed. However, unlike
that in GR,  any extension must be restricted to the Diff($M, \; {\cal{F}}$) of Eq.~(\ref{1.2}). Otherwise,
the resulting solutions do not satisfy the field equations. Explicit examples of this kind were given in~\cite{CW}. 

\begin{figure}[tbp]
\centering
\includegraphics[width=8cm]{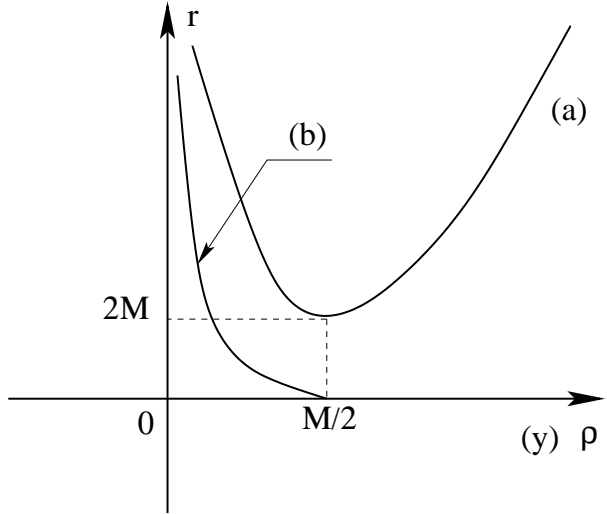}
\caption{The function $r$ defined: (a)  by Eq.~(\ref{4.15}); and (b)  by Eq.~(\ref{4.16}). } 
\label{fig2}
\vspace{0.75cm}
\end{figure}

In~\cite{HMT}, the isotropic coordinate $\rho$ was introduced, 
\bq
\lb{4.15}
r = \rho\left(1 + \frac{M}{2\rho}\right)^{2},
\eq
in terms of which  
 the metric (\ref{4.5a}) takes the form,
 \bq
\lb{4.5c}
ds^{2} = - dt^{2} + \left(1 + \frac{M}{2\rho}\right)^{4}\Big(d\rho^{2} + \rho^{2}d^{2}\Omega\Big),
\eq 
which is non-singular for $\rho > 0$. However, this cannot be considered as an extension to the region $r < 2M$, 
as now the geometrical radius $r$ is still restricted to $r \in (2M, \infty)$ for $\rho > 0$, as shown by Curve (a)  in Fig. \ref{fig2}. 
Instead, it connects two asymptotic regions, where $r = 2M$ acts as a throat, a situation  quite similar
to the Einstein-Rosen bridge~\cite{MTW}.  However, a fundamental difference of the metric (\ref{4.5c}) from the corresponding one 
in General Relativity  is that  it is not singular for 
any $\rho \in (0, \infty)$, while in General Relativity the metric still has a coordinate singularity at $\rho = M/2$ (or $r = 2M$) 
\cite{MTW}.  Therefore, in the HL theory Eq.~(\ref{4.5c}) already represents   an extension of the metric (\ref{4.5a}) 
 beyond the surface $r = 2M$. Since this extension is   analytical, it is  unique. It is remarkable to note that in this
extension the metric has the correct signature.  

It should be noted that  the Einstein-Rosen bridge is not stable in General Relativity ~\cite{MTW}.
Therefore, it would be  very interesting to know if this is still the case  in the HL theory.

\begin{figure}[tbp]
\centering
\includegraphics[width=8cm]{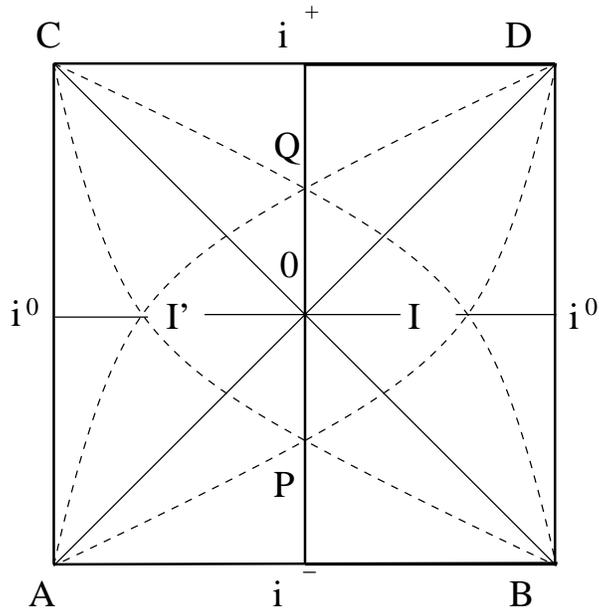}
\caption{The global structure of the spacetime  for $N^{r} = 0, \; C = - 2M <  0$ and $\Lambda_{g} = 0$.    The vertical line $i^{+}i^{-}$ represents the Einstein-Rosen throat 
($r = r_{g} \equiv 2M$), which is non-singular and connects the two asymptotically-flat  regions $I$ and $I'$.  The horizontal line $AB\; (CD)$ is the line where $t = -\infty \; (\infty)$, while the 
vertical lines $CA$ and $DB$ are the lines where $r = \infty$. The lines $t = $ constant are the straight lines parallel to $i^{0}i^{0}$, 
while the ones $r = $ constant are the straight lines parallel to $i^{-}i^{+}$.  The curved dotted lines $AD$ and $BC$, as well as the solid straight lines $AD$ and $BC$,  are the radial null geodesics.}
\label{fig3a}
\vspace{0.75cm}
\end{figure}

To study its global structure, we introduce the coordinate $r^{*}$ by
\bqn
\lb{4.5ca}
r^{*} &\equiv&  \int{\left(1 + \frac{M}{2\rho}\right)^{2}d\rho} = M\ln\left(\frac{2\rho}{M}\right)\nb\\
& & + \rho\left(1- \frac{M^{2}}{4\rho^{2}}\right) = 
\begin{cases}
-\infty, & \rho = 0,\\
\infty, & \rho = \infty.\\
\end{cases}
\eqn
Then, in terms of $r^{*}$  the metric can be also cast in the form of Eq.~(\ref{4.8b}). Following what was done in that case, one can see that the global structure of the 
spacetime is   given by Fig. \ref{fig3a}. 

\begin{figure}[tbp]
\centering
\includegraphics[width=8cm]{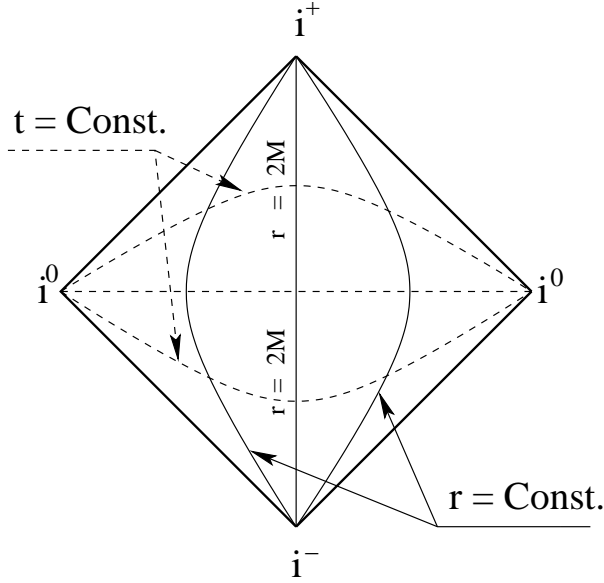}
\caption{The Penrose diagram for $N^{r} = 0, \; C = - 2M <  0$ and $\Lambda_{g} = 0$.  The straight lines $i^{+}i^{0}$ represent the future null infinities
at which we have $r = \infty$ and $t = \infty$, while the ones  $i^{-}i^{0}$ represent the past null infinities where
$r = \infty$ and $t = - \infty$. The vertical line $i^{+}i^{-}$ represents the Einstein-Rosen throat ($r = 2M$), which is non-singular and connects the 
two asymptotically-flat
 regions. }
\label{fig3}
\vspace{0.75cm}
\end{figure}

To compare it with that  given in GR,  the corresponding Penrose diagram is presented  in Fig. \ref{fig3}, although it is forbidden in the HL theory
by the  foliation-preserving diffeomorphisms Diff($M, \; {\cal{F}}$) of Eq.~(\ref{1.2}), as mentioned above.

It is interesting to see which kind of matter fields can give rise to such a spacetime in GR. To this purpose, we first
calculate the corresponding 4-dimensional Einstein tensor, 
\bq
\lb{4.17b}
{}^{(4)}G_{\mu\nu} = \frac{2M}{r^{3} f} \delta^{r}_{\mu} \delta^{r}_{\nu}  - \frac{M}{r}\left( \delta^{\theta}_{\mu} \delta^{\theta}_{\nu}
+ \sin^{2}\theta \delta^{\phi}_{\mu} \delta^{\phi}_{\nu} \right), 
\eq
which  corresponds to an anisotropic fluid, $T^{GR}_{\mu\nu} = \rho^{GR}u_{\mu}u_{\nu} + p^{GR}_{r}r_{\mu}r_{\nu}
+ p^{GR}_{\theta}\left(\delta^{\theta}_{\mu} \delta^{\theta}_{\nu}
+ \sin^{2}\theta \delta^{\phi}_{\mu} \delta^{\phi}_{\nu}\right)$,  with $\rho^{GR} = 0,\; p^{GR}_{r} = M/(4\pi Gr^{3})$ and $p^{GR}_{\theta} = - M r/(8\pi G)$,
where $u_{\mu} = \delta^{t}_{\mu}$ and $r_{\mu} = f^{-1/2}\delta^{r}_{\mu}$. Clearly,
such a fluid does not satisfy any of the energy conditions~\cite{HE73}. In particular, when $r \gg 1$  the tangential pressure becomes unbounded 
from below, while the radial pressure vanishes. Such a fluid is usually considered as non-physical in GR. However, in the current setup the spacetime 
is vacuum, 
and one cannot eliminate it by simply considering the energy conditions. Then, if the configuration
is stable, one can use it to construct time-machines~\cite{Visser}.

Inserting Eq.~(\ref{4.5}) into Eq.~(\ref{4.3c}), and considering  the fact that the range of  $r$ now is  $ r \in (2M, \infty)$,
 we find that the Hamiltonian constraint is satisfied, provided that
\bq
\lb{4.17}
 \Lambda = 0, \;\;\; 20\big(g_{6} - 3g_{8}\big) - {231}g_{3} \zeta^{2}M^{2} = 0.
 \eq
Then, Eqs.~(\ref{4.3a})  and (\ref{4.3b}) have the solution,
\bqn
\lb{4.17a}
A &=& 1 + A_{0} \sqrt{1 - \frac{2M}{r}} + \frac{g_{3}}{40\zeta^{2}M^{2}r^{6}}\Big[16\big(r-M\big)r^{5} \nb\\
& & - 8M^{2}\big(r + M\big)r^{3} - 3M^{3}\big(5r^{2} + 7 Mr + 1050 M^{2}\big)\Big],\nb\\
g_{5} &=& g_{8} = 0.  
\eqn

It is interesting to note that, replacing $\rho$ by $-y$ we find that in terms of $y$  metric (\ref{4.5c})
 takes the form, 
 \bq
\lb{4.5d}
ds^{2} = - dt^{2} + \left(1 - \frac{M}{2y}\right)^{4}\Big(dy^{2} + y^{2}d^{2}\Omega\Big),
\eq
 from which we can see that the geometrical radius now is given by
\bq
\lb{4.16}
r = y\left(1 - \frac{M}{2y}\right)^{2}.
\eq
Clearly, the whole region $ 0 \le r < \infty$ now is mapped to $  0 < y \le M/2$, as shown by Curve (b) in Fig. \ref{fig2}.
Metric (\ref{4.5d}) can be also obtained from metric (\ref{4.5c}) by the replacement,
$M \rightarrow -M$ and $\rho \rightarrow
y$. So, it must correspond to the case $C > 0$, i.e., the one with a negative mass, described
in the previous sub-case.

\subsection{$C = 0,\;\;\; \Lambda_{g}  \not= 0$}

We have
\bq
\lb{4.18a}
\nu = - \frac{1}{2}\ln\left(1 - \frac{1}{3}\Lambda_{g} r^{2}\right),
\eq
for which we find that
\bqn
\lb{4.19a}
{\cal{L}}_{V} &=& 2\big(\Lambda - \Lambda_{g}\big) + \frac{4(3g_{2} + g_{3})}{3\zeta^{2}}\Lambda^{2}_{g} 
 + \frac{8(9g_{4} + 3g_{5} + g_{6})}{9\zeta^{4}}\Lambda^{3}_{g},\nb\\
F_{ij} &=& \frac{g_{ij}}{9\zeta^{4}}\Big[3\zeta^{4}\big(\Lambda_{g} - 3\Lambda\big) + 2\zeta^{2}\big(3g_{2} + g_{3}\big)\Lambda_{g}^{2} 
+ 4\big(9g_{4} + 3g_{5}+ g_{6}\big)\Lambda_{g}^{3}\Big].
\eqn
To study the solutions further, we consider the cases $\Lambda_{g} > 0 $ and  $\Lambda_{g} < 0$, separately.

\subsubsection{$\Lambda_{g} < 0.$}  In this case, defining 
$r_{g} \equiv \sqrt{{3}/{\left|\Lambda_{g}\right|}}$,
we find that the corresponding metric takes the form,
\bq
\lb{4.21}
ds^{2} = - d{t}^{2} + \frac{dr^{2}}{1 + \left(\frac{r}{r_{g}}\right)^{2}} + r^{2} d^{2}\Omega,
\eq
which shows  that the metric is not singular except  at $r = 0$. 
But, it  can be shown
that this is a coordinate singularity. Setting
\bqn
\lb{4.22}
r^{*} &\equiv& \int{\frac{dr}{\sqrt{1 +  \left(\frac{r}{r_{g}}\right)^{2}}}} \nb\\
&=& r_{g}\ln\left\{\frac{r}{r_{g}} + \sqrt{1 +  \left(\frac{r}{r_{g}}\right)^{2}}\right\},
\eqn
one can cast the metric (\ref{4.21}) exactly in the form of Eq.~(\ref{4.8b}). Then, its global structure is that of Fig. \ref{fig1a}, and 
the corresponding Penrose diagram is  given by Fig. \ref{fig1}, but now   the center $r = 0$ is free of any spacetime singularity. 
Thus, the range of $r$ now is  $r \in [0, \; \infty)$. 
We then find that the Hamiltonian constraint (\ref{4.3c})  is satisfied, provided that ${\cal{L}}_{V} = 0$, i.e., 
\bqn
\lb{4.22a}
& & \Lambda \zeta^{4}r_{g}^{6} +   6\big(3g_{2} + g_{3}\big) \zeta^{2}r_{g}^{2}  - 12\big(9g_{4} + 3g_{5}  + g_{6}\big) 
 =  -3\zeta^{4}r_{g}^{4}. 
\eqn
Inserting the above into   Eqs.~(\ref{4.3a})  and (\ref{4.3b}), we obtain the solution, 
\bq
\lb{4.22b}
A = A_{0}\sqrt{1 + \left(\frac{r}{r_{g}}\right)^{2}} + A_{1},
\eq
where $A_{1}$ is a constant, given by
\bq
\lb{4.22c}
A_{1} \equiv  1- \Lambda r_{g}^{2} - \frac{3 - 3g_{2} - g_{3}}{\zeta^{2}r^{2}_{g}}.
\eq


\vspace*{-.4cm}
\subsubsection{$ \Lambda_{g} > 0.$} 
\begin{figure}[tbp]
\centering
\includegraphics[width=8cm]{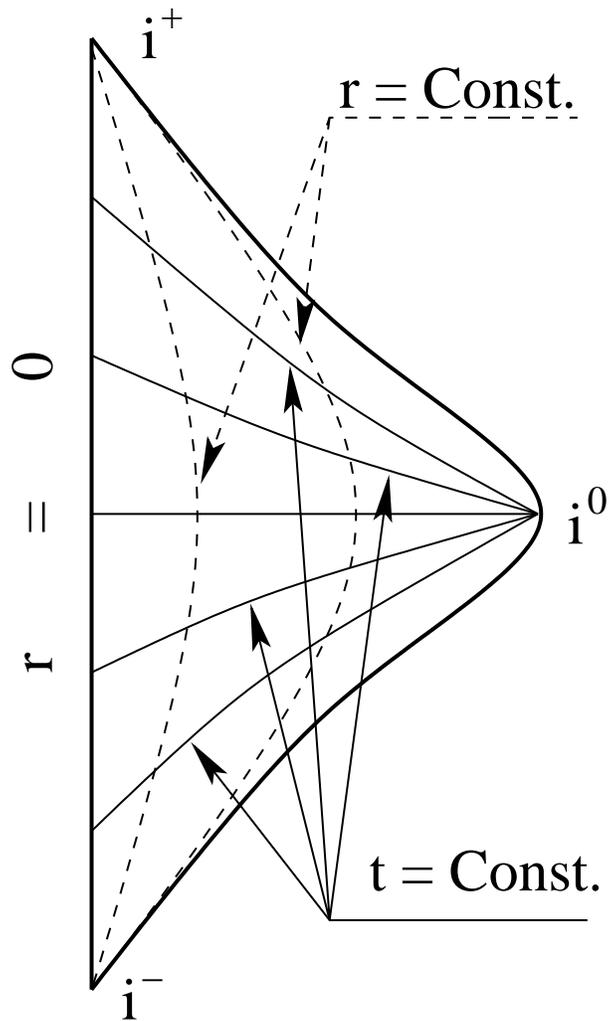}
\caption{The Penrose diagram for $N^{r} = 0,\; C =0$ and $\Lambda_{g} > 0$, which is the Einstein static universe.    The curves $i^{-}i^{0}$ and
$i^{+}i^{0}$ are, respectively, the lines where $t = -\infty,\; x = \pi$, and $t = + \infty,\; x = \pi$.}
\label{fig5}
\vspace{0.75cm}
\end{figure}

In this case,  the corresponding metric takes the form,
\bq
\lb{4.23}
ds^{2} = - dt^{2} + \frac{dr^{2}}{1 - \left(\frac{r}{r_{g}}\right)^{2}} + r^{2} d^{2}\Omega.
\eq
Clearly,  the metric has wrong signature in the region $r > r_{g}$. In fact, the hypersurface $r = r_{g}$ already represents the geometrical boundary
of the spacetime, and any extension beyond it is not needed. To see this clearly, we first introduce the coordinate $r^{*}$ via the relation,
\bq
\lb{4.24}
r^{*} \equiv \int{\frac{dr}{\sqrt{1 -  \left(\frac{r}{r_{g}}\right)^{2}}}}
= r_{g}\arcsin\left(\frac{r}{r_{g}}\right).
\eq
Then, in terms of $r^{*}$ the corresponding metric can be cast in the form $ds^{2} = r_{g}^{2}d\bar{s}^{2}$, where
\bq
\lb{4.25}
d\bar{s}^{2} = - d\bar{t}^{2} + dx^{2} +  \sin^{2}x d^{2}\Omega,
\eq
with $\bar{t} = t/r_{g},\; x = r^{*}/r_{g}$. But, this is exactly the homogeneous and isotropic Einstein static universe, which is geodesically    complete  for
$ - \infty < \bar{t} < \infty,\; 0 \le x \le \pi,\;
0 \le \theta \le \pi$ and $0 \le \phi \le 2\pi$,  with an $R\times{S^{3}}$ topology~\cite{HE73}. Then, it is easy to see that its global structure is given by Fig. \ref{fig1a}, but
now the vertical line $i^{-}i^{+}$ is free of spacetime singularity, and the line $AB$ is the one where $r = r_{g}$ (or $x = \pi$). The corresponding Penrose diagram is given by Fig. \ref{fig5}.

Therefore, in this case the range of $r$ is $r \in [0, \; r_{g}]$. Then, the Hamiltonian constraint (\ref{4.3c})
requires,
\bqn
\lb{4.26}
& & \Lambda \zeta^{4}r_{g}^{6} +   6\big(3g_{2} + g_{3}\big) \zeta^{2}r_{g}^{2}  +12\big(9g_{4} + 3g_{5}  + g_{6}\big) 
 =  3\zeta^{4}r_{g}^{4}. 
\eqn
Hence,  Eqs.~(\ref{4.3a})  and (\ref{4.3b}) have the solution,
\bq
\lb{4.27}
A = A_{0}\sqrt{1 - \left(\frac{r}{r_{g}}\right)^{2}} + A_{2},
\eq
where $A_{2}$ is another integration constant, given by
\bq
\lb{4.28}
A_{2} \equiv  1+ \Lambda r_{g}^{2} + \frac{3 - 3g_{2} - g_{3}}{\zeta^{2}r^{2}_{g}}.
\eq

It should be noted that in General Relativity the Einstein static universe is obtained by the exact balance between the gravitational attraction of matter ($\rho_{m} = \rho_{c},\; p_{m} = 0$)
and the cosmic repulsion ($\Lambda = \Lambda_{c}$), where $\Lambda_{c} = 4\pi G \rho_{c}$. As a result, the configuration is not stable against small perturbations
\cite{Hart}. However, in the present case since the spacetime is vacuum, Eq.~(\ref{4.26}) suggests that the balance is made by the attraction of the high-order curvature 
derivatives and the cosmic  repulsion, produced by both $\Lambda$ and $\Lambda_{g}$. Then, it would be very interesting to know whether it is stable or not in the current setup.

\subsection{$C \not= 0,\;\;\;  \Lambda_{g}  \not= 0$}

When $\Lambda_{g}, \ C \not= 0$, we find that
\bqn
\lb{4.29a}
 & & R^{ij}R_{ij} =  \frac{9 C^{2} + 8 \Lambda_{g}^{2} r^{6}}{6r^{6}},\nb\\
& & R^{i}_{j}R^{j}_{k}R^{k}_{i} = \frac{1}{36 r^{9}}\Big(27 C^{3} + 108 \Lambda_{g}C^{2} r^{3} + 32 \Lambda_{g}^{3}r^{9}\Big), \nb\\
& & \left(\nabla_{i}R_{jk}\right) \left(\nabla^{i}R^{jk}\right) =    \frac{45C^{2}}{2r^{8}}\Bigg(1 + \frac{C}{r} - \frac{1}{3}\Lambda_{g}r^{2}\Bigg),~~~~~~
\eqn
from which one can see that the spacetime is singular at $r = 0$. Moreover, we find from (\ref{A.2}) that 
\bqn
\label{Frrgeneral}
F_{rr}&= &\frac{1}{36 r^8 \zeta^4 F(r)}  \biggl\{
-27 C^3 (22 g_5+25 g_6-20 g_8) 
 -81 C^2 r (8 g_5+9 g_6-7 g_8 )
	 \nonumber\\
& & -9 C^2  r^3 \Bigl[\Lambda_g(-26 g_5-30 g_6+25 g_8) + \zeta ^2 g_3 \Bigr]
	 \nonumber\\
& & +12 C r^6 \Bigl[-3 \zeta ^4 +\Lambda_g \zeta ^2 (12 g_2+5 g_3)
 +\Lambda_g^2 (36 g_4+14 g_5+6g_6-g_8)\Bigr]	 \nonumber\\
& &+4 r^9 \Bigl[-3 \zeta ^4 (3 \Lambda - \Lambda_g)+2 \zeta ^2 \Lambda_g^2 (3 g_2+g_3)
+4 \Lambda_g^3 (9 g_4+3 g_5+g_6)\Bigr]\biggr\},
\eqn
where the third-order polynomial $F(r)$ is defined by 
$$F(r) = C + r - \frac{\Lambda_g}{3}r^3 \qquad \text{i.e.} \qquad e^{2\nu} = \frac{r}{F(r)}.$$
The function $\mathcal{L}_V$ is given by
\begin{equation}\label{LVintegrand}
  \mathcal{L}_V = \frac{\alpha + \beta r + \gamma r^3 + \delta r^9}{36 r^9 \zeta ^4},
  \end{equation}
where
\begin{eqnarray*}
  \alpha & =& 27 C^3 g_6+810 C^3 g_8,
  	\\
  \beta &=& 810 C^2 g_8,
  	\\
   \gamma &=& 108 C^2 g_5 \Lambda_g+108 C^2 g_6 \Lambda_g-270 C^2 g_8 \Lambda_g 
  +54 C^2 g_3 \zeta ^2,
  	\\
   \delta &=& 144 g_2 \zeta ^2 \Lambda_g^2+288 g_4 \Lambda_g^3+96 g_5 \Lambda_g^3+32 g_6 \Lambda_g^3
 +48 g_3 \zeta^2\Lambda_g^2+72 \zeta ^4 \Lambda -72 \zeta ^4 \Lambda_g.
\end{eqnarray*}

All the quantities in (\ref{4.29a}) are finite for any $r \neq 0$. On the other hand,
from Eq.~(\ref{4.1}) one can see that the metric coefficient $g_{rr}$ could become singular at some points. 
To study the nature of these singularities, we distinguish the four cases, $C > 0,\; \Lambda_{g} > 0$;  $C > 0,\; \Lambda_{g} < 0$;  $C < 0,\; \Lambda_{g} > 0$; 
and $C < 0,\; \Lambda_{g} < 0$. 

\subsubsection{$C > 0,\; \Lambda_{g} > 0.$} 
\begin{figure}[tbp]
\centering
\includegraphics[width=8cm]{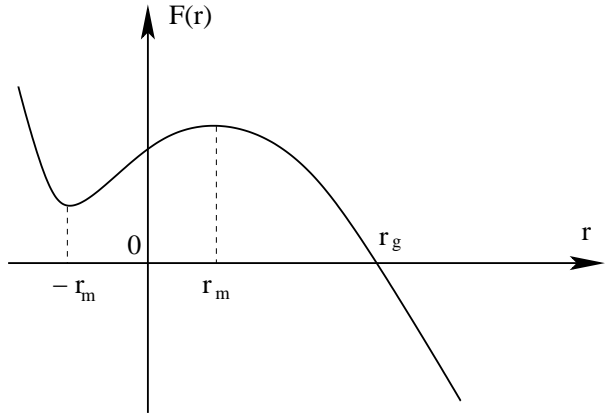}
\caption{ The function $F(r) \equiv re^{-2\nu}$ for $N^{r}  = 0,\; C > 0$ and $\Lambda_{g} > 0$, where $r_{m} = 1/\sqrt{\Lambda_{g}}$.}
\label{fig5a}
\vspace{0.75cm}
\end{figure}

\begin{figure}[tbp]
\centering
\includegraphics[width=8cm]{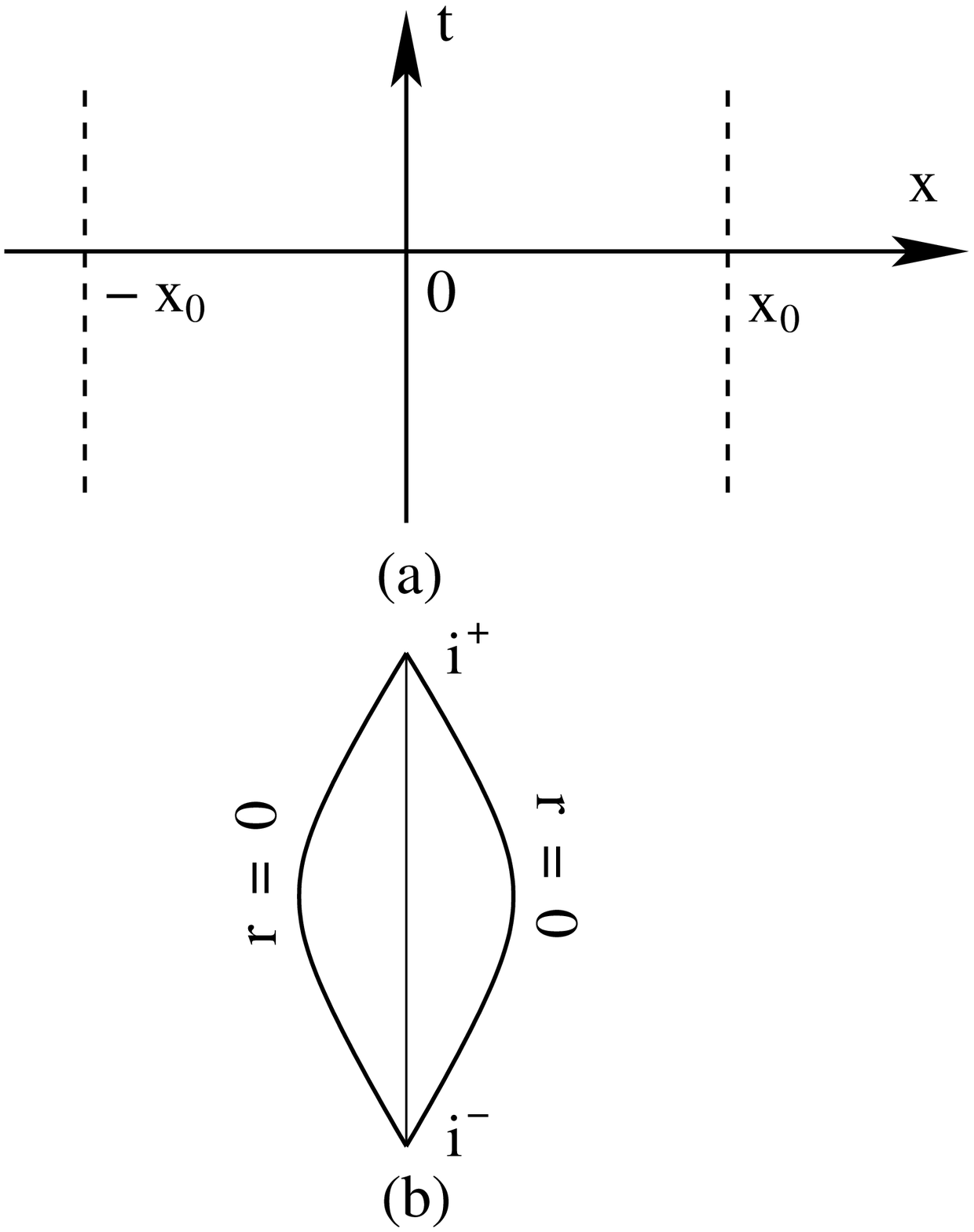}
\caption{(a) The spacetime in the ($t, x$)-plane, where $x_{0} \equiv \sqrt{r_{g}}$. (b) The Penrose diagram for $N^{r} = 0,\; C > 0 ,\; \Lambda_{g} > 0$.    The curves $i^{-}i^{+}$  
are the lines where  $r = 0$, at which the spacetime is singular. The straight line $i^{-}i^{+}$ represents the surface $r = r_{g}$.}
\label{fig6}
\vspace{0.75cm}
\end{figure}

In this case, the polynomial $F(r)$ has exactly one real positive root at, say, $r = r_{g}(C, \Lambda_{g})> 0$, as shown in Fig. \ref{fig5a}. We find that
\bq
\lb{4.30}
e^{2\nu} = \frac{r}{D(r)(r_{g} - r)},
\eq
where $D(r) \equiv \Lambda_{g}(r^{2} + r_{g} r + d)/3$, $d = r_{g}^{2} - 3/\Lambda_{g}$, and
$D(r) > 0$ for all $r > 0$.   Introducing the coordinate $x$ via the relation
\bq\label{xrrelation}
x = \int{\frac{dr}{2\sqrt{r_{g} - r}}} = - \sqrt{r_{g} - r}, 
\eq
or, inversely, $r = r_{g} - x^{2}$, 
the corresponding metric in terms of $x$ takes the form
\bq
\lb{4.31}
ds^{2} = - dt^{2} + \frac{4(r_{g} - x^{2})}{D(x)} d^{2}{x} +  \big(r_{g} - x^{2}\big)^{2} d^{2}\Omega, 
\eq
where $D(x) = \Lambda_{g}(x^{4} - 3r_{g}x^{2} + 3r_{g}^{2} - 3/\Lambda_{g})/3 > 0$ for $|x| < \sqrt{r_{g}}$. Clearly, the coordinate singularity at $r = r_{g}$ (or
$x = 0$) now is removed, and the metric is well defined for $|x| < \sqrt{r_{g}}$. At  the points, $ x = \pm  \sqrt{r_{g}}$ (or $r = 0$), the spacetime is singular, as
shown by Eq.~(\ref{4.29a}). Thus, in the present case the spacetime is restricted to the region $|x| <  \sqrt{r_{g}}$, $ - \infty < t < \infty$ in the ($t, x$)-plane, with the two
spacetime singularities located at $x = \pm \sqrt{r_{g}}$ as its boundaries.
The global structure of the spacetime and  the corresponding Penrose diagram are  shown in Fig. \ref{fig6}. 

The change of variables (\ref{xrrelation}) can be understood by considering the one-form
$$e^\nu dr = \frac{\sqrt{r}dr}{\sqrt{D(r) (r_g -r)}}.$$
Even though the denominator of the right-hand side vanishes at $r = r_g$, we can turn $e^\nu dr$ into a nonsingular one-form by introducing a Riemann surface. Indeed, if we promote $r$ to a complex variable and define the genus $1$ Riemann surface $\Sigma$ as the two-sheeted cover of the complex $r$-plane obtained by introducing two branch cuts along the intervals $[0, r_g]$ and $[r_1, r_2]$, where $r_1$ and $r_2$ are the two (possibly complex) zeros of $D(r)$, $e^\nu dr$ is a holomorphic one-form on $\Sigma$. 
Letting $(0,r_g]_1$ and $(0, r_g]_2$ denote the covers of the interval $(0, r_g]$ in the first and second sheets of $\Sigma$, respectively, the spacetime consists of points $(r, \theta, \phi, t)$ with $r \in (0, r_g]_1 \cup (0, r_g]_2$.
The variable $x = - \sqrt{r_{g} - r}$ introduced in (\ref{xrrelation}) is analytic near the branch point at $r = r_g$ and $r \in (0, r_g]_1 \cup (0, r_g]_2$ corresponds to $x \in (-\sqrt{r_g}, \sqrt{r_g})$. We can fix the definition of $x$ by choosing the branch of the square root so that, say, $x \geq 0$ for $r \in (0, r_g]_1$. Thus, in terms of the variable $x$, the spacetime manifold can be covered by a single global chart (no double cover is necessary) and the metric $ds^2$, which involves the square of the differential $e^{\nu}dr$, is manifestly nonsingular at $r = r_g$. 
In particular, the metric of the extended spacetime is analytic, which ensures that the extension is unique.

The Hamiltonian constraint is
\begin{equation}\label{hamiltonianconstraintrg}  
  \int_0^{r_g} \mathcal{L}_V e^\nu r^2 dr = 0.
\end{equation}
Indeed, the Hamiltonian constraint (\ref{eq1}) should be interpreted as
\bq
  \int  \mathcal{L}_V \text{Vol}_g = 0,
\eq
where $\text{Vol}_g$ is the volume form induced by the metric $g_{ij}$ and the integration extends over a spatial slice of the spacetime. Using the variables $(r,\theta, \phi)$, we have
$$\text{Vol}_g = e^\nu r^2 \sin\theta dr d\theta d\phi,$$
and the integration extends over $\theta \in [0, \pi]$, $\phi \in [0, 2\pi]$, and $r \in [0, r_g]_1 \cup [0, r_g]_2$.
By symmetry, the contributions from the sets where $r \in [0, r_g]_1$ and $r \in [0, r_g]_2$ are equal. Since each contribution is proportional to the left-hand side of (\ref{hamiltonianconstraintrg}), the constraint reduces to (\ref{hamiltonianconstraintrg}). 

In view of (\ref{LVintegrand}), the constraint (\ref{hamiltonianconstraintrg}) becomes
\begin{equation}\label{Hamconstraintalphabeta}
\int_0^{r_g}
\frac{\alpha + \beta r + \gamma r^3 + \delta r^9}{36 r^7 \zeta^4 \sqrt{F(r)}} \sqrt{r} dr = 0.
\end{equation}
Denoting the integrand in (\ref{Hamconstraintalphabeta}) by $I(r)$, we see that $|I(r)|$ is bounded by a constant times $1/\sqrt{r_g -r}$ as $r \rightarrow r_g$. Thus, the integral converges near $r_g$. On the other hand, as $r \rightarrow 0$,
$$I(r) = \frac{\alpha}{36r^{\frac{13}{2}}\zeta^4\sqrt{C}} + O\left(\frac{1}{r^{\frac{11}{2}}}\right),$$
so that (\ref{Hamconstraintalphabeta}) can only be satisfied if $\alpha = 0$. Using similar arguments, we infer that the coefficients $\beta, \gamma$, $\delta$ must also vanish, i.e., 
$$\alpha = \beta = \gamma = \delta = 0.$$
Solving these equations, we conclude that the Hamiltonian constraint is satisfied if and only if the $g_j$'s satisfy the following four conditions:
\begin{eqnarray}\label{g6845conditions}
&&g_4 = \frac{\zeta ^4 (\Lambda_g-\Lambda )-2 g_2
   \zeta ^2 \Lambda_g^2}{4 \Lambda_g^3}, 
	\\ \nonumber 
&& g_5 =  -\frac{g_3 \zeta ^2}{2 \Lambda_g}, \qquad g_6 = 0, \qquad g_8 = 0.
\end{eqnarray}

Using the conditions (\ref{g6845conditions}) in the expression (\ref{Frrgeneral}) for $F_{rr}$, we find that Eqs.~(\ref{4.3a}) and (\ref{4.3b}) have the solution
\begin{equation}\label{Aexpression}
A(r) = - \frac{\sqrt{F(r)}}{2\sqrt{r}} 
\int_{r_0}^r \frac{F_{rr}(r')(r')^{3/2} dr' }{\sqrt{F(r')}},
\end{equation}
where
\begin{eqnarray} \nonumber
 \label{Frrexpression}
 F_{rr} =&&\; 
-\frac{1} {36 r^8 \zeta ^2 \Lambda_g F(r)} 
 \Bigg\{-297 C^3 g_3  
 -324C^2r g_3 + 126C^2 r^3\Lambda_g g_3   	 \nonumber\\
&&   +12 C r^6 \Big[2\Lambda_g^2 (3 g_2+g_3)+\zeta ^2 (9 \Lambda -6\Lambda_g)\Big]
	\nonumber\\
&&  +8 r^9 \Lambda_g \Big[2\Lambda_g^2 (3 g_2+g_3)+\zeta ^2 (9 \Lambda -6 \Lambda_g)\Big]\Bigg\},
 \end{eqnarray}  
 and $r_0 \in (0, r_g)$ is a constant.
The integrand in (\ref{Aexpression}) is smooth for $0 < r < r_g$. Thus, $A(r)$ is a smooth function of $r \in (0, r_g)$. Unless   $g_3 = 0$, the integral diverges as $r \rightarrow 0$, so that $A(r)$ has a singularity at $r = 0$. As $r \rightarrow r_g$, the integrand is bounded by $\text{const}\times(r_g - r)^{-3/2}$. This implies that $A(r)$ is bounded as $r \rightarrow r_g$. In fact, viewed as a function on the Riemann surface $\Sigma$, $A(r)$ is analytic near $r = r_g$. This follows since the integrand in (\ref{Aexpression}) is a meromorphic one-form with a pole of at most second order at $r = r_g$. Thus, the integral has a pole of at most order one at $r_g$, which is cancelled by the simple zero of the prefactor $\sqrt{F(r)} = \sqrt{D(r)(r_g - r)}$.
In conclusion, the gauge field $A$ given by (\ref{Aexpression}) is a smooth function everywhere on the extended spacetime away from the singularity at $r = 0$.

 \subsubsection{$C > 0,\; \Lambda_{g} < 0.$} In this case, $F(r) > 0$ for $r > 0$ and the metric coefficient $g_{rr}$ is positive and non-singular except at the point $r = 0$, at which a naked spacetime singularity appears. The  corresponding Penrose diagram is given by Fig. \ref{fig1} with $ r \in (0, \infty)$. The Hamiltonian constraint (\ref{4.3c}) requires that
\begin{equation}\label{constraintcase2}  
  \int_0^\infty \mathcal{L}_V e^{\nu} r^2 dr = 0.
\end{equation}
As in the previous subsection, this constraint is equivalent to the conditions given in (\ref{g6845conditions}). 

The function $A(r)$ is again given by the formulas (\ref{Aexpression})-(\ref{Frrexpression}) and is a smooth function of $r \in (0, \infty)$. As $r \to \infty$, the absolute value of the integrand is bounded by $\text{constant} \times r^{-2}$. Thus, choosing $r_0 = \infty$ in (\ref{Aexpression}), we find that $A(r)$ is bounded as $r \to \infty$.
Unless  $g_3 = 0$, the integral diverges as $r \to 0$, so that $A(r)$ has a singularity at $r = 0$.

\subsubsection{$C < 0,\; \Lambda_{g} > 0.$} 
\begin{figure}[tbp]
\centering
\includegraphics[width=8cm]{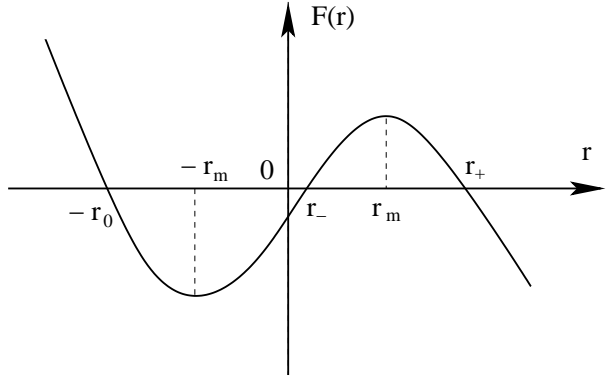}
\caption{ The function $F(r) = re^{-2\nu}$ for $C < 0$ and $\Lambda_{g} > 0$, where $r _{m} \equiv 1/\sqrt{\Lambda_{g}}$. $F(r) = 0$ has two positive roots $r_{\pm}$ only for $\Lambda_{g} < {4}/{(9C^{2})}$. When 
$\Lambda_{g} \ge {4}/{(9C^{2})}$,  $F(r)$ is always non-positive for any $r > 0$. }
\label{fig7}
\vspace{0.75cm}
\end{figure}

\begin{figure}[tbp]
\centering
\includegraphics[width=12cm]{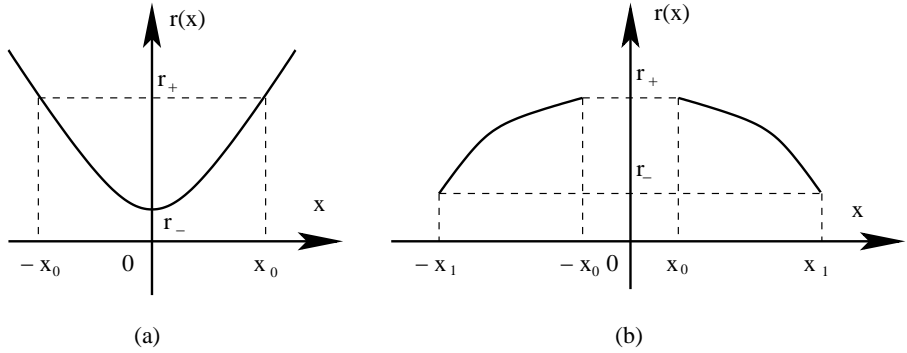}
\caption{ (a) The function $r$ vs $x$ given by Eq.~(\ref{4.38a}), where $x_{0} \equiv \sqrt{r_{+}  - r_{-}}$. (b) The function $r$ vs $x$ given by Eq.~(\ref{4.38b}), where $x_{1} \equiv  x_{0} + \sqrt{x_{0}}$.}
\label{fig8}
\vspace{0.75cm}
\end{figure}

In this case, if $\Lambda_{g} > 4/(9C^{2})$, $e^{2\nu} = r/F(r)$ is strictly negative  for all $r > 0$, so that, in addition to $t$, the coordinate $r$ is also timelike. The physics of such a spacetime
 is unclear,  if there is any. Therefore, in the following we consider only the case
\bq
\lb{4.36}
0 < \Lambda_{g} < \frac{4}{9C^{2}}.
\eq
 Then, we find that $F(r)$ is positive only for $0 < r_{-} < r < r_{+}$, where $r_{\pm}(\Lambda_{g}, C)$ are the two positive roots of   $F(r) = 0$, as shown in Fig. \ref{fig7}. We write $e^{2\nu}$ as
 \bq
 \lb{4.37}
 e^{2\nu} = \frac{r}{(r + r_{0})(r - r_{-})(r_{+} - r)},
 \eq
 where $r_{0}(\Lambda_{g}, C) > 0$. To extend the solution beyond $r = r_{\pm}$, we shall first consider the extension beyond $r = r_{-}$. Such an extension can be obtained via
 \bq
 \lb{4.38a}
 x = \int{\frac{dr}{2\sqrt{r - r_{-}}}} = \sqrt{r - r_{-}},
 \eq
 or inversely, $r = x^{2} + r_{-}$. Since $r < r_{+}$, we find that $ - x_{0} < x < x_{0}$ with  $x_{0} \equiv \sqrt{r_{+} - r_{-}}$. It can be seen that the coordinate singularity at $r = r_{-}$
 disappears, and the extended region is given by $ |x| < x_{0}$, as shown by Fig. \ref{fig8} (a).
 
\begin{figure}[tbp]
\centering
\includegraphics[width=8cm]{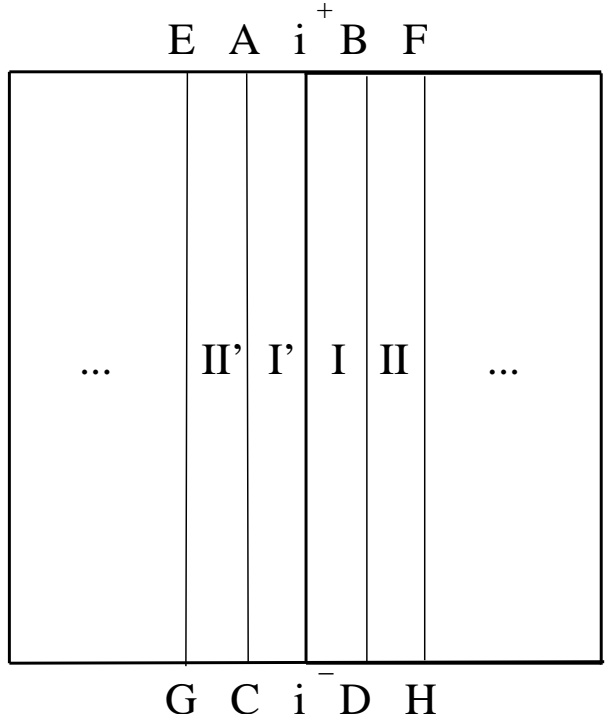}
\caption{ The global structure of the spacetime for $C < 0,\; \Lambda_{g} > 0$ and $\Lambda_{g} < 4/(9C^{2})$.  The vertical line $i^{+}i^{-}$ is the one where $r = r_{-}$, and the ones
$AC$ and $BD$ represent the lines where $r =r_{+}$, while on the lines $EG$ and $FH$ we have $r =r_{-}$. The spacetime   repeats itself  infinitely many times
in both directions of the $x$-axis. }
\label{fig9a}
\vspace{0.75cm}
\end{figure}

\begin{figure}[tbp]
\centering
\includegraphics[width=8cm]{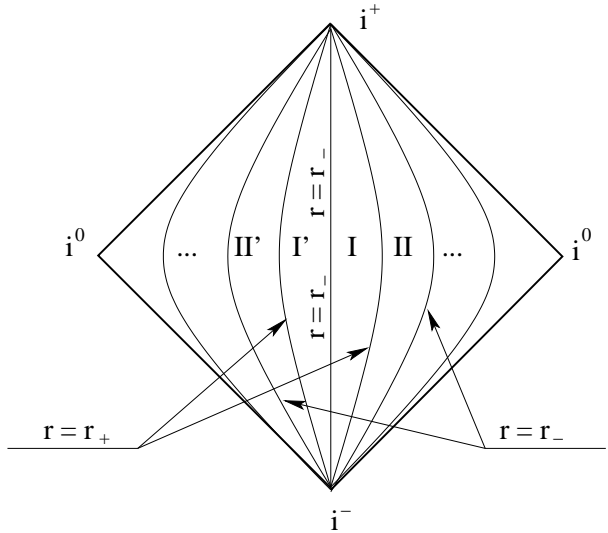}
\caption{ The Penrose diagram for $C < 0,\; \Lambda_{g} > 0$ and $\Lambda_{g} < 4/(9C^{2})$.  }
\label{fig9}
\vspace{0.75cm}
\end{figure}

To extend the solution beyond $r_{+}$, we introduce $x$ via the relation
 \bq
 \lb{4.38b}
 r = r_{+} - (x \mp x_{0})^{2}, 
 \eq
 where the ``$-$" sign applies when $x > x_0$ and the ``$+$" sign applies when $x < - x_{0}$. Fig. \ref{fig8} (b) shows the graph of $r$ as a function of $x$. From Fig. \ref{fig8}   we can see that the extension along  both the positive and the negative 
 directions of $x$ need to continue in order to have a    maximal spacetime. This can be done by repeating the above process infinitely many times, so finally   the whole ($t, x$)-plane 
 is covered by an infinite number of finite strips, in each of which we have $ r_{-} \le r \le r_{+}$. The global structure is that of
 Fig. \ref{fig9a} and the corresponding Penrose diagram is given by Fig. \ref{fig9}. Thus, in this case we have $r \in[r_{-}, r_{+}]$.

The Hamiltonian constraint (\ref{4.3c}) requires that
\begin{equation}\label{constraintcase3}
\int_{r_-}^{r_+}
\frac{\alpha + \beta r + \gamma r^3 + \delta r^9}{36 r^7 \zeta^4 \sqrt{F(r)}} \sqrt{r} dr = 0.
\end{equation}
Geometrically, this condition can be understood by introducing a Riemann surface $\Sigma$ as a double cover of the complex $r$-plane with two branch cuts along $[r_-, r_+]$ and $[-r_0, 0]$. The integrand in (\ref{constraintcase3}) is a one-form $\omega$ on $\Sigma$ which is holomorphic in a neighborhood of the closed curve $a_1 \equiv [r_-, r_+]_1 \cup [r_+, r_-]_2$. Topologically, the elliptic curve $\Sigma$ is a torus, $a_1$ is a nontrivial cycle, and the condition (\ref{constraintcase3}) states that the integral of $\omega$ along the cycle $a_1$ vanishes. 
This imposes a constraint on the coefficients $\alpha, \beta, \gamma, \delta$, which translates into a condition on the $g_j$'s involving elliptic integrals. Assuming this condition to hold, the function $A(r)$ is given by (\ref{Aexpression}) with $r_0 \in (r_-, r_+)$ and $F_{rr}$ as in (\ref{Frrgeneral}).

\subsubsection{$C < 0,\; \Lambda_{g} < 0.$}  
\begin{figure}[tbp]
\centering
\includegraphics[width=8cm]{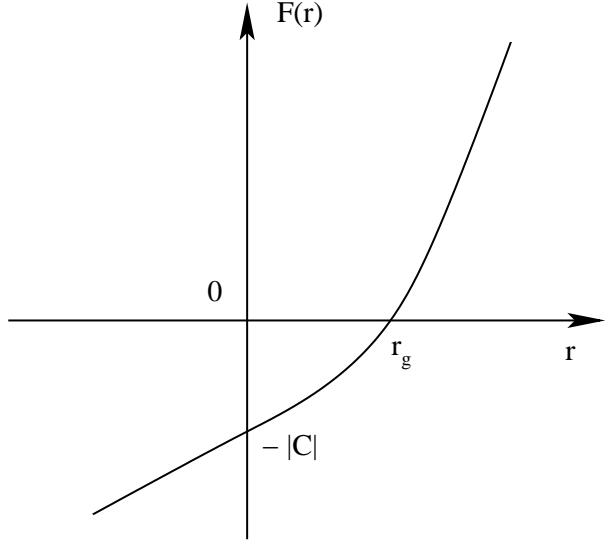}
\caption{ The function $F(r) = re^{-2\nu}$ for $C < 0$ and $\Lambda_{g} < 0$, where $r_{g}$ is the only positive root of $F(r) = 0$. }
\label{fig10}
\vspace{0.75cm}
\end{figure}

In this case,  the function $F(r) =  re^{-2\nu}$ is positive only for $r > r_{g}$, as shown in Fig. \ref{fig10}. Thus,   $e^{2\nu}$ can be written in the form,
\bq
\lb{4.40}
e^{2\nu} = \frac{r}{D(r) (r - r_{g})},
\eq
where $D(r) > 0$ for $r > 0$. The extension can be carried out by introducing a new coordinate $x$
via the relation,
\bq
\lb{4.41}
r = x^{2} + r_{g}.
\eq
In terms of $x$ the coordinate singularity at $r = r_{g}$ disappears, and the extended spacetime is given by $ - \infty < t, x < \infty$ in the ($t, x$)-plane. Its global structure
is given by Fig. \ref{fig3a}, while 
the  corresponding Penrose diagram  is given by Fig. \ref{fig3}. Thus, in this case the range of $r$ is  $r \in [r_{g}, \infty)$. 

The Hamiltonian constraint (\ref{4.3c}) requires that
$$\int_{r_g}^{\infty}
\frac{\alpha + \beta r + \gamma r^3 + \delta r^9}{36 r^7 \zeta ^4 \sqrt{F(r)}} \sqrt{r} dr = 0.$$
The behavior of the integrand as $r \to \infty$ implies that
$$\alpha = \beta = \gamma = \delta = 0,$$
so that the constraint reduces to (\ref{g6845conditions}) and the function $A(r)$ is given by (\ref{Aexpression})-(\ref{Frrexpression}),
which  is  not singular  everywhere in the extended spacetime.

\section{Vacuum Solutions with $N^{r} \not= 0$}

When $N^{r} \not= 0$, the vacuum solutions are given by ~\cite{GSW},  
\bq
\lb{5.0}
ds^{2} = - dt^{2} + e^{2\nu}\left(dr + e^{\mu-\nu} dt\right)^{2} + r^{2}d^{2}\Omega,
\eq
  with
 \bqn
 \lb{5.1}
 \mu &=& \frac{1}{2}\ln\Bigg(\frac{2m}{r} + \frac{1}{3}\Lambda r^2 - 2A(r) + \frac{2}{r}\int^{r}{A(r') dr'}\Bigg),\nb\\
  \nu &=&   \varphi = \Lambda_{g} = 0,
 \eqn
 where the gauge field $A$ must satisfy   the  Hamiltonian constraint,
\bq
\lb{5.2}
\int_{0}^{\infty}{r A'(r) dr} = 0.
\eq
Otherwise, it is free. However, as shown in~\cite{GSW}, the solar system tests seem uniquely to choose the Schwarzschild 
solution $A = 0$. Therefore, in the following we shall consider only this case, 
 \bqn
 \lb{5.1a}
 \mu &=& \frac{1}{2}\ln\Bigg(\frac{2m}{r} + \frac{1}{3}\Lambda r^2\Bigg),\nb\\
  \nu &=&   \varphi = \Lambda_{g} = A = 0.
 \eqn
 It should be noted that if $(N, \nu, N^{r})$ is a solution of the vacuum equations, so is $(N, \nu, - N^{r})$. The latter can be
 easily obtained by the replacement  $t \rightarrow - t$. With such changes,  we have $K_{ij} \rightarrow - K_{ij}$ (in the static case).
Clearly, these   do not affect the singularity behavior. 
 We then obtain~\cite{CW,GPW} \footnote{There is a typo in the expression of K given by Eq.~(3.2) in~\cite{CW}. Although it propagates 
to other places,  this does not affect our main conclusions, as $K$ and $K_{ij}K^{ij}$ have similar singularity behavior. },
\bqn
\lb{5.1b}
R_{ij} &=& 0,\nb\\
K &=& \epsilon_{1} \sqrt{\frac{3}{r^{3}\left(6m + \Lambda r^{3}\right)}} \; \left(3m + \Lambda r^{3}\right),\nb\\
K_{ij}K^{ij} &=&  \frac{27m^{2} + 6m\Lambda r^{3} + \Lambda^{2}r^{6}}{r^{3}\left(6m + \Lambda r^{3}\right)},
\eqn
where $\epsilon_{1} (= \pm 1)$ originates from  the expression $N^{r} = \epsilon_{1} e^{\mu}$, obtained by the 
replacement  $t \rightarrow - t$, as mentioned above.
To further study  the above solutions,  let us  consider the cases
(1) $\; m = 0, \Lambda \not=0$;  (2) $\; m \not= 0, \Lambda =0$; and (3) $\; m \not= 0, \Lambda \not=0$ separately. We shall
assume that $m \ge 0$, while $\Lambda$ can take any values. 

\subsection{$m = 0, \;\;\; \Lambda \not=0$}

In this case, only $\Lambda > 0$ is allowed~\cite{GSW}, as can be seen from
Eq.~(\ref{5.1}). That implies that the anti-de Sitter spacetime cannot be written in the static form of Eq.~(\ref{3.1b})
with the  projectability condition. Then, we have $ N^{2} = f = 1, \; N^{r} = \epsilon_{1} {r}/{\ell}$, or
\bq
\lb{5.3}
ds^2 = - dt^{2} + \left(dr + \epsilon_{1} \frac{r}{\ell} dt\right)^{2} + r^{2}d^{2}\Omega,
\eq
where $\ell \equiv \sqrt{3/|\Lambda|}$.  
Without loss of generality, we shall consider only the case $\epsilon_{1} = -1$, as the case   
 $\epsilon_{1} = 1$ can be simply obtained from the one $\epsilon_{1} = -1$ by inverting the time coordinate. 
In terms of $N, N^{i}, g_{ij}$ or their inverses, $N_{i}, g^{ij}$, the metric is non-singular, except for the trivial  $r = 0$ and
$\theta = 0, \pi$. In addition,   from Eq.~(\ref{5.1b}) we also find that 
\bq
\lb{5.5}
K = -\sqrt{2\Lambda},\; \;\; K_{ij}K^{ij} = \Lambda, \; (m = 0).
\eq
On the other hand,  in terms of the 4-dimensional metric, $g_{\mu\nu}$ and $g^{\mu\nu}$, it is  not singular either, as
one can see from the expressions,
\bqn
\lb{5.3a}
\left({}^{(4)}g_{\mu\nu}\right) &=&  \left(\begin{matrix}
 - \frac{\ell^{2} - r^{2}}{\ell^{2}}, &- \frac{r}{\ell}\delta^{r}_{i}\\
-\frac{r}{\ell}\delta^{r}_{i}, & g_{ij} \end{matrix}\right),\nb\\
\left({}^{(4)}g^{\mu\nu}\right) &=& 
\left(\begin{matrix}
 - 1, & -\frac{r}{\ell}\delta_{r}^{i}\\
-\frac{r}{\ell}\delta_{r}^{i}, & g^{ij} - \frac{r^{2}}{\ell^{2}}\delta_{r}^{i}\delta_{r}^{j}\\
\end{matrix}\right),
\eqn
although the nature of the radial coordinate does change,
\bq
\lb{5.3b}
g^{\mu\nu}r_{,\mu} r_{,\nu} = 1 - \frac{r^{2}}{\ell^{2}} = 
\begin{cases}
{\mbox{ timelike}}, &  r > \ell, \\
{\mbox{ null}}, &  r = \ell, \\
{\mbox{ spacelike}}, &  r < \ell. \\
\end{cases}
\eq
To study  the   solution further in the HL theory, we consider two different regimes,
$E \ll M_{*}$ and  $E \gg M_{*}$,  where $M_{*} = {\mbox{min.}}\left\{M_{A}, \; M_{B}, ... \right\}$ and $M_{n}$'s are the energy scales
appearing in the dispersion relation (\ref{Poly}).

\subsubsection{$E \ll M_{*}.$} When the energy $E$ of the test particle is much less than $M_{*}$, from Eq.~(\ref{Poly}) one can see that $F(\zeta) \simeq \zeta$. 
This corresponds to the relativistic case (n = 1). Then, for the ingoing test particles ($\epsilon = -1$), we have
\bq
\lb{5.4}
H = N\sqrt{f} + N^{r} = \frac{\ell - r}{\ell} .
\eq
Thus, the hypersurface $r = \ell$ is indeed a horizon. In fact, it represents a cosmological horizon, as first found in General Relativity~\cite{GH}. 

However, because of the restricted diffeomorphisms (\ref{1.2}), it is very interesting to see the global structure of the 
de Sitter spacetime in the HL theory. To this purpose, let us consider the coordinate transformations,
\bq
\lb{5.6}
t' = \ell e^{-t/\ell},\;\;\;
r' = r e^{-t/\ell},
\eq
in terms of which the corresponding metric takes the form,
\bqn
\lb{5.7}
ds^{2}&=& - d{t}^{2} + e^{2{t}/\ell}\left(dr'^{2} + r'^{2}d^{2}\Omega\right).\nb\\
&=&  \left(\frac{\ell}{t'}\right)^{2}\left(- dt'^{2} + dr'^{2} + r'^{2}d^{2}\Omega\right).
\eqn
From Eq.~(\ref{5.6}) we can see that the whole $(t, r)$-plane, $ -\infty < t < \infty,\; r \ge 0$, is mapped to the region $t', \; r ' \ge 0$.  However, the metric
now becomes singular at $t' = 0, \infty$ (or $t = \pm \infty$). To see the nature of these singularities, one may recall the 5-dimensional embedding of
the de Sitter spacetime in General Relativity~\cite{HE73}, from which we find that in terms of the 5-dimensional coordinates $v$ and $w$, $t'$ is given by
 $t' = \ell^{2}/(v + w)$. Therefore, $t'  \ge 0$ corresponds to $v + w \ge 0$.  Thus,  the region $t', \; r ' \ge 0$  only represents the half hyperboloid 
 $v + w \ge 0$, as shown by Fig. 16 (ii) in~\cite{HE73}.  In particular,  $t' = 0$ represents the boundary of the
 spacelike infinity, so extension beyond this surface may not be needed.  Although
 the extension given in~\cite{HE73} in terms of the  static Einstein universe coordinates
 $(\bar{t}, \bar\chi, \bar{\theta},\bar{\phi})$  is forbidden here by the restricted diffeomorphisms (\ref{1.2}),
  as that extension requires, 
 $$
 t = \ell \ln\left[\cosh\left(\frac{\bar{t}}{\ell}\right)\cos(\bar\chi) +  \sinh\left(\frac{\bar{t}}{\ell}\right)\right],
 $$
  the extension across $t' = \infty$ (or $v + w = 0^{+}$)  seems necessary. 

\begin{figure}[tbp]
\centering
\includegraphics[width=8cm]{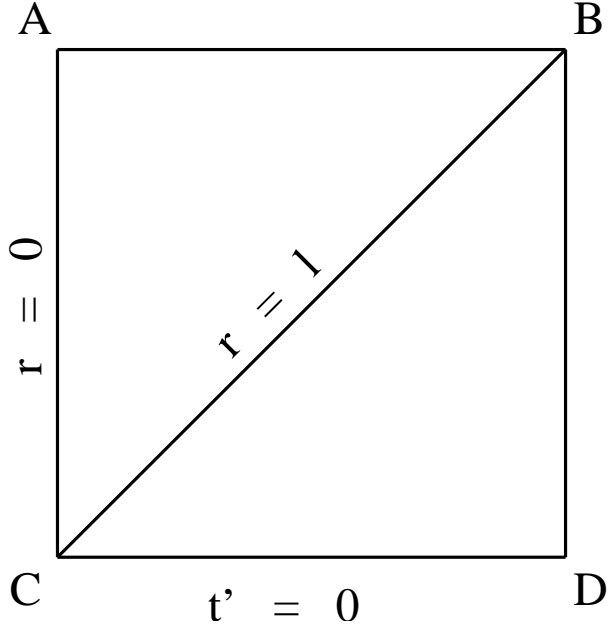}
\caption{ The global structure of the de Sitter solution $N^{2} = f = 1, \; N^{r} = - \sqrt{r/\ell}$ in the HL theory with the restricted diffeomorphisms (\ref{1.2}) for the 
region $t' \ge 0$. The horizontal line  $AB$ corresponds to $t' = \infty$ (or $t = -\infty$), while the vertical line $BD$ to $r' = \infty$ (or $r = \infty$).}
\label{fig11}
\vspace{0.75cm}
\end{figure}

 Another way to see the need of an extension beyond $t' = 0$ is that the metric (\ref{5.7}) is well-defined for $t' < 0$. So, one may  simply  take   $ - \infty < t' < \infty$. 
 But, this cannot be considered as an extension, as   the metric (\ref{5.7}) is singular at $t' = 0$, and the two regions $t' > 0$ and $t' < 0$ are not smoothly connected
 in the $t', r'$-coordinates.  In this sense, a proper extension is still needed. However, due to the  restricted diffeomorphisms (\ref{1.2}), it is not clear if such extensions
 exist or not.  Fig. \ref{fig11} shows the global structure of the region $t' \ge 0$, which is quite different from its corresponding Penrose diagram~\cite{GH}.

\subsubsection{$E \gg M_{*}.$} When the energy $E$ of the test particle is greater than  $M_{*}$,  from Eq.~(\ref{Poly}) one can see that high order momentum terms become important, and
$F(\zeta) \simeq \zeta^{n}, \; (n \ge 2)$. For the sake of simplicity,  we consider the case with $n = 2$ only. Then, from Eqs.~(\ref{eqq16}) and 
(\ref{4.19}) we find that 
\bqn
\lb{5.8}
X &=& \frac{2\ell E}{\sqrt{r^{2} + 4 \ell^{2}E} + r},\nb\\
H &=& \frac{r}{\ell} -  \frac{4\ell E}{\sqrt{r^{2} + 4 \ell^{2}E} + r}.
\eqn
Thus, $H(r, E) = 0$ has only one real root,
\bq
\lb{5.9}
r_{H} = \left(\frac{4\ell^{2}E}{3}\right)^{1/2},
\eq
at which we find that
\bq
\lb{5.10}
H(r_{H}, E) =  \frac{12\ell E}{4 \ell^{2}E + 3r^{2}_{H}} > 0. 
\eq
Eqs.~(\ref{CT}) and (\ref{delta}) then tell us that the surface $r = r_{H}$ is a horizon for a test particle with energy $E$. It should be noted that, in contrast to the Schwarzschild case  [cf. Eq.~(\ref{root})],  $r_{H}$ now is proportional to $E$, that is, the higher the energy of the test particle, the lager  the radius of the horizon. To understand
this, let us consider the acceleration of a test particle with its four-velocity $u_{\lambda} = - \delta^{t}_{r}$,  located on a surface $r$. Then, we find that
\bq
\lb{acc}
a_{\mu} \equiv u_{\mu;\lambda} u^{\lambda} = 
\begin{cases}
- \frac{m}{r^{2}}\delta^{r}_{\mu}, & Schwarzschild,\\
\frac{r}{\ell^{2}}\delta^{r}_{\mu}, & de Sitter.\\
\end{cases}
\eq
That is, for the Schwarzschild solution, the test particle feels an attractive force, while for the de Sitter solution, it feels a repulsive one. Because of this difference,  in the 
de Sitter spacetime $r_{H}$ is proportional to $E$, in contrast to the Schwarzschild one, where it is inversely proportional to $E$, as shown explicitly in Eq.~(\ref{root}).

\subsection{$m > 0, \;\;\; \Lambda =0$}

When $\Lambda = 0$ and $m > 0$, it is the Schwarzschild solution. In particular, in the IR, the surface $r = 2m$ represents a
horizon, while for high energy particles, the radius of the horizon is energy-dependent, as explicitly given by Eq.~(\ref{root}) for $n = 2$.
So, we shall not repeat these studies, but simply note that
now the solution takes the form,
\bq
\lb{5.11}
ds^{2} = - dt^{2} + \left(dr - \sqrt{\frac{2m}{r}}dt\right)^{2} + r^{2}d\Omega^{2},
\eq
which is singular only at $r = 0$, as can be seen from Eq.~(\ref{5.1b}). So, it already represents a maximal spacetime
in the HL theory. 

It is interesting to note that the above metric covers only  half  of the maximally extended spacetime given in GR. This can be seen easily by introducing 
the coordinate $\tau$~\cite{GPW},
\bqn
\lb{5.12}
\tau & \equiv& t - \int{\frac{\sqrt{2mr}}{r - 2m} dr}\nb\\
&=& t - 2\sqrt{2mr}  - 2m \ln\left(\frac{r - 2m}{\left(\sqrt{r} + \sqrt{2m}\right)^{2}}\right),~~~
\eqn
in terms of which, the solution takes the standard Schwarzschild form,
$ds^{2} = - f(r)d\tau^{2} + f^{-1}(r)dr^{2} + r^{2}d\Omega^{2}$ with $f(r) = 1 - 2m/r$. Of course, the above transformations are forbidden by Eq.~(\ref{1.2}).

 \subsection{$m > 0, \;\;\; \Lambda \not= 0$}
 
 In this case, it is convenient to further distinguish the two subcases $\Lambda > 0$ and $\Lambda < 0$.
 
 \subsubsection{$\Lambda > 0.$} In this case, the metric takes the form, 
 \bq
 \lb{5.13a}
 ds^{2} = -dt^{2} + \left(dr - \sqrt{\frac{2m}{r} + \frac{r^{2}}{\ell^{2}}} \;dt\right)^{2} + r^{2}d\Omega^{2}.
 \eq
 
 When $E \ll M_{*}$, as in the last case the dispersion relation becomes relativistic, and $F(\zeta) \simeq \zeta$, for which we have $n = 1$. Then, we find that
 \bqn
 \lb{5.13}
 H(r)  &=& 1 + N^{r} = 1 - \sqrt{\frac{2m}{r} + \frac{r^{2}}{\ell^{2}}} \nb\\
 &=& \frac{F(r)}{\ell^{2}\left(1 + \sqrt{\frac{2m}{r} + \frac{r^{2}}{\ell^{2}}}\right)},
 \eqn
 but now $F(r) \equiv - \left(r^{3} - \ell^{2}r + 2m \ell^{2}\right)$. Clearly, $F(r)$ has one maximum and one minimum, respectively, at $r = \pm r_{m}$, where
 $r_{m} =  \ell/\sqrt{3}$ and $F(r_{m}) = -2\ell^{2}(m - 1/(3\sqrt{\Lambda})$, as shown in Fig. \ref{fig7}.
 Thus, when $ m^{2} > 1/(9\Lambda^{2})$, $H(r) = 0$ has no real positive root, and a horizon does not exist even in the IR. Therefore, the singularity at $r = 0$ is
 naked. When $ m^{2} < 1/(9\Lambda^{2})$, $H(r) = 0$ has two real and positive roots,  $r_{\pm},\; (r_{+} > r_{-})$, where $r = r_{+}$ is often referred to as
 the cosmological horizon and $r = r_{-}$ the black hole event horizon~\cite{GH}. When $m^{2} = 1/(9\Lambda^{2})$, the two horizons coincide. In GR, the 
 corresponding Penrose diagrams were given in~\cite{GH}. However, as argued above, in the HL theory these diagrams are not allowed, as they are obtained
 by coordinate  transformations that violate the restricted diffeomorphisms (\ref{1.2}). Nevertheless, since the metric is not singular in the current form, it already 
 represents a maximal  spacetime. 
 
 When $E \gg M_{*}$, the high momentum terms dominate, and for $n = 2$, we find that
  \bqn
 \lb{5.14}
 X(r)  &=&  \frac{2E}{\sqrt{\frac{2m}{r} + \frac{r^{2}}{\ell^{2}}+ 4E} + \sqrt{\frac{2m}{r} + \frac{r^{2}}{\ell^{2}}}},  \nb\\
H(r) &=& \sqrt{\frac{2m}{r} + \frac{r^{2}}{\ell^{2}}} -  2X = \frac{F(r)}{\Delta(r)},
 \eqn
 where $\Delta(r) > 0$ for any $r \in (0, \infty)$, and $F(r) \equiv r^{3} - 4E\ell^{2}r/3 + 2m \ell^{2}$. 
 It can be shown that when $ m^{2} > 8\ell E^{3/2}/27$,  $H(r) = 0$ has no real and positive roots. Thus, in this case there are no  horizons, and the singularity at
 $r = 0$ must be naked.  When $ m^{2} < 8\ell E^{3/2}/27$,  $H(r) = 0$ has two real and positive roots, say, $r_{1, 2} \; (r_{2} > r_{1})$, but now $r_{1, 2} = r_{1,2}(E, m, \ell)$.
 Thus,  in this case there also exists two horizons, but each of them depends on $E$.  When $ m^{2} = 8\ell E^{3/2}/27$, we have $r_{1} = r_{2}$, and the two horizons 
 coincide.

\subsubsection{$\Lambda < 0.$} In this case, the metric takes the form, 
 \bq
 \lb{5.15}
 ds^{2} = -dt^{2} + \left(dr - \sqrt{\frac{2m}{r} - \frac{r^{2}}{\ell^{2}}} \;dt\right)^{2} + r^{2}d\Omega^{2},
 \eq
 where $\ell \equiv \sqrt{3/|\Lambda|}$. Then, from Eq.~(\ref{5.1b}), it can be seen that the spacetime is singular at
 $r_{s} \equiv (2m\ell^{2})^{1/3}$~\cite{CW}. This is different from GR, in which the only singularity of the anti-de Sitter Schwarzschild solution is at $r = 0$. 
 
  When $E \ll M_{*}$, as in the last case the dispersion relation becomes relativistic. Then, we find that
 \bqn
 \lb{5.16}
 H(r)  &=& 1 + N^{r} = 1 - \sqrt{\frac{2m}{r} - \frac{r^{2}}{\ell^{2}}} \nb\\
 &=&  \frac{F(r)}{r\ell^{2}\left(1 + \sqrt{\frac{2m}{r} - \frac{r^{2}}{\ell^{2}}}\right)},
 \eqn
 but now with $F(r) \equiv r^{3} + \ell^{2}r - 2m \ell^{2}$, which is a monotonically increasing function, as shown by Fig. \ref{fig10}. 
 Thus, $H(r) = 0$  has one and only one  real and positive root $r_{H} = r_{H}(m, \ell)$. But, $r_{H}$ is 
 always less than $r_{s}$, i.e., $r_{H} < r_{s}$. Thus, the singularity 
 at $r= r_{s}$ is a naked singularity.
 
  When $E \gg M_{*}$, let us consider only the case $n = 2$. Then,  we find that
   \bqn
 \lb{5.17}
 X(r)  &=&  \frac{2E}{\sqrt{\frac{2m}{r} - \frac{r^{2}}{\ell^{2}}+ 4E} + \sqrt{\frac{2m}{r} - \frac{r^{2}}{\ell^{2}}}},  \nb\\
H(r) &=& \sqrt{\frac{2m}{r} - \frac{r^{2}}{\ell^{2}}} -  2X = \frac{F(r)}{\Delta(r)},
 \eqn
 where $\Delta(r) > 0$ for any $r \in (0, \infty)$, and $F(r) \equiv r^{3} + 4E\ell^{2}r/3 - 2m \ell^{2}$. 
 It can be shown that this $F(r)$ is also a monotonically increasing function, as shown by Fig. \ref{fig10}, and $F(r) = 0$ has only one real and positive root,
  $r_{H} = r_{H}(m, E, \ell)$. Again, since $H(r_{s}) = 1$ and $H(r_{H}) = 0$, we find that $r_{H}$ is 
 also always less than $r_{s}$,  although now  $r_{H}$ depends on $E$. Thus, the singularity 
 at $r= r_{s}$ is a naked singularity.

\section{Slowly Rotating Vacuum Solutions}
 Slowly rotating vacuum solutions in other versions of the HL theory
have been studied by several authors~\cite{rbhs}. The goal of this section is to derive slowly rotating black hole solutions in the HMT setup. We will seek a solution of the form
\begin{eqnarray}\label{ansatz1}
 && ds^2 = - dt^2 + r^2(d\theta^2 + \sin^2 \theta d\phi^2)
  	\\ \nonumber
&&  + e^{2\nu(r)}\left[dr + e^{\mu(r) - \nu(r)}(dt - a \omega(r) \sin^2\theta d\phi)\right]^2,
\end{eqnarray}
where the functions $\nu(r), \mu(r)$, and $\omega(r)$ are independent of $(t, \theta, \phi)$. By requiring that the metric satisfy the equations to first order in the small rotation parameter $a$, we will be able to determine $\nu, \mu$, and $\omega$. 

The ansatz (\ref{ansatz1}) is motivated by the fact that it agrees with the Kerr solution to first order in $a$. Indeed, the Kerr line element expressed in Doran coordinates~\cite{Doran2000} is given by
\begin{eqnarray}\label{Doran}
  ds_{\text{Kerr}}^2 &=&  -dt^2 + (r^2 + a^2 \cos^2\theta)d\theta^2 	\\ \nonumber
&&   + (r^2 + a^2)\sin^2\theta d\phi^2 + \frac{r^2 + a^2 \cos^2\theta}{r^2 + a^2}
	\\ \nonumber
&& \times \left[dr + \frac{\sqrt{2mr(r^2 + a^2)}}{r^2 + a^2 \cos^2\theta} (dt - a\sin^2\theta d\phi)\right]^2,
\end{eqnarray}
where $m$ and $a$ are parameters. As $a \to 0$, this metric coincides with (\ref{ansatz1}) to first order in the rotation parameter $a$,  provided that
$$\nu(r) = 0, \qquad \mu(r) = \log \sqrt{\frac{2m}{r}}, \qquad \omega(r) = 1.$$
In particular, when $a = 0$, it reduces to the Schwarzschild metric in Painlev\'e-Gullstrand form.

Note that the form (\ref{ansatz1}) of the line element is compatible with the projectability condition $N = N(t)$; its ADM coefficients are
$$N = 1, \qquad N^i = (e^{\mu(r) - \nu(r)}, 0, 0).$$
Working in the gauge $\varphi = 0$, the momentum constraint (\ref{eq2}) for the metric (\ref{ansatz1}) reduces to
\begin{eqnarray} \label{momentumconstraint3}
&&- \frac{2e^{\mu - 3 \nu} \nu'}{r} + O(a^2) = 0,
	\\ \nonumber
&&\frac{a e^{2\mu-2\nu}}{2 r^4} \Bigl[r^2  \left(\omega''+\omega' \left(4 \mu'-\nu'\right)\right)
	\\ \nonumber
&& -2 \omega \bigl(1-r^2  \mu''
+r^2 \mu' \nu'-2r^2  \mu'^2 +r \nu'\bigr)\Bigr]+ O(a^2) = 0,
\end{eqnarray}
while the equation (\ref{eq4b}) obtained from variation with respect to $A$ yields
\begin{equation}\label{varyAeq}  
  \left(1-r^2 \Lambda_g \right) e^{2 \nu}+2 r \nu' - 1 + O(a^2) = 0.
\end{equation}
The first equation in (\ref{momentumconstraint3}) implies that $\nu$ is constant, and then (\ref{varyAeq}) shows that
$$\nu = 0, \qquad \Lambda_g = 0.$$
This yields
\begin{eqnarray}\label{RijLKetc}
&& R_{ij} = O(a^2), 
	\\ \nonumber
&& \mathcal{L}_K = -\frac{2}{r^2}e^{2\mu}(1 + 2r\mu') + O(a^2), 
	\\ \nonumber
&& \mathcal{L}_V = 2\Lambda + O(a^2).
\end{eqnarray}

The ($rr$)-component of the dynamical equations (\ref{eq3}) gives
\bqn
\label{rrdynamical}
&& \frac{2 r A_0' - r^2\Lambda +2 r e^{2 \mu} \mu'+e^{2 \mu}}{r^2} +\frac{2 A_1'}{r} a\nb\\
& & ~~~~~~~~~~~~~~
+\mathcal{O}(a^2) = 0,  
\eqn
where we have assumed that $A(r)$ has the form
\begin{equation}\label{Aansatz}
A(r) = A_0(r) + A_1(r) a + \mathcal{O}(a^2).
\end{equation}
The terms of $\mathcal{O}(1)$ in (\ref{rrdynamical}) imply that
\begin{equation}\label{murexp}
\mu(r) = \frac{1}{2} \ln \left(\frac{2 m}{r}+\frac{1}{3} r^2 \Lambda-2
   A_0(r)+ \frac{2}{r} \int_{r_0}^{r}{A_0(s) \, ds}\right),
   \end{equation}
where $r_0 > 0$ is a constant, while the terms of $\mathcal{O}(a)$ imply that $A_1$ is a constant. With these choices, all the components of the dynamical equations as well as the equations obtained from variation with respect to $A$ and $\varphi$ are satisfied to first order in $a$, and the Hamiltonian constraint (\ref{eq1}) becomes
$$\int_0^\infty r A_0'(r) dr + \mathcal{O}(a^2) = 0.$$
Finally, the second equation in the momentum constraint (\ref{momentumconstraint3}) is satisfied to $\mathcal{O}(a)$ provided that
\bqn
\label{lambdaexpression}  
  \omega(r) &=&e^{-2\mu}\left(\frac{d_1}{r} + d_2 r^2\right)
         \\ \nonumber
& = &\frac{d_1 + d_2r^3}{2m + 2\int_{r_0}^r A_0(s) ds - 2rA_0 + \frac{\Lambda}{3}r^3}, 
 \eqn
where  $ d_1$ and $d_2$  are the integration constants.

In summary, the ansatz (\ref{ansatz1})  gives a solution to first order in $a$ provided that $\mu(r)$ is given by (\ref{murexp}), $\omega(r)$ is given by (\ref{lambdaexpression}), and
\bq\lb{Kerr}
 \nu = 0, \qquad  A(r) = A_0(r) + aA_1 + \mathcal{O}(a^2),
\eq 
where $r_0 >0, m, \Lambda, A_1, d_1, d_2$ are arbitrary constants and $A_0(r)$ can be freely chosen as long as
$$
\int_0^\infty r A_0'(r) dr = 0.
$$
We recover the slowly rotating version of the Kerr solution by taking $A_0 = 0$, $\Lambda =0$, and $d_2 = 0$. Setting $a = 0$, on the other hand, we recover the static solutions obtained
in~\cite{GSW}. 

Let us point out that the standard Einstein equations also allow for a nonzero value of $d_2$ in the slowly rotating limit. Indeed, substituting the ansatz (\ref{ansatz1}) with $\nu = 0$ into the vacuum Einstein equations
$$R_{\alpha\beta} - \frac{1}{2}g_{\alpha\beta} R = 0, \qquad \alpha,\beta = 0, 1,2,3,$$
we find that they are satisfied to order $\mathcal{O}(a)$ if and only if
$$\mu(r) = \frac{1}{2}\ln \left(\frac{2m}{r}\right),$$
where $m > 0$ is a constant, and $\omega(r)$ is given by (\ref{lambdaexpression}) with arbitrary constants $d_1$ and $d_2$.

\section{Conclusions}

 In this chapter, we have systematically studied black holes in the HL theory, using the kinematic method of test particles provided by  KK  in~\cite{KKb}, in which a horizon 
 is defined as the surface at which massless test particles are infinitely redshifted. Because of the nonrelativistic  dispersion relations (\ref{1.4}),  we have shown explicitly the 
 difference between   black holes defined in General Relativity and the ones defined here. In particular, the radius of the horizon  usually depends on the energy of the test particles.

 When applying this definition to the spherically symmetric and static vacuum solutions found recently in~\cite{HMT,AP,GSW},   we 
 have found  that   for test particles with sufficiently high energy, the radius of the horizon can be made arbitrarily small,
 although the singularities at the center can be seen in principle only by test particles with  infinitely high energy. 
 
 We  paid particular attention to the global structures of the static solutions. Because of the restricted diffeomorphisms (\ref{1.4res}), they
 are dramatically different from the corresponding ones given in GR, even  the solutions are the same. In particular, the restricted diffeomorphisms (\ref{1.4res}) do not allow us to draw Penrose
 diagrams, although one can create  something similar to them,  for example, see Figs. \ref{fig1a}, \ref{fig3a}, \ref{fig6}, \ref{fig9a}, \ref{fig11}. 
 But,  it must be noted that, since the speed of the test particles in the HL theory can 
 be infinitely large, the causality in this theory is also dramatically different from that of General Relativity [cf. Fig.\ref{HLcone}]. In particular, the light-cone structure in General Relativity does not apply to the
 HL theory.  Among the static solutions, a very interesting case is the one given by Fig. \ref{fig3a}, which corresponds to an Einstein-Rosen bridge. In GR, this solution
is made of an exotic fluid as one can see from Eq.~(\ref{4.17b}), which is clearly unphysical, and most likely unstable, too. However, in the HL theory, the solution is a vacuum one,
and it would be very interesting to see if this configuration is stable or not in the HMT setup.

 Finally, we have studied the slowly rotating solutions in the HMT setup~\cite{HMT}, and found explicitly all such solutions, which are
  characterized by an arbitrary function $A_{0}(r)$. When   $A_{0} = 0$ they reduce    to the slowly rotating Kerr solution obtained in GR. 
  When the rotation is switched off, they reduce to the static solutions obtained in~\cite{GSW}.

\chapter{Summary and Outlook}
\renewcommand{\theequation}{6.\arabic{equation}} \setcounter{equation}{0}

\section{Summary}

Isaac Newton was the first person to give a mathematical theory of gravity, which he first published in his \textit{Philosophi\ae~Naturalis Principia Mathematica} on July 5th, 1687~\cite{IsaacNewton}. It successfully predicted the return of Halley's comet, and later many planets were discovered using this law of gravitation.  Although there were a few inconsistencies found in the theory toward the end of nineteenth century, no one dared to suggest that a better theory of gravitation was required to account for those discrepancies. After the success of the Special Theory of Relativity, Albert Einstein saw that Newtonian gravity contradicts relativity. After ten years of work using the best mathematical tools of the day, Einstein formulated General Relativity in its final form a century ago on November 25th, 1915. At that time and for many years later, Einstein's theory was better than Newton's only to account for three very tiny effects. 

Einstein's Classical Field Theory is the best description of gravity we have today and it describes the universe at large. However, the structure of bodies themselves are held together by other fundamental interactions in nature, which are best described by Quantum Field Theory. The final stages of gravitational collapse and earliest moments of the birth of the universe require us to have the same description of all the fundamental interactions. This is a daunting task and our current best hope is String Theory. The heart of the problem is to formulate gravity as a quantum field theory and this problem is often simply called quantum gravity. Although at such high energies one cannot separate pure gravity contributions from that of other interactions, it is still a big step forward just to quantize pure gravity. Many such efforts in this direction are Loop Quantum Gravity, Causal Dynamical Triangulations, Asymptotic Safety and the latest Ho\v{r}ava-Lifshitz gravity which is investigated in this dissertation.  

Ho\v{r}ava-Lifshitz Theory of quantum gravity is motivated by the Lifshitz theory in solid state physics.  One of the essential ingredients of the theory is the inclusion of higher-dimensional spatial derivative operators which dominate in the ultraviolet, making the theory power-counting renormalizable. The exclusion of higher-dimensional time derivative operators, on the other hand, guarantees that the theory is unitary. The problem of non-unitarity has plagued the quantization of gravity for a long time. However, this asymmetrical treatment of the space and time variables inevitably leads to the breaking of Lorentz and Diffeomorphism symmetries. 

We studied spherically symmetric static spacetimes generally filled with anisotropic fluid. We found the vacuum solutions and used them to understand the solar system tests. We found that the theory is consistent with solar system tests, by properly choosing the parameter involved in the theory. We considered the gravitational collapse of spherical fluid and derived the junction conditions across the collapsing star under the assumption that they be mathematically meaningful in terms of distribution theory. We  systematically studied black holes in the Ho\v{r}ava-Lifshitz Theory by following  the kinematic approach, in which a horizon is defined as the surface at which  massless test particles are infinitely redshifted. Because of the nonrelativistic  dispersion relations, the speed of particles is unlimited, and test particles do not follow geodesics. As a   result, there are  significant differences in causal structures and black holes between General Relativity and HL theory. In particular, the horizon radii generically depend on the energies of test particles.  By applying them to the spherical static vacuum solutions, we find that, for test particles with sufficiently high energy, the radius of the horizon can be made as small as desired, although the singularities can be seen in principle only by observers with  infinitely high energy. In these studies, we pay  particular attention to the global structure of the solutions, and find that, because of the foliation-preserving diffeomorphism symmetry, they are quite different from the corresponding ones given in GR, even though the solutions are the same. 

\section{Outlook}

So far the studies show that Ho\v{r}ava-Lifshitz Theory is not ruled out by any experiment and at the same time, it has made predictions that deviate from GR. Of the four possible models, the original version with projectability is not favored by observations. On the other hand, the non-projectable version with $U(1)$ symmetry proposed by Zhu, Wu, Wang and Shu~\cite{ZSWW, ZWWS} is free of theoretical inconsistencies and compatible with all the observations and in addition, it is embedded into String Theory, which was the original purpose of the HL theory.

Some of the important questions concerning the foundational aspects of the theory are worth pursuing. They are the following: 
\begin{description}
\item[] Renormalizability: Although the theory is shown to be power-counting renormalizable, establishing  full renormalizability has not yet been accomplished. 

\item[] Infrared limit: One should rigorously show that HL theory reduces to General Relativity in the low energy limit using renormalization group flow study.

\item[] Lorentz violating effects: It is interesting to study how the Lorentz violations in the gravitational sector propagate to the matter sector and if they have any observable effects at low energies. This problem might also help to address the question of coupling matter to the theory.
\end{description}

Here is the list of some important problems that might point us to future directions for the development of the theory. 
\begin{description}

\item[] Quantization: Since HL theory is supposed to be a candidate for quantum gravity, it is natural to quantize the theory. This task has already been taken up for some special cases in~\cite{Anderson:2011bj, Li:2014bla}. It will be very interesting to study the graviton-graviton scattering problem and compare it with the results obtained in the case of General Relativity by DeWitt. 

\item[] Black hole entropy: One obstacle to studying black entropy in HL theory was the lack of proper understanding of the black hole horizons, as the theory allows superluminal particles that can escape from the Killing horizon. This has been remedied with the discovery of the Universal Horizon~\cite{Blas:2011ni, Lin:2014ija, Lin:2014eaa}. Now the next step is to compute the black hole entropy and see how it compares with the standard results of Bekenstein and Hawking~\cite{Berglund:2012bu, Berglund:2012fk}. 
 
\end{description}


\end{document}